\documentclass[acmtog]{acmart}

\usepackage{booktabs} 
\usepackage{mathtools,stmaryrd,nicefrac}
\usepackage{ifthen,comment,xspace}
\usepackage{overpic}
\usepackage{amsmath,bm}
\usepackage{pifont}
\usepackage{amsfonts}
\usepackage{subcaption}
\usepackage{multirow}
\usepackage{xspace} 

\usepackage[ruled]{algorithm2e} 

\AtBeginDocument{%
  \providecommand\BibTeX{{%
    \normalfont B\kern-0.5em{\scshape i\kern-0.25em b}\kern-0.8em\TeX}}}

\begin{document}
\newcommand{\todo}[1]{{\color{red}[TODO: #1]}}
\newcommand{\revise}[1]{{\color{blue}#1}}
\newcommand{\zilin}[1]{{\color{blue}#1}}
\newcommand{\dd}{\,\mathrm{d}}

\newcommand{\btf}{\mathbf{BTF}}
\newcommand{\bH}{\mathbf{H}}
\newcommand{\bU}{\mathbf{U}}
\newcommand{\bD}{\mathbf{D}}

\newcommand{\bx}{\mathbf{x}}
\newcommand{\by}{\mathbf{y}}
\newcommand{\bz}{\mathbf{z}}
\newcommand{\bh}{\mathbf{h}}
\newcommand{\bd}{\mathbf{d}}

\newcommand{\R}{\mathbb{R}}
\newcommand{\wi}{\mathbf{\bm{\omega}_i}}
\newcommand{\wo}{\mathbf{\bm{\omega}_o}}
\newcommand{\uv}{\mathbf{u}}

\newcommand{\FLIP}{\protect\reflectbox{F}LIP\xspace}

\title{A Dynamic By-example BTF Synthesis Scheme}


\author{Zilin Xu}
\affiliation{%
 \institution{University of California, Santa Barbara}
 \country{USA}
 \postcode{93106}
}
\email{zilinxu@ucsb.edu}

\author{Zahra Montazeri}
\affiliation{%
 \institution{ University of Manchester}
 \country{United Kingdom}
}
\email{zahra.montazeri@manchester.ac.uk}

\author{Beibei Wang}
\affiliation{%
 \institution{Nanjing University}
 \country{China}
}
\email{beibei.wang@nju.edu.cn}

\author{Ling-Qi Yan}
\affiliation{%
 \institution{University of California, Santa Barbara}
 \country{USA}
  \postcode{93106}
}
\email{lingqi@cs.ucsb.edu}

\begin{abstract}
Measured Bidirectional Texture Function (BTF) can faithfully reproduce a realistic appearance but is costly to acquire and store due to its 6D nature (2D spatial and 4D angular). Therefore, it is practical and necessary for rendering to synthesize BTFs from a small example patch. While previous methods managed to produce plausible results, we find that they seldomly take into consideration the property of being \emph{dynamic}, so a BTF must be synthesized before the rendering process, resulting in limited size, costly pre-generation and storage issues. In this paper, we propose a dynamic BTF synthesis scheme, where a BTF at any position only needs to be synthesized when being queried. Our insight is that, with the recent advances in neural dimension reduction methods, a BTF can be decomposed into disjoint low-dimensional components. We can perform dynamic synthesis only on the positional dimensions, and during rendering, recover the BTF by querying and combining these low-dimensional functions with the help of a lightweight Multilayer Perceptron (MLP). Consequently, we obtain a fully dynamic 6D BTF synthesis scheme that does not require any pre-generation, which enables efficient rendering of our infinitely large and non-repetitive BTFs on the fly.  We demonstrate the effectiveness of our method through various types of BTFs taken from UBO2014~\cite{weinmann:2014:ubo14}.

\end{abstract}




\begin{teaserfigure}
\addtolength{\tabcolsep}{-5.0pt}
    \begin{tabular}{ccccc}
        \begin{overpic}[width=0.25\columnwidth]{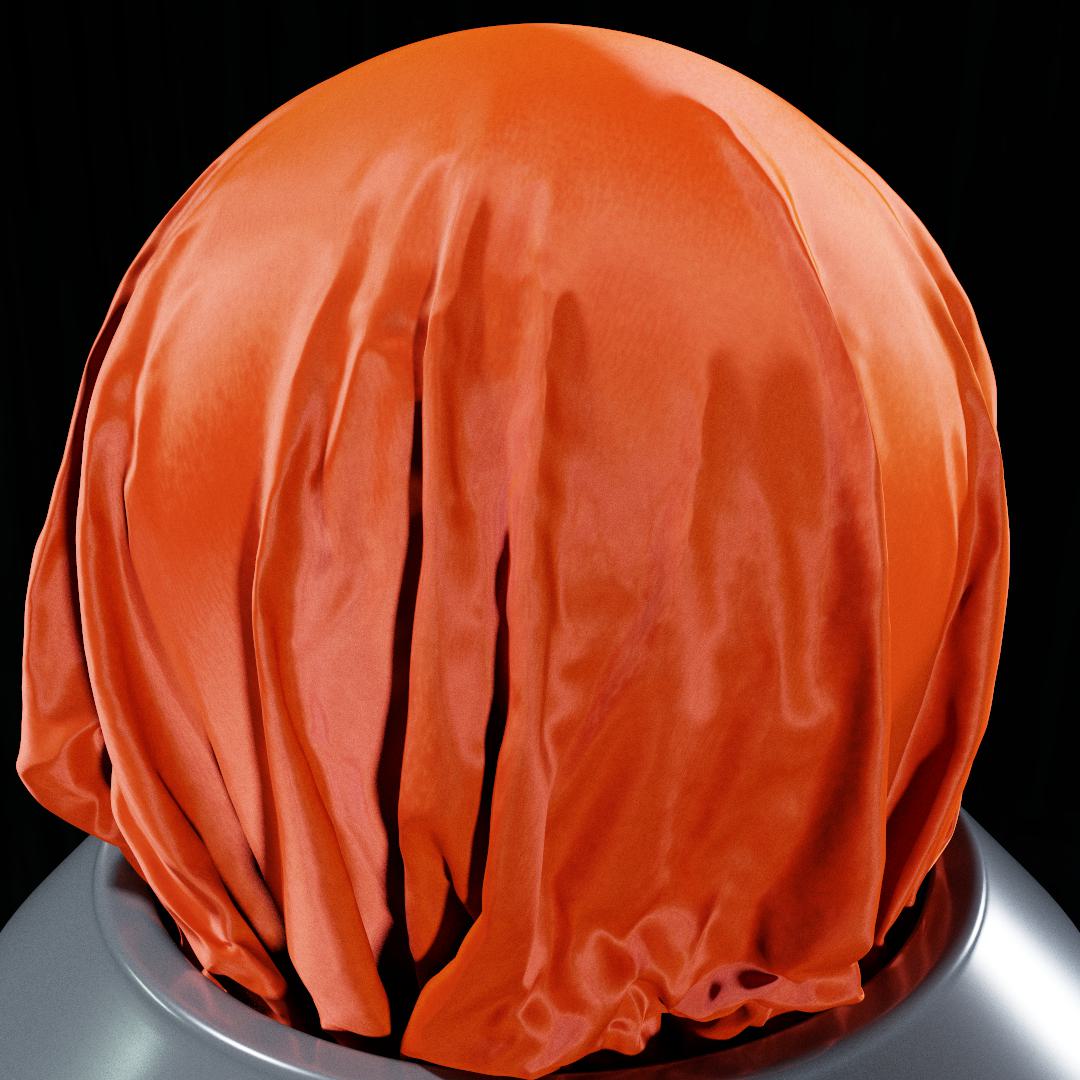}
        \put(2, 2){\color{white}\large{\textsc{Fabric07}}}
        \end{overpic} &
        \begin{overpic}[width=0.25\columnwidth]{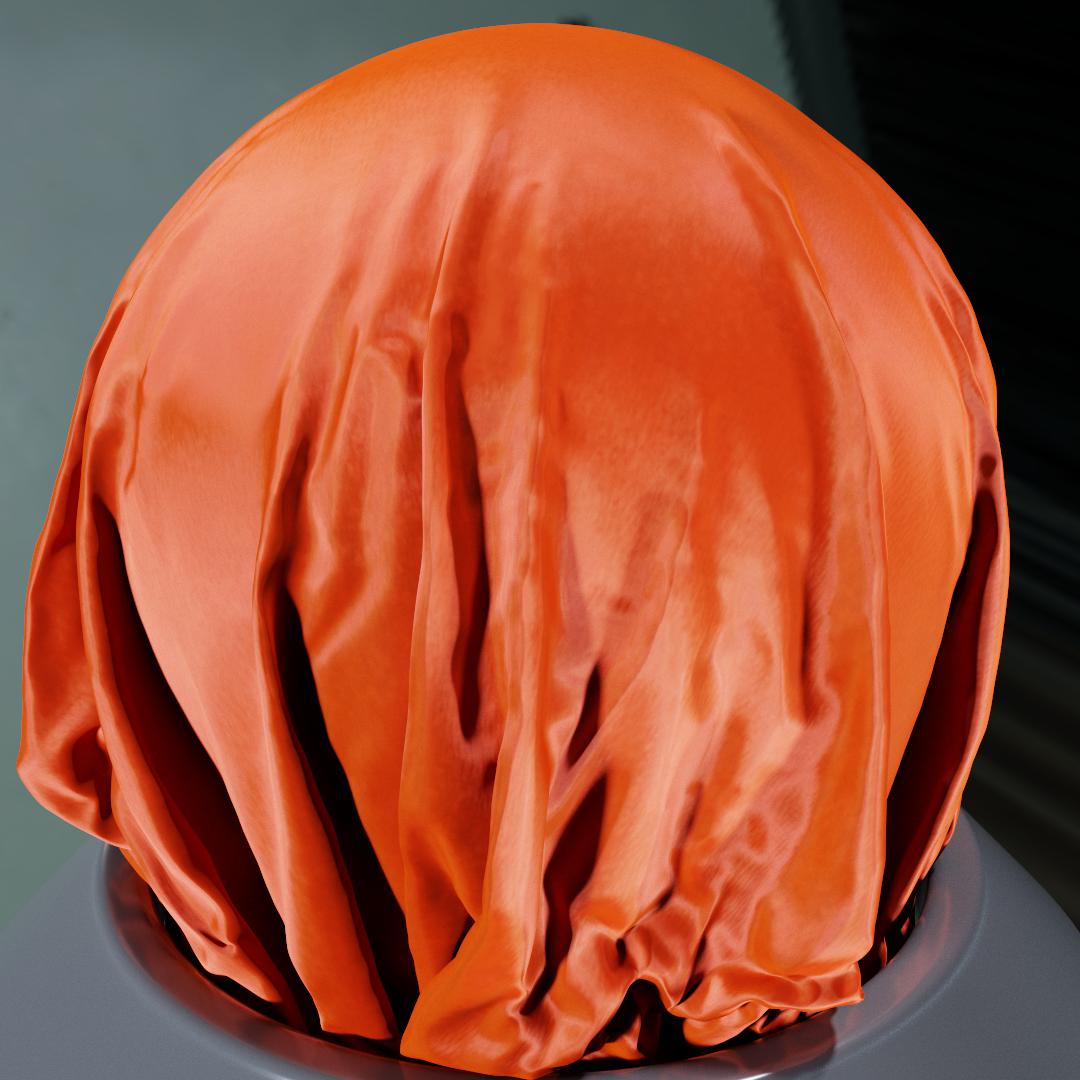}
        \put(79, 79){\includegraphics[width=0.05\columnwidth]{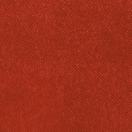}}
        \put(-23, 92){\color{white}\large{Strong Sheen}}
        \end{overpic}  &
         \, &
        \begin{overpic}[width=0.25\columnwidth]{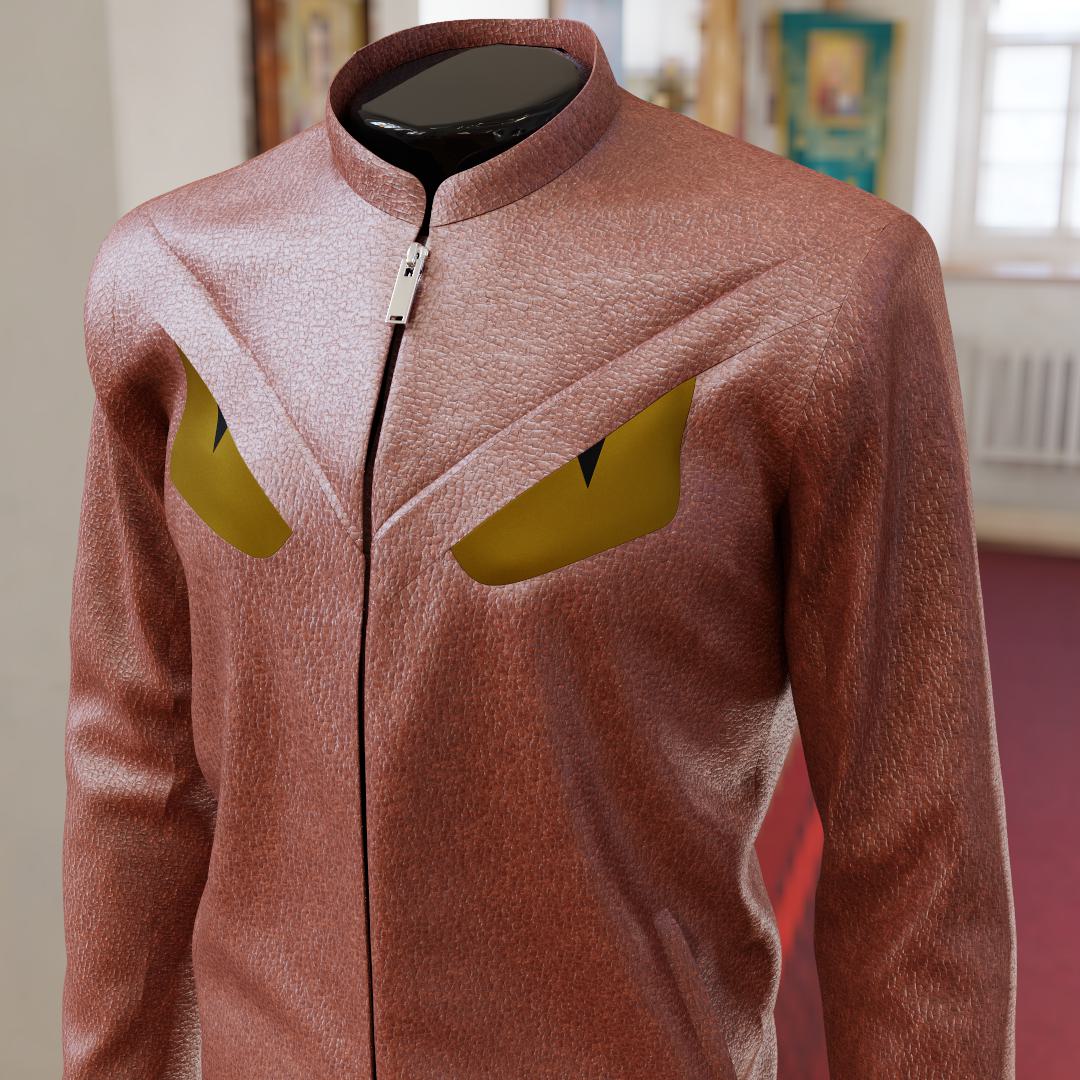}
        \put(2, 2){\color{white}\large{\textsc{Leather11}}}
        \end{overpic} &
        \begin{overpic}[width=0.25\columnwidth]{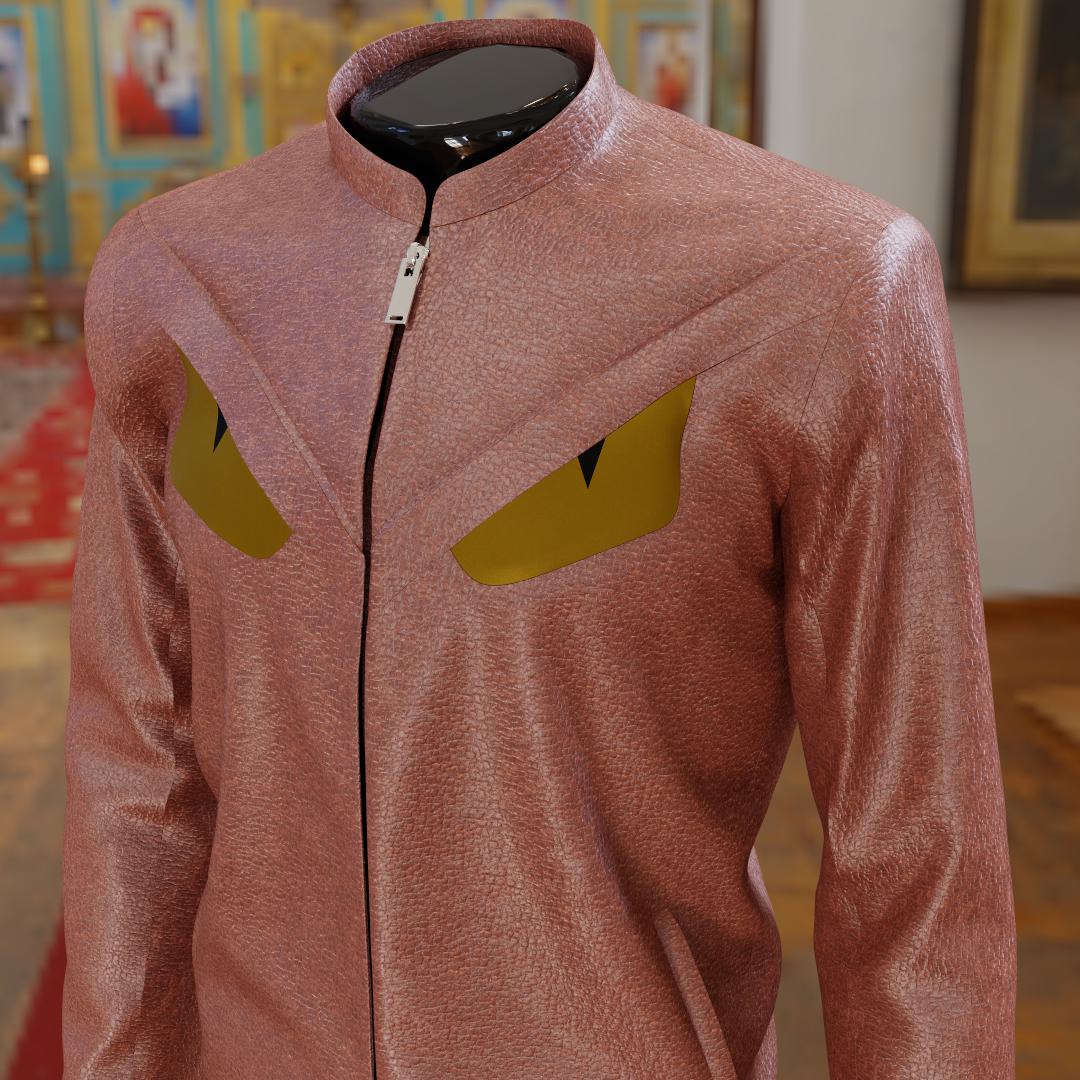}
        \put(79, 79){\includegraphics[width=0.05\columnwidth]{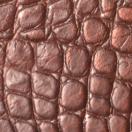}}
        \put(-30, 92){\color{white}\large{Glos}\color{black}\large{sy \& } \color{white}\large{Complex Pattern}}
        \end{overpic} 
        \\
        \begin{overpic}[width=0.25\columnwidth]{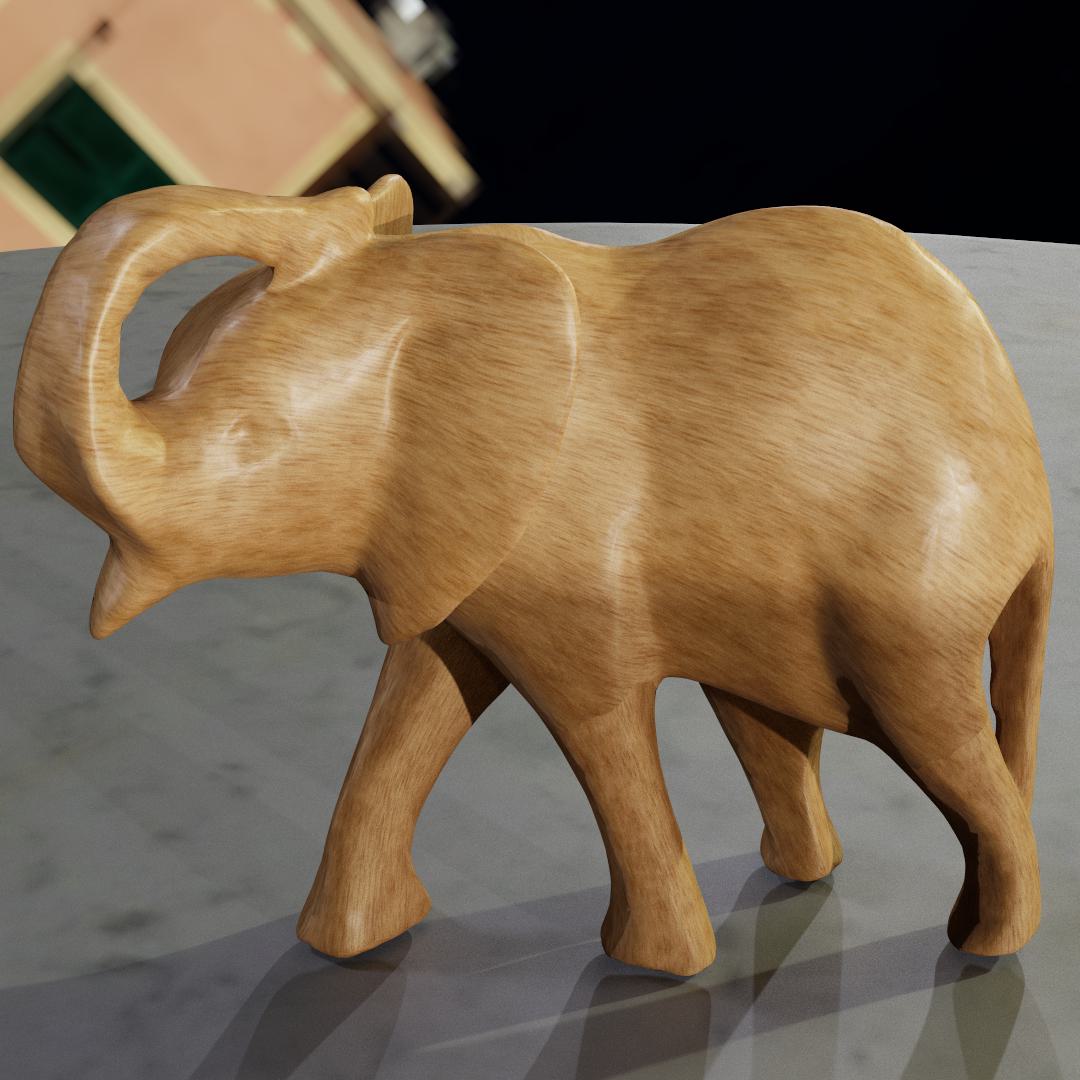}
        \put(2, 2){\color{white}\large{\textsc{Wood06}}}
        \end{overpic} &
        \begin{overpic}[width=0.25\columnwidth]{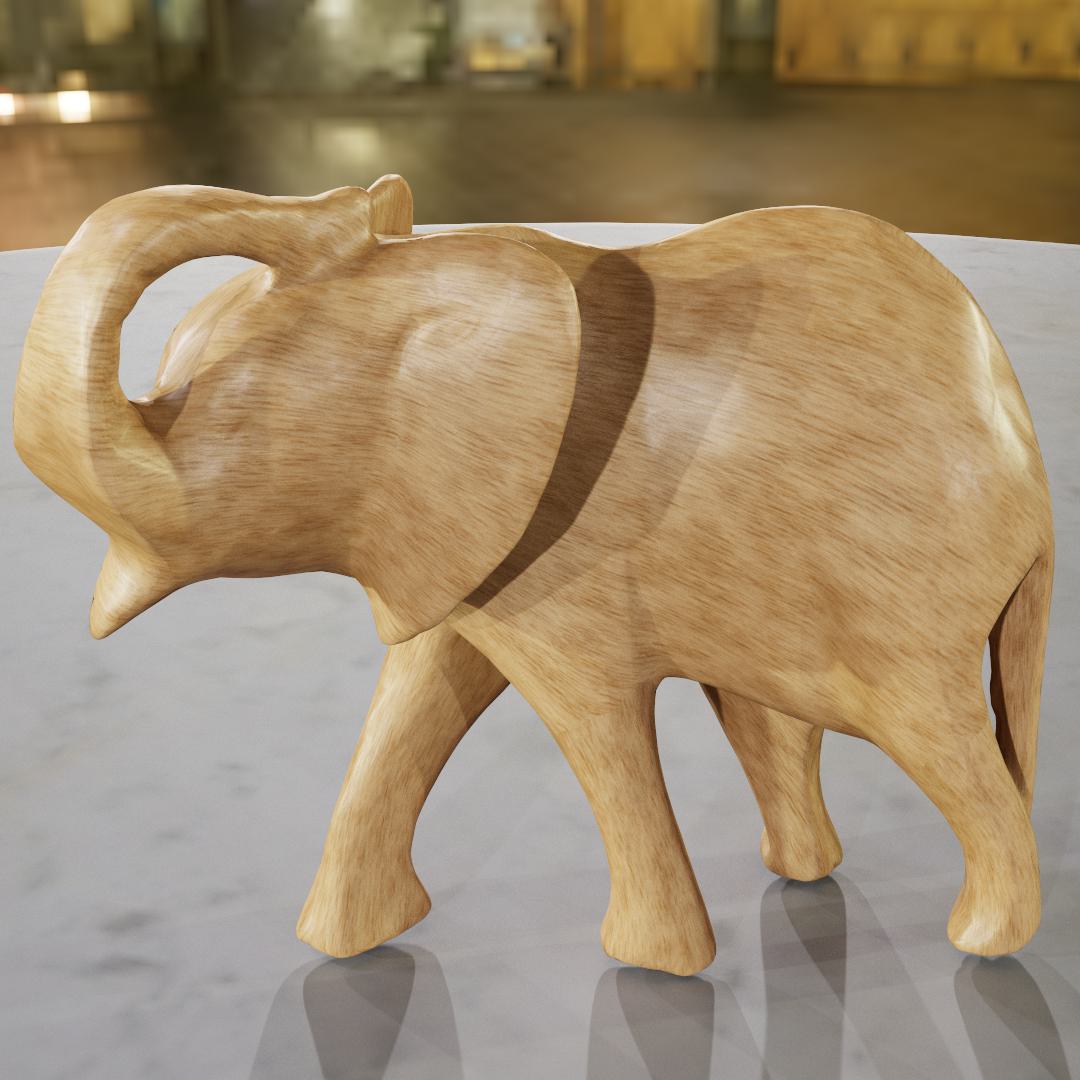}
        \put(79, 79){\includegraphics[width=0.05\columnwidth]{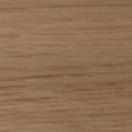}}
        \put(-23, 92){\color{white}\large{Strong Specular}}
        \end{overpic} & \, &
        \begin{overpic}[width=0.25\columnwidth]{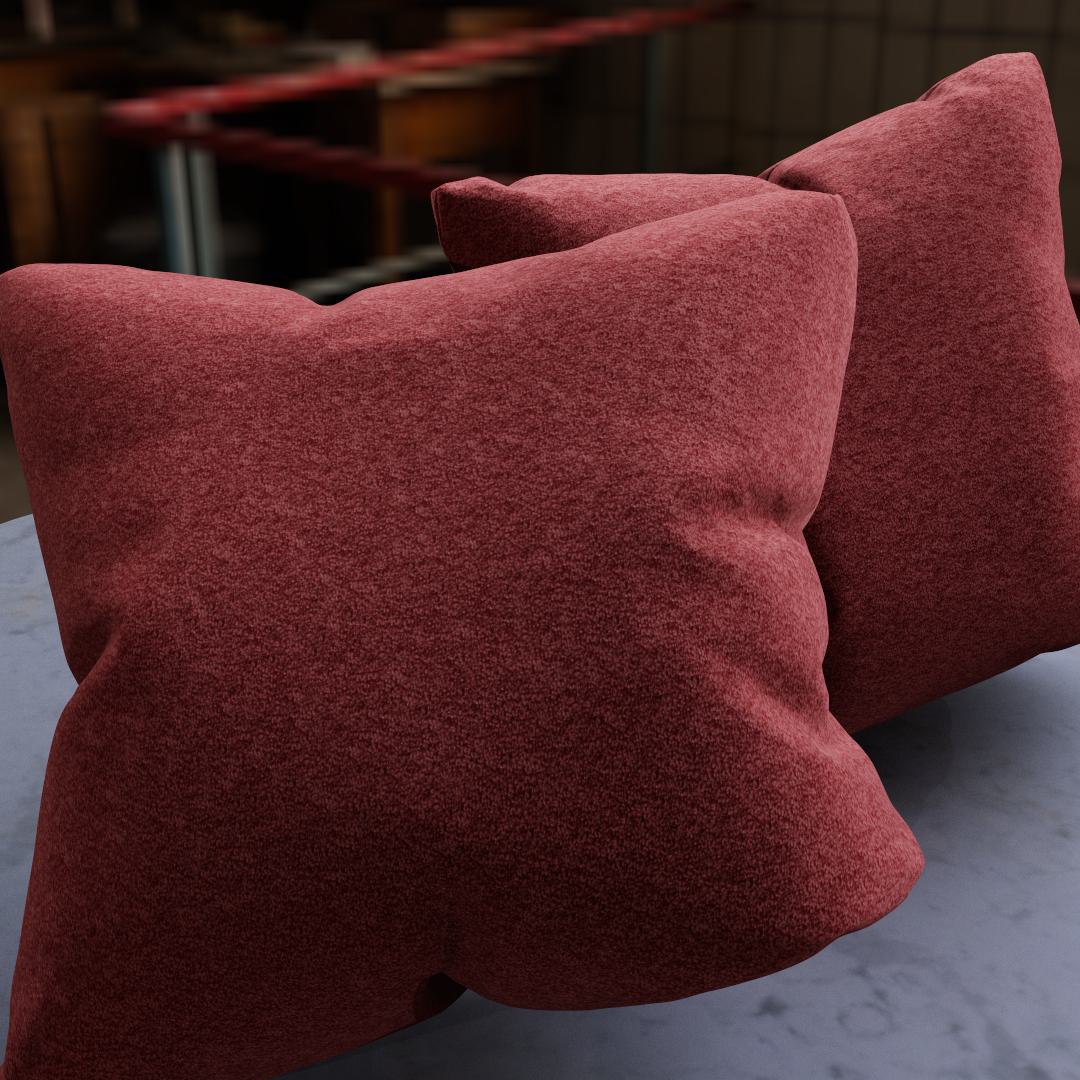}
        \put(2, 2){\color{white}\large{\textsc{Carpet02}}}
        \end{overpic} &
        \begin{overpic}[width=0.25\columnwidth]{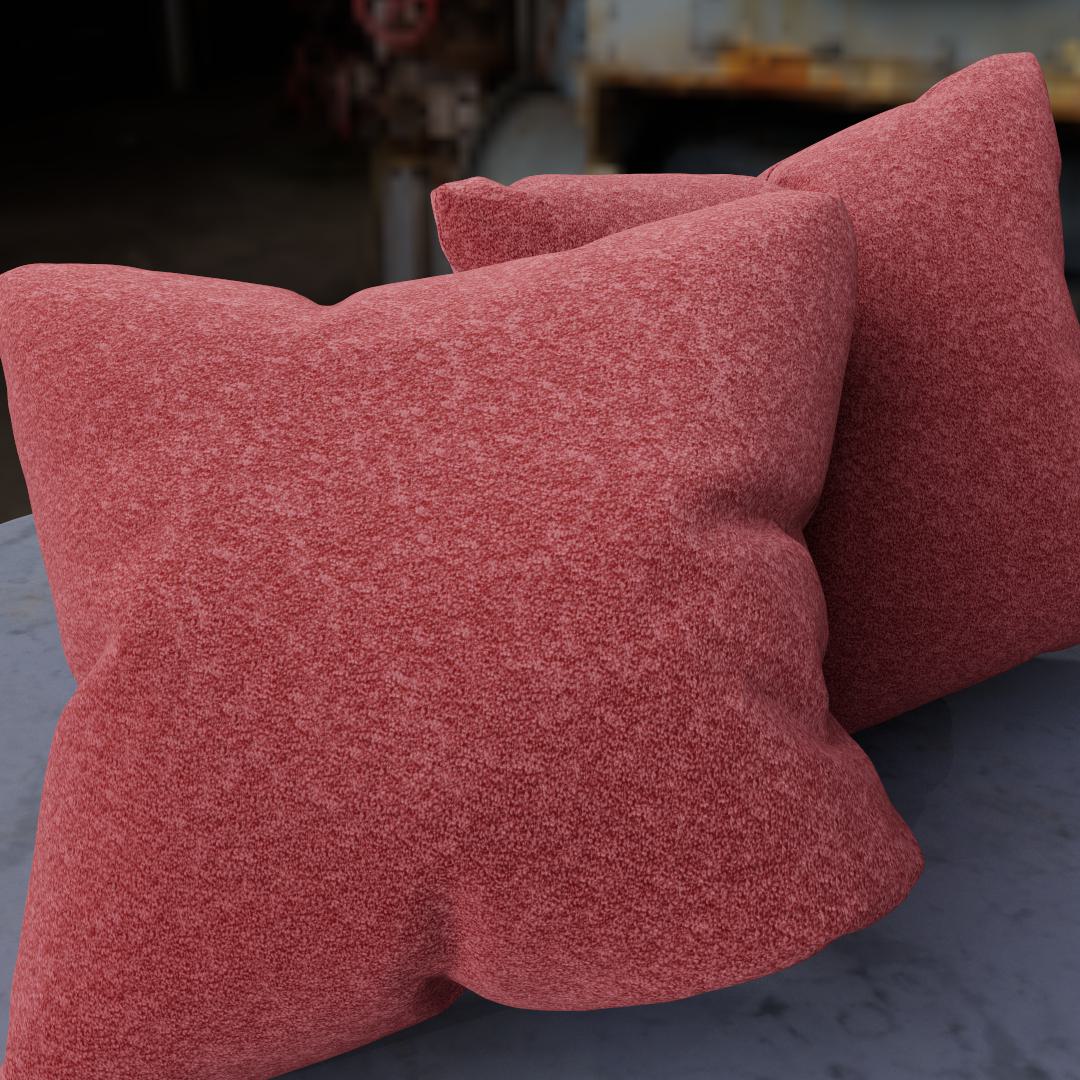}
        \put(79, 79){\includegraphics[width=0.05\columnwidth]{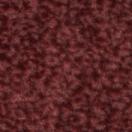}}
        \put(-14, 92){\color{white}\large{Fine Details}}
        \end{overpic} 
	\end{tabular}
  \caption{We propose a novel by-example BTF synthesis scheme that can dynamically synthesize a non-repetitive, infinitely large BTF from a small example BTF ($400 \times 400$). 
  To demonstrate our method's effectiveness in both synthesis and rendering, we select four representative example BTFs (shown at the top-right corners) with unique visual effects from UBO2014~\cite{weinmann:2014:ubo14}.
  Each of them uses our Triple Plane with histogram-preserving blending~\cite{heitz:2018:histo} and is rendered under two different lighting conditions. Our method managed to synthesize a non-repetitive BTF while faithfully capturing the complex visual effects of each BTF.
  }
  \Description{BTF Synthesis.}
  \label{fig:teaser}
\end{teaserfigure}

\maketitle
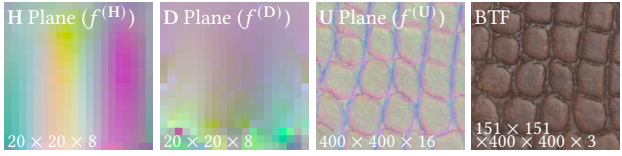
\begin{figure}[t]
	\centering
	\addtolength{\tabcolsep}{-3.5pt}
	\small
	\begin{tabular}{cccc}
    	\begin{overpic}[width=0.23\columnwidth]{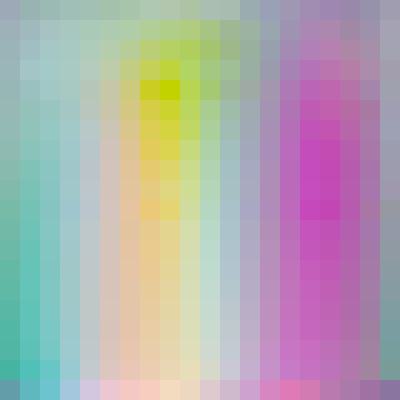}
            \put(2,2){\color{white}\footnotesize{$20 \times 20 \times 8$}}
            \put(2,85){\color{white}$\bH$ Plane ($ f^{(\bH)}$)}
            \end{overpic} &
            \begin{overpic}[width=0.23\columnwidth]{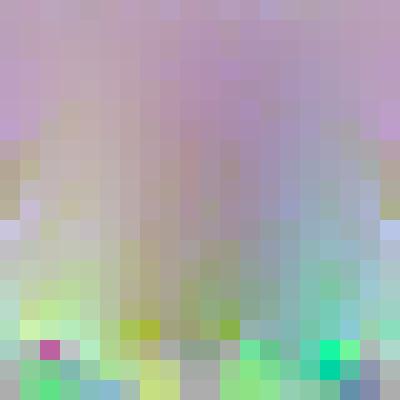}
            \put(2,2){\color{white}\footnotesize{$20 \times 20 \times 8$}}
            \put(2,85){\color{white}$\bD$ Plane ($ f^{(\bD)}$)}
            \end{overpic} &
            \begin{overpic}[width=0.23\columnwidth]{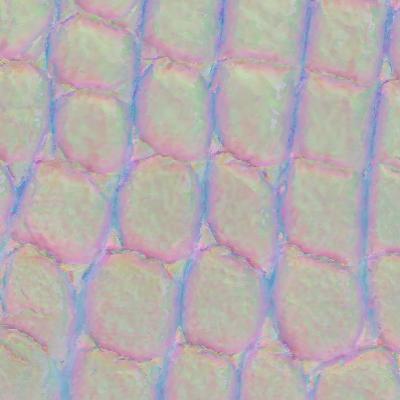}
            \put(2,2){\color{white}\footnotesize{$400 \times 400 \times 16$}}
            \put(2,85){\color{white}$\bU$ Plane ($ f^{(\bU)}$)}
            \end{overpic} &
            \begin{overpic}[width=0.23\columnwidth]{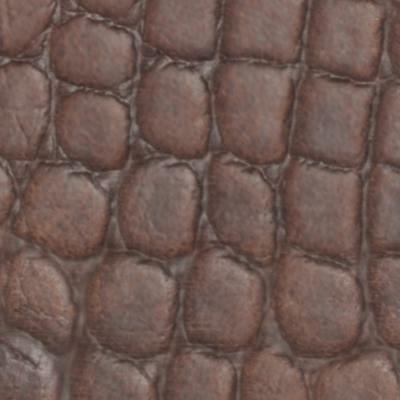}
             \put(2,85){\color{white}BTF}
            \put(2,11){\color{white}\footnotesize{$151\times 151$}}
            \put(2,2){\color{white}\footnotesize{$\times 400 \times 400 \times3$}}
            \end{overpic} 
	\end{tabular}
          \vspace{-3mm}
	\caption{\label{fig:visual}
        The direct visualization of our feature planes.
        The positional feature plane shows highly semantic characteristics with detailed structures and variation information close to the BTF. The resolution of each feature plane follows Biplane~\cite{fan:2023:BTF}'s setting. 
	}
        \vspace{-3mm}
\end{figure}

\section{Introduction}
\label{sec:intro}
Photorealistic rendering places great emphasis on accurately reproducing the appearance of real-world objects.
In general, the appearance of a virtual object's surface is established by a Bidirectional Reflectance Distribution Function (BRDF). The BRDFs can be empirical, analytical, data-driven, or hybrid.
While the empirical and analytical BRDF models can produce a plausible appearance, they are essentially simulations of real-world appearance and may disregard certain intricate visual effects. 
The data-driven appearance models are usually derived from direct measurements of a piece of a real-world object, resulting in measured Spatially Varying Bidirectional Reflectance Distribution Functions (SVBRDFs) or Bidirectional Texture Functions (BTFs) --- the faithful replications of the real-world appearance. 
However, the data-driven materials demand a substantial amount of measured data since the SVBRDFs or BTFs are typically 6D functions with 2D spatial and 4D angular, creating difficulties in both data acquisition and utilization during rendering. Therefore, leveraging dimension reduction or decomposition is essential.
Nevertheless, due to the high-dimensional nature of materials and the common assumption of far-field illumination during the measurement process, the feasible measurement area is restricted to a small patch (e.g., $400 \times 400$ texels) of the example object, as the exhaustive measurement on an entire object (e.g., a garment) is hard to accomplish.
Applying a material with only such a small example patch in rendering will produce undesirable spatial repetitive patterns despite its high fidelity.
Therefore, a faithful BTF synthesis scheme is important to practically bring the real-world appearance into rendering. 

Previous work on material synthesis often involves creating node graphs of procedural materials, or generating G-buffers/2D parametric maps of the BRDF models (usually based on microfacet models), creating an analytical BTF representation (e.g., MATch~\cite{Shi:2020:MATch}). 
These methods can be efficient, but still \emph{synthetic}, 
thus inherently inaccurate compared to the measured BTFs because the real-world appearance has a geometric and reflectance complexity that is difficult to describe using analytic BRDFs, let alone that some of those BRDFs are further simplified (uniform, isotropic, etc.) such as  PhotoMat~\cite{zhou:2023:PhotoMat}.
Moreover, those synthesis methods are usually coupled with reconstruction --- they first \emph{estimate} the appearance from some inputs, commonly one or a few photographs taken by a cellphone with a flashlight. Consequently, their quality is further compromised, often losing positional details and angular resolution.
Therefore, they are unsuitable for achieving accurate appearances compared to properly using the measured BTFs.

Many BTF synthesis methods have been proposed in the past decades to generate a large BTF from a small measured BTF patch. These methods (e.g., Tong et al.~\shortcite{Tong:2002:btfsyn} and Koudelka et al.~\shortcite{koudelka:2003:acquisition}) may produce a plausible synthesis result with a better-preserved structure due to the quilting-based (i.e., similar to texture quilting~\cite{Efros:2001:quilting}) synthesis. However, they rely on highly compressed BTF representation such as principal component analysis (PCA) and Spherical Harmonics (SH), which introduces significant quality loss. Moreover, to the best of our knowledge, all of these methods are \emph{non-dynamic} and cannot be tiled infinitely. The large BTF must be first generated to a proper size (e.g., $4K \times 4K$) and then be used in rendering process.
Even with their dimension reduction methods (e.g., PCA) applying to the BTF, pre-generating a very large BTF on the compressed low-dimension domain and then storing it is still not practical.
That is why a dynamic property is in high demand, yet it has been ignored in past decades.

In this paper, we challenge the problem of \emph{dynamically} synthesizing an infinitely large BTF without spatial repetition from a given example of measured BTF. 
Our insight is that with the advance of previous dimension reduction methods, it is possible to decompose a 6D BTF into the outer product ($\circ$) of 2D neural feature planes (one positional and two directional planes, possibly multi-channel). We call this decomposition approach \textit{Triple Plane}\footnote{We name it Triple Plane to distinguish from the Triplane method~\cite{chan:2022:efficient} that uses three intersecting planes to represent a 3D function instead of our three disjoint 2D planes that decompose a 6D BTF.}. 
After the decomposition, we find the decomposed 2D positional neural feature plane still exhibiting semantic information that has a similar visual appearance to the BTF, as shown in Fig.~\ref{fig:visual}, which inspires us to perform texture synthesis in the position domain \emph{as if they are textures}. In this way, the synthesized BTF can be recovered from the synthesized positional feature and the direction features via a lightweight Multilayer Perceptron (MLP). With the analysis, we reduce the BTF synthesis task into two key sub-tasks: a faithful decomposition capturing the full 6D BTF without losing details, and a synthesis method that is able to preserve content and allow dynamic queries, i.e., query without pre-generating a large texture, to steer away from heavy storage and runtime rendering cost. 

\begin{figure}[t]
	\centering
        \includegraphics[width=0.4\textwidth]{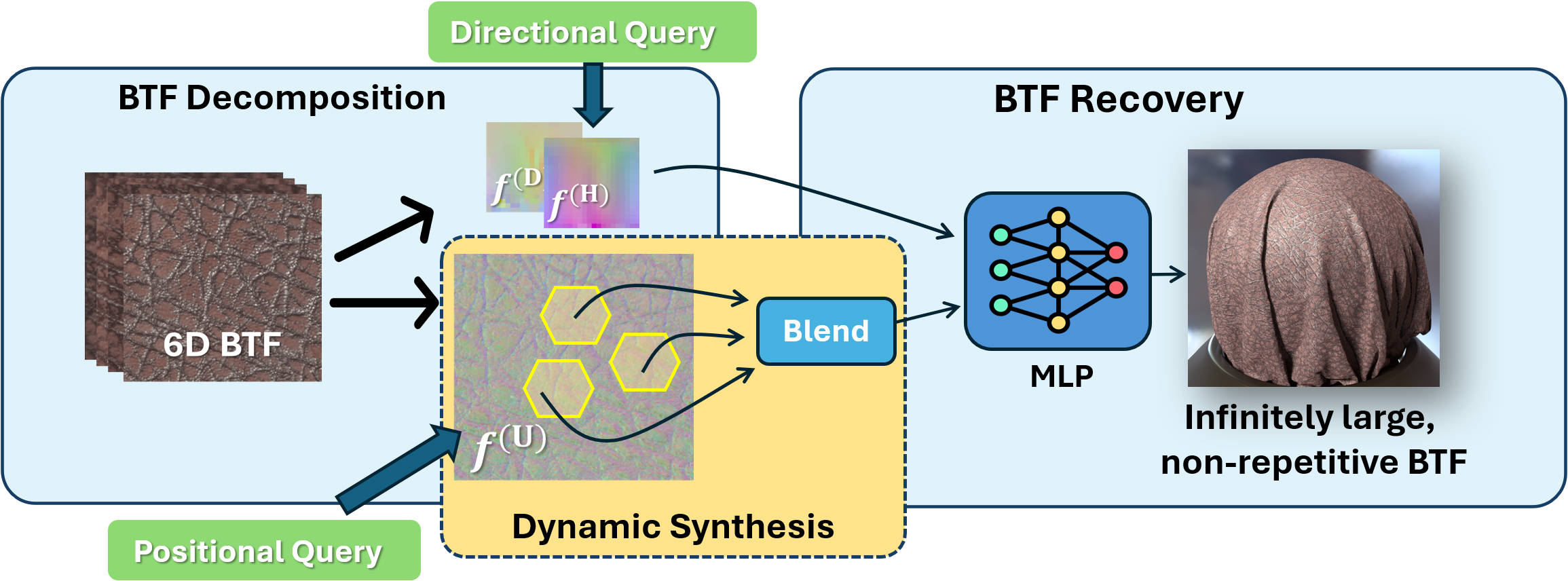}
         \vspace{-3mm}
	\caption{\label{fig:pipeline}
        Our method's pipeline. We first decompose 6D BTF into three 2D planes ($f^{(\bU)}, f^{(\bH)}$ and $f^{(\bD)}$), then perform dynamic synthesis on the positional plane as if it is a 2D texture. The synthesized reflectance is recovered via a lightweight MLP with the input of positional and directional features.	}
         \vspace{-3mm}
\end{figure}

Consequently, we are able to essentially generate infinitely large, non-repetitive BTF that can be directly queried during rendering without any storage overhead --- all we need as input is just a measured BTF example patch. 
Our scheme is general and compatible with both different neural-based BTF representation methods and texture synthesis (we specifically focus on dynamic by-example synthesis).
Fig.~\ref{fig:pipeline} shows the overall pipeline of our scheme, and in Sec.~\ref{sec:res}, we demonstrate the high-quality BTFs we synthesized on a variety of objects with a fast, lightweight MLP. It demonstrates not only the novelty of our idea to be the first of its kind to dynamically generate BTFs on a decomposed low-dimension but also the practicality of our scheme.
In summary, our main contributions are as follows:

\begin{itemize}
    \item a novel, simple but effective approach that enables dynamic, infinitely large, and non-repetitive BTF synthesis,
    \item a general BTF synthesis scheme that is compatible with different neural dimensional decomposition techniques and different texture synthesis methods, and,
    \item a Triple Plane approach for BTF representation that decomposes the 6D BTF into neural outer products of three 2D functions and achieves high fidelity in real-time performance.
\end{itemize}

\section{Related Work}

\subsection{BTF Acquisition}
BTF can be accurately measured through specialized equipment that traverses and measures each pair of the incident and outgoing angles individually (very precise, but producing a large amount of data), or it can be roughly estimated from one or a few photographs using neural networks.
Neural-based BTF acquisition methods have become popular in recent years because of their convenience. 
However, the quality of these methods is far from comparable to measured BTFs since they are either capturing 4D isotropic BTF~\cite{zhou:2023:PhotoMat} or estimating 2D parametric maps (e.g., normal, albedo, etc.) for the analytical appearance model~\cite{guo:2020:MatGAN,zhou:2022:tilegen,deschaintre:2018:single,gao:2019:deep,guo:2021:highlight,henzler:2021:generative}.
To achieve a highly realistic appearance, utilizing the measured BTFs is essential. 
The first measured BTF dataset (CUReT) was built by Dana et al.~\shortcite{dana:1999:reflectance}.
There are several open-source measured BTF datasets and UBO2014~\cite{weinmann:2014:ubo14} is the most commonly used one with $400 \times 400$ spatial and $151 \times 151$ angular resolution.
Unfortunately, due to the high-dimensional nature of BTF and the physical limitations of the measurement technique, the measured BTF is restricted in small size, making it unsuitable for many scenarios, as it would produce unsatisfactory repetitive tiling patterns on objects. 
To overcome this limitation, we propose a novel BTF synthesis approach, which generates an infinitely large without repetition from only a small example patch.
\vspace{-1mm}

\subsection{BTF Compression \& Dimension Decomposition}
Due to the high dimensionality of BTF, dimension decomposition methods like Principal Component Analysis (PCA) and Tensor Decomposition are essential in compressing BTF. 
There is a substantial amount of work attempting to apply PCA~\cite{koudelka:2003:acquisition,ruiters:2009:parallelized,guthe:2009:btf} and tensor decomposition~\cite{vasilescu:2004:tensortextures, Ruiters:2009:BTFTensorDecom} for BTF compression.
With the rapid development of deep learning, neural compression became popular for material representation~\cite{hu:2020:deepbrdf,zheng:2021:compact, Fan:2022:NLBRDF} because they frequently outperformed classic methods. 
Rainer et al.~\shortcite{Rainer:2019:Neural} proposed the first neural BTF compression method with an auto-encoder architecture. Later, Rainer et al.~\shortcite{Rainer:2020:Unified} introduce a unified network that projects different BTF onto a shared latent space.

The aforementioned works are only designed to compress the BTF. From another perspective, neural dimension decomposition methods that expose the 2D positional neural texture are more effective for BTF representation recently.
NeuMIP~\cite{kuznetsov:2021:neumip} and its later extended works~\cite{kuznetsov:2022:neumip2, xue2024hierarchical} can also be used for BTF compression. They have single or multiple neural textures progressively learned during training.
The closest work to us is Biplane~\cite{fan:2023:BTF}, which decomposes BTF into a 2D positional plane and a 2D half-vector plane. With a conditional input of a 2D difference vector, it can recover the reflectance of different BTF from a universal but large MLP.  
However, none of these methods completely decomposed the directional and position dimensions, i.e., the direction as a condition input to the positional feature instead of an independent feature.
That means the directional information is implicitly mixed and stored with the positional features (or latent vectors). 
In this paper, we focus on neural dimension decomposition instead of the BTF compression methods since it can expose the 2D positional semantic information, which can be subsequently used for BTF synthesis. To achieve effective and efficient BTF decomposition, we made a modification (we call it a Triple Plane) on the top of Biplane~\cite{fan:2023:BTF} that fully separates the positional and directional dimensions and leverages a lightweight MLP to recover the reflectance for one particular BTF.
\vspace{-1mm}

\subsection{Texture Synthesis}
\label{subsec:texturesyn}
In many applications, there is a need to bind textures to large-scale surfaces, e.g., the entire surface of a mountain. However, textures have a limited size of representational scale that is not typically large, even for $4K$ textures.
Directly tiling them will cause an undesirable repetitive pattern. Therefore, the texture synthesis methods are proposed to solve this problem.

Quilting and optimization based texture synthesis methods~\cite{efros:1999:texture,Efros:2001:quilting,kwatra:2005:texture,barnes:2009:patchmatch,kaspar:2015:self} commonly produce a plausible result but involve an offline process that finds the best-matched candidate patches (or texels) from the original texture by neighborhood searching or iterative optimization. It is like "growing" the texture from the synthesized area to the unsynthesized area region by region. Thus, they usually have difficulty supporting dynamic queries, and the new texture must be synthesized before use. As mentioned in Sec.~\ref{sec:intro}, the dynamic property is more desirable and practical for BTF. Therefore, we don't overly focus on such methods. In this paper, we refer to any synthesis method, including Wang tiling~\cite{wang:1961:tiling}, that needs neighboring information for synthesis as a quilting-based method for simplicity.

In another aspect, dynamic by-example texture synthesis methods~\cite{heitz:2018:histo,mikkelsen:2022:hextile} are designed to generate a (commonly infinitely large) new texture from a small texture patch called ``example'' or ``exemplar'' on the fly. New texture at any place can be instantly queried without pre-generation. It typically involves a random selection step that selects some texture patches from the example texture and a blending step that blends those example patches together. However, due to their fully dynamic design, they have difficulty achieving the same high quality as quilting-based methods.

Thanks to our dimensional decomposition design for the BTFs, in an orthogonal approach, we leverage the previous texture synthesis method to synthesize BTFs on the position domain. 
To achieve dynamic property, by-example texture synthesis methods are our preferred approach, although quilting-based methods may produce better synthesized BTFs. Note that both quilting-based and by-example synthesis have their own advantages and disadvantages. We do not aim to improve the previous texture synthesis work, but we expect to find some synthesis methods that have spatial variations and do not strictly repeat, from which our method can be further improved.

\subsection{BTF synthesis}
\label{subsec:btfsyn}

Many BTF synthesis methods have been developed in the past decades. Since BTF is a 6D function, people usually first compress its angular dimension and then perform synthesis. 
Previous BTF synthesis methods can mainly divided into two categories. The first is the quilting-based methods ~\cite{Liu:2001:btfSyn,Tong:2002:btfsyn,koudelka:2003:acquisition,zhou:2005:decorating,kawasaki:2005:patch,Lefebvre:2006:appearance,Ruiters:2013:BTFInterpolation,steinhausen:2015:extrapolation}, which rely on the texture quilting which we introduced in the Sec.~\ref{subsec:texturesyn}. The second is tiling-based methods~\cite{Leung:2007:tileableBTF,zhang:2008:fast} that leverage the idea of Wang Tiling~\cite{wang:1961:tiling}. Apart from them, NeuBTF~\cite{rodriguez:2023:neubtf} is a neural method that synthesizes BTF with a given guide map.

Those methods may produce spatially good results, but in order to support BTF synthesis, they rely on highly compressed BTF representation, e.g., PCA~\cite{koudelka:2003:acquisition,zhou:2005:decorating,zhang:2008:fast, Ruiters:2013:BTFInterpolation,steinhausen:2015:extrapolation} or Spherical Harmonics (SH)~\cite{kawasaki:2005:patch, Leung:2007:tileableBTF,Lefebvre:2006:appearance} which are well-known to lose the high-frequency in angular domain.
Moreover, none of these methods is \emph{dynamic}, and it is impossible for them to generate an infinitely large texture on the fly.

\section{Method}
\label{sec:overview}
\subsection{Problem Analysis}
Our objective is to propose an effective scheme for \emph{dynamic}, \emph{infinitely large} BTF synthesis, where the BTF is generated only upon query. The immediate challenge is clear: BTFs are inherently 6D functions, thus occupying a large amount of storage. Together with the complexity of measuring accurate BTFs, it is usually difficult to acquire BTFs with high spatial resolutions. Nevertheless, high-resolution BTFs are required to realistically depict not only large-scale objects, such as an entire terrain, but also small objects with abundant detail, such as leather, cloth and wood. Therefore, synthesizing high-resolution BTFs is necessary.

The key to synthesizing BTFs, as we emphasize throughout this paper, is the property of being \emph{dynamic}. Consider a toy example: given a small texture of resolution $512^2$ to start with, how to find a texel's value at a very large coordinate, such as $(100000, 100000)$, on the synthesized texture. Those methods in Sec.~\ref{subsec:btfsyn}, either by quilting patches or growing pixels, must compute from around the small example to that large coordinate, then answer this question. In rendering, such a query can happen per shader thread. Therefore, the entire BTF must be synthesized and stored prior to rendering\footnote{..., unless one can afford to repeat the same query-after-synthesis process for each thread, which is impossible in practice, no matter how fast previous methods perform. Therefore, it further distinguishes the concepts between dynamic and fast/real-time.}. This is already an issue for large textures. For BTFs, this issue is further magnified, since even with heavy compression (thus significant quality loss), large BTFs still suffer more than large textures. As a result, what we need is the ability to query an arbitrary location on a synthesized BTF but without actually synthesizing it beforehand, a.k.a., the property of being dynamic.

However, dynamic synthesis only existed in 2D textures so far. Therefore, our insight is to reduce the BTF synthesis process into texture synthesis tasks, so it can benefit from the advancement of dynamic synthesis. Following this idea, we propose our full scheme that decomposes a BTF into three disjoint 2D functions, synthesizes in the 2D position domain dynamically, and reconstructs the 6D BTF. Therefore, our method supports plugging in different methods for BTF compression (as long as 2D positional semantic information is exposed, see Sec.~\ref{sec:res:compBTF}) and texture synthesis (even including those non-dynamic, e.g.,~\cite{Efros:2001:quilting}). We clarify that we do not intend to resolve these methods' own disadvantages, but we do evaluate combinations of representative methods comparatively to determine how they fit/benefit our general scheme, as shown in Fig.~\ref{fig:comp_syn}.

\subsection{Theoretical Formulation}

To achieve the aforementioned goal, we have set out to accomplish three tasks: constructing a faithful dimension decomposition method for BTF which decomposes the 6D BTF into a 2D positional function and two 2D angular functions; performing texture synthesis on the position domain based on an input spatial query; finally recovering the new 6D BTF from the synthesized function and the angular functions. Fig.~\ref{fig:pipeline} shows our scheme's overall pipeline.

\paragraph{BTF \& Parameterization} A $\btf(\cdot)$ is a 6D function representing the reflectance (usually a RGB value) of the surface with an input of 2D spatial coordinates $\uv \in \R^2 $ and a pair of 2D incident (viewing) and outgoing (lighting) directions ($\wi \in \R^2$, $\wo \in \R^2$):
\begin{equation}
\btf(\uv, \wi, \wo).
\end{equation}
It can be considered as a texture at which each texel is storing a 4D BRDF. 
Comparing to incident and outgoing directions, it is better to reparameterized the BTFs with Rusinkiewicz parameterization~\cite{Rusinkiewicz:2001:rusinkiewiczCoordinate} which describes the angular dimension using the half vector $\bh \in \R^2$ of ($\wi$, $\wo$) and a difference vector $\bd \in \R^2$ between the outgoing direction $\wo$ and half vector $\bh$:
\begin{equation}
\btf(\uv, \bh, \bd).
\end{equation}
\paragraph{Dimension Decomposition}
Due to the 6D nature of BTFs, storing even a small patch of BTF requires substantial memory space. Utilizing compression methods, e.g., PCA is essential in reducing runtime memory usage.
Our target is to find a way to decompose the 6D BTF into a conceptual ``outer product'' ($\circ$) of three independent 2D functions $\{f^{(i)} \vert i \in (\bU, \bH, \bD)\}$ without losing details:
\begin{equation}
\btf(\uv, \bh, \bd) = f^{(\bU)}(\uv) \circ f^{(\bH)}(\bh) \circ f^{(\bD)}(\bd).
\end{equation}
However, finding such a purely mathematical solution is difficult. Inspired by the recent advances in dimension reduction methods, we turn to seek out a data-driven approach to achieve our goal.
The ``outer product'' of those 2D functions is achieved by a neural operator $\mathcal{N}(\cdot)$:
\begin{equation}
\label{eq:neuOp}
\btf(\uv, \bh, \bd) = \mathcal{N}(f^{(\bU)}(\uv), f^{(\bH)}(\bh), f^{(\bD)}(\bd)).
\end{equation}
Generally, dimensional decomposition methods like Tensor Decomposition aim to decompose the high-dimensional functions into 1Ds. But we are decomposing the 6D BTF into a series of 2D functions. It brings convenience in performing synthesis on the position domain, which will be introduced next.

\paragraph{By-example Synthesis}
We noticed that the decomposed positional feature plane exhibits a highly semantic characteristic with detailed structures and variation information close to the BTF, as shown in Fig.~\ref{fig:visual}. 
It gives us the idea to perform BTF synthesis directly on the positional domain via texture synthesis methods.
There are numerous available texture synthesis methods, among them, the by-example texture synthesis methods fit our goal best, which allows dynamic query and synthesis without pre-generation.
Specifically, with a target query position ($\uv^* \in \R^2$, typically from an infinitely large planar domain), the by-example texture synthesis process mainly involves two steps: finding multiple example patches from the example texture $f$ corresponding to the target query; blending those example patches while keeping some statistic properties (e.g., histogram~\cite{heitz:2018:histo}) of the example texture to obtain the synthesized texture $f^*$. Generally, the texture synthesis method can be formulated as follows:
\begin{equation}
f^*(\uv^*) =  \mathbf{Syn}(f, \uv^*).
\end{equation}
The synthesis function $\mathbf{Syn}(\cdot)$ only corresponds to the example texture $f$ and the target query $\uv^*$. Notably, it does not rely on the local neighborhood of the target query, eliminating the need for 'voting' from previously synthesized areas. This approach differs from quilting-based methods, which are not dynamic; however, it does not prevent the use of such methods in our framework. We assume the positional feature plane can be treated as a 2D texture:
\begin{equation}
f^{*(\bU)}(\uv^*) = \mathbf{Syn}(f^{(\bU)}, \uv^*).
\end{equation}
\paragraph{BTF Recovery}
After first decomposing 6D BTF into a series of 2D functions and performing synthesis in the position domain, the next thing is to recover the 6D function from the 2D functions. As illustrated in Eqn.~\ref{eq:neuOp}. We train a lightweight MLP as the neural operator to obtain the reflectance of the synthesized BTF:
\begin{equation}
\btf^*(\uv^*, \bh, \bd) = \mathcal{N}(f^{*(\bU)}(\uv^*), f^{(\bH)}(\bh), f^{(\bD)}(\bd)).
\end{equation}
Moreover, a general model is unnecessary in our case, so one MLP and the corresponding feature planes are only responsible for one specific BTF.

\subsection{Implementation}
\label{sec:method}
\paragraph{BTF Decomposition}
As described in Sec.~\ref{sec:overview}, one of our goals is to decompose 6D BTF function into an outer product of three 2D functions. As our method is designed to be general to neural dimensional decomposition methods, we modified Biplane~\cite{fan:2023:BTF} with a 4-layer lightweight MLP and fully decomposed 2D functions $\{f^{(i)} \vert i \in (\bU, \bH, \bD)\}$.
Those 2D functions are obtained by initializing three 2D discrete feature planes with learnable parameters. 
The specific values in the planes are progressively learned during training. 
This idea is inspired by Biplane~\cite{fan:2023:BTF}, but we decompose 6D BTF completely into three 2D functions, and we call it a \emph{Triple Plane}. 

In practice, given a 6D query of $(\uv, \wi, \wo)$, we first convert it onto Rusinkiewicz coordinates system~\cite{Rusinkiewicz:2001:rusinkiewiczCoordinate}: $(\uv, \bh, \bd)$.
Then, we use them as texture coordinates to bilinearly fetch features from each corresponding plane.

\paragraph{By-example BTF Synthesis}
As visualized in Fig.~\ref{fig:visual}, the decomposed positional plane shows a highly semantic characteristic with detailed structures and variation information close to the original BTF.
We found that directly extending the texture-based by-example synthesis methods to our positional feature "texture", i.e., feature plane $f^{(\mathbf{U})}$ produces a plausible result. 
Our by-example BTF synthesis scheme shows strong compatibility with different texture synthesis methods, which we will demonstrate in Sec.~\ref{sec:res}. 
The specific synthesis method used can be very flexible.
We emphasize the histogram-preserving blending~\cite{heitz:2018:histo} since we found our method can also benefit from preserving the histogram of the feature plane, though the feature plane is in a latent space. 

\paragraph{BTF Recovery}
Finally, to obtain the BTF reflectance from the decomposed feature planes, we employ a lightweight MLP $\mathcal{N}$ with $4$ layers of $32$ hidden nodes (except the last outputting layer): 
\begin{equation}
\btf(\uv, \bh, \bd) = \mathcal{N}(f^{(\bU)}(\uv), f^{(\bH)}(\bh), f^{(\bD)}(\bd)),
\end{equation}
where $f^{(\bU)}$ can be the synthesized $f^{*(\bU)}$. For simplicity, we do not strictly distinguish between these two terms. 
The feature planes and the lightweight MLP are jointly trained with a simple $\ell_1$ loss:  
\begin{equation}
\begin{split}
\mathcal{L} = & \ell_1(\mathcal{N}(f^{(\bU)}(\uv), f^{(\bH)}(\bh), f^{(\bD)}(\bd)), \btf(\uv, \bh, \bd)).
\end{split}
\end{equation}
The BTF synthesis is not performed during training. We expect our method to support synthesis automatically.

For more detailed implementation information, please refer to supplementary material.

\section{Results}
\label{sec:res}

Our proposed method introduces a general BTF synthesis scheme that allows flexibility in selecting each component, including neural BTF representation and texture synthesis. To illustrate this versatility, we first validate each component independently by comparing various methods. Subsequently, we integrate these components to assess the overall effectiveness of our dynamic BTF synthesis approach. For the neural BTF representation, we implement a modified NeuMIP~\cite{kuznetsov:2021:neumip} without the offset module and only use the finest layer of the feature pyramid to demonstrate the advance of our Triple Plane. As for the texture synthesis, we compare histogram-preserving blending~\cite{heitz:2018:histo}, another dynamic by-example synthesis method Hex-Tiling~\cite{mikkelsen:2022:hextile} and a non-dynamic texture quilting~\cite{Efros:2001:quilting}. Unless specifically stated, the term "ours" refers to the Triple Plane with histogram-preserving blending (if performing synthesis).

\subsection{Validation and Comparison of BTF Representation}
\begin{figure}[t]
	\centering
	\addtolength{\tabcolsep}{-3.5pt}
	\small
	\begin{tabular}{ccccc}
        \raisebox{0.35in}{\rotatebox[origin=c]{90}{\textsc{Wood06}}} &
    	\begin{overpic}[width=0.23\columnwidth]{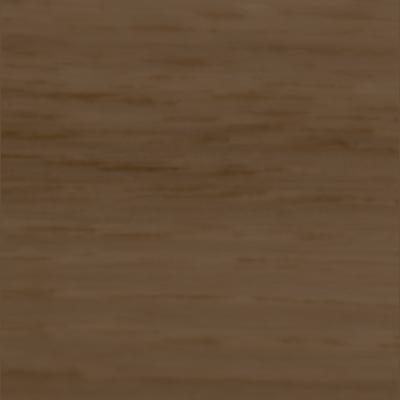}
            \put(2, 2){\color{white}\small{0.00725}}
            \end{overpic} &
            \begin{overpic}[width=0.23\columnwidth]{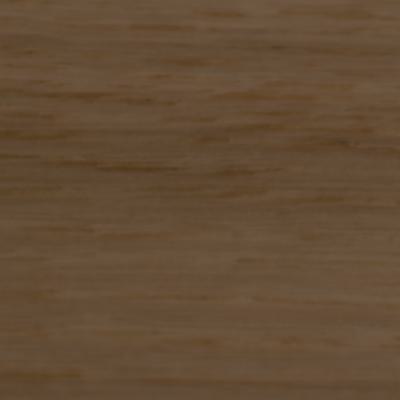}
            \end{overpic} &
            \begin{overpic}[width=0.23\columnwidth]{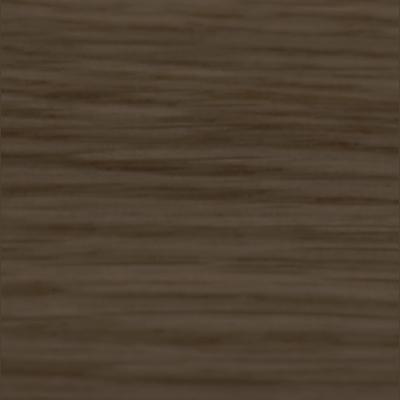}
            \put(2, 2){\color{white}\small{0.00991}}
            \end{overpic} &
            \begin{overpic}[width=0.23\columnwidth]{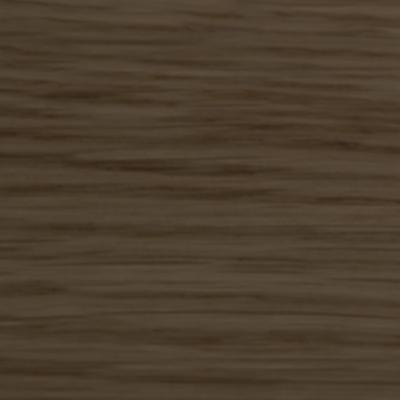}
            \end{overpic}
        \\
        \raisebox{0.35in}{\rotatebox[origin=c]{90}{\textsc{Leather08}}} &
    	\begin{overpic}[width=0.23\columnwidth]{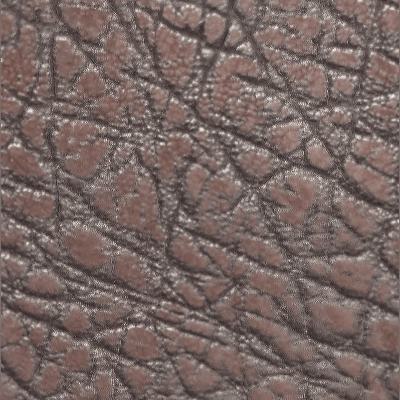}
            \put(2, 2){\color{white}\small{0.12887}}
            \end{overpic} &
            \begin{overpic}[width=0.23\columnwidth]{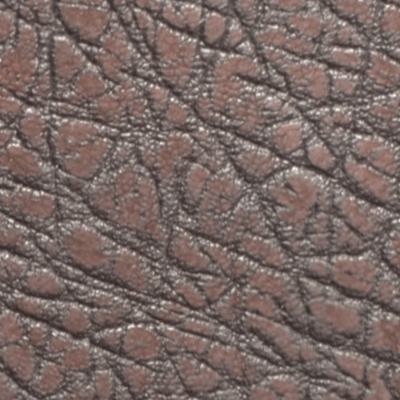}
            \end{overpic} &
            \begin{overpic}[width=0.23\columnwidth]{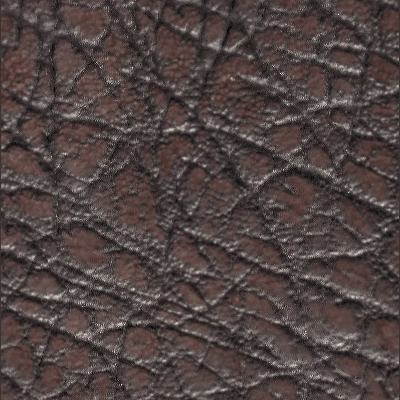}
            \put(2, 2){\color{white}\small{0.10547}}
            \end{overpic} &
            \begin{overpic}[width=0.23\columnwidth]{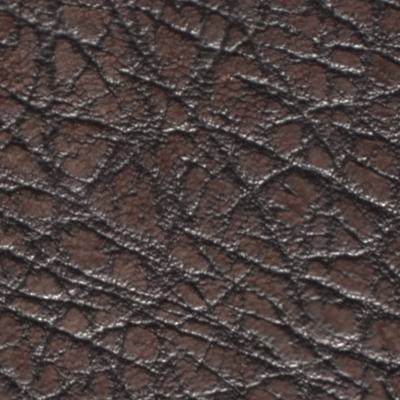}
            \end{overpic}
        \\
        & Ours (Angle $1$) & GT (Angle $1$) & Ours (Angle $2$) & GT (Angle $2$)
	\end{tabular}
          \vspace{-3mm}
	\caption{\label{fig:valid_rep}
        Validation of our method's representation capability for different BTFs.  
        We visualize our results along with ground truth (GT) BTFs with two different pairs of directions for each different material. We use structural dissimilarity (DSSIM$\downarrow$) as the error metric. 
	}
        \vspace{-4mm}
\end{figure}

In Fig.~\ref{fig:valid_rep}, we visualize our BTF recovery results along with ground truth BTFs with two different pairs of directions for each material. 
As shown in the figure, our method faithfully captures the complex patterns and the highly specular reflection.

\begin{figure}[t]
	\addtolength{\tabcolsep}{-4.5pt}
	\begin{tabular}{cccc}
   	\raisebox{0.5in}{\rotatebox[origin=c]{90} {\textsc{Wood06}}} &
        \begin{overpic}[width=0.32\columnwidth]{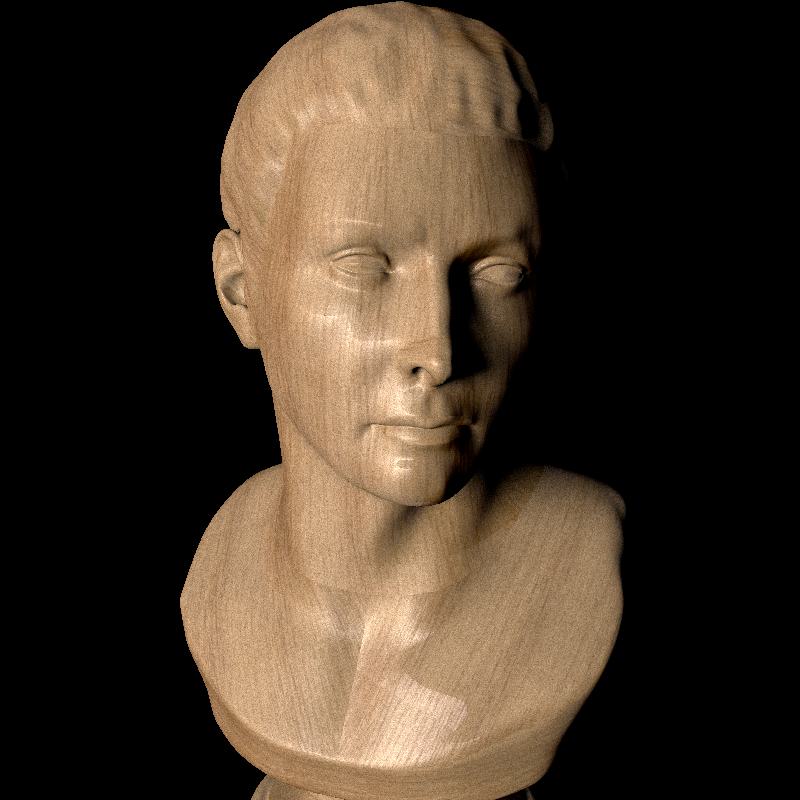}
             \put(65, 2){\includegraphics[width=0.1\columnwidth]{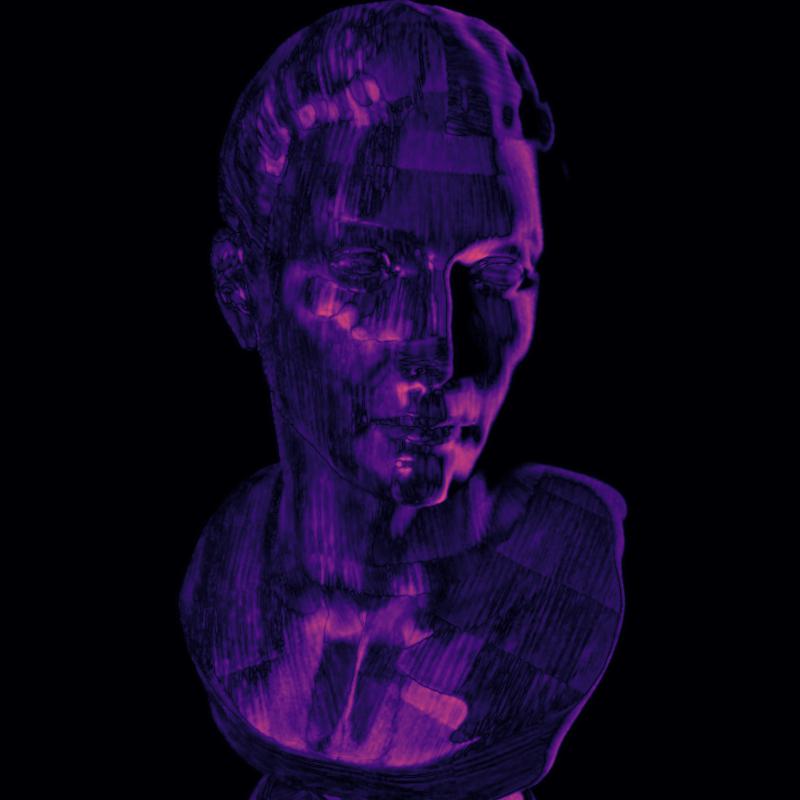}}
                   \put(2, 2){\color{white}\small{ 0.05951}}
            \end{overpic} &
   	\begin{overpic}[width=0.32\columnwidth]{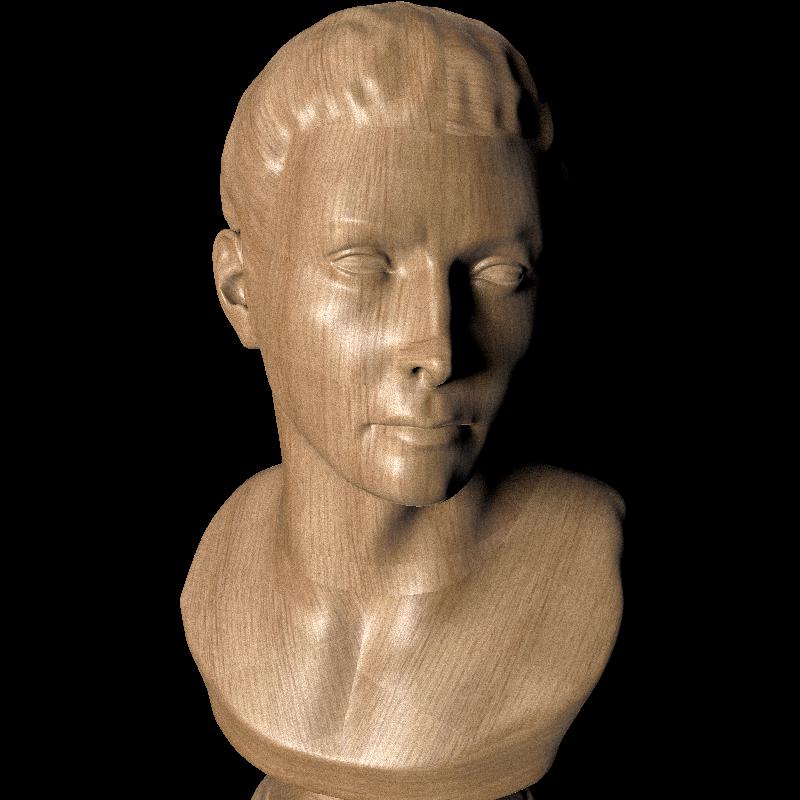}
         \put(65, 2){\includegraphics[width=0.1\columnwidth]{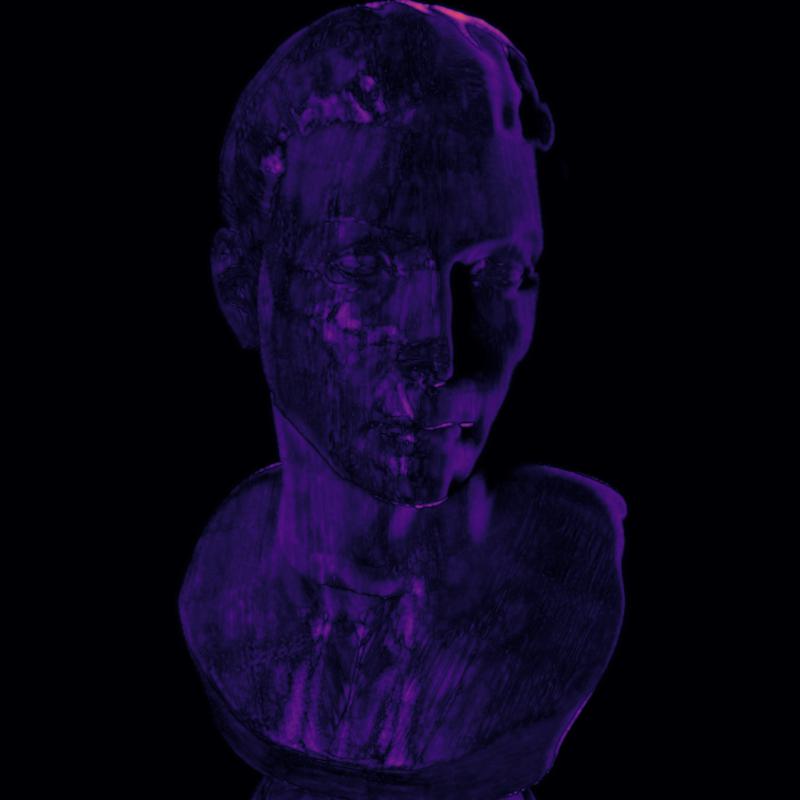}}
              \put(2, 2){\color{white}\small{0.04365}}
            \end{overpic} &
            \begin{overpic}[width=0.32\columnwidth]{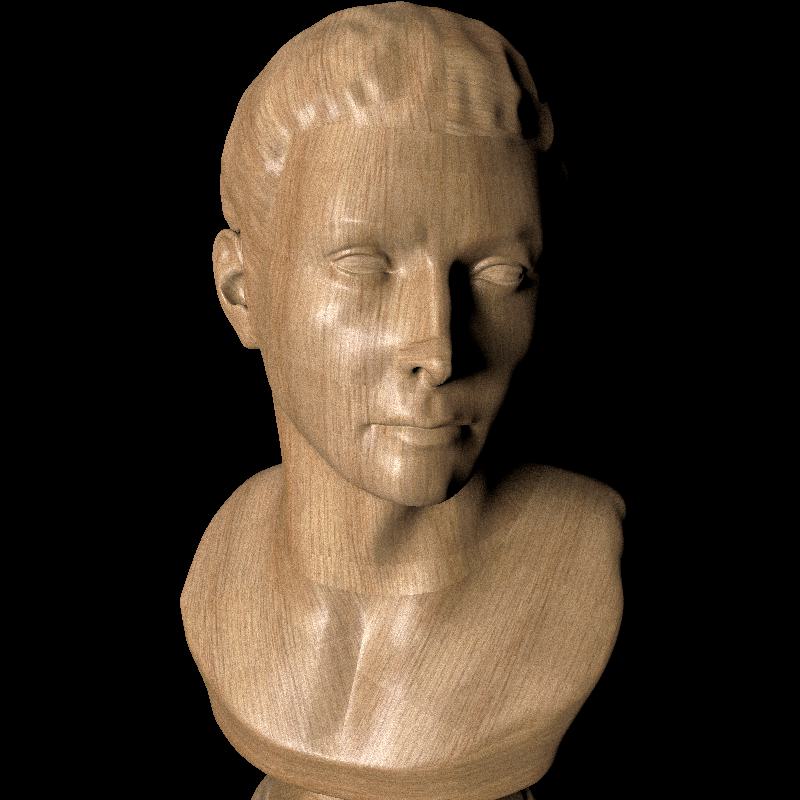}
            \put(74, 74){\includegraphics[width=0.08\columnwidth]{pics/teaser/wood06.jpg}}
            \end{overpic}  
	\\
    	\raisebox{0.5in}{\rotatebox[origin=c]{90} {\textsc{Leather08}}} &
 	\begin{overpic}[width=0.32\columnwidth]{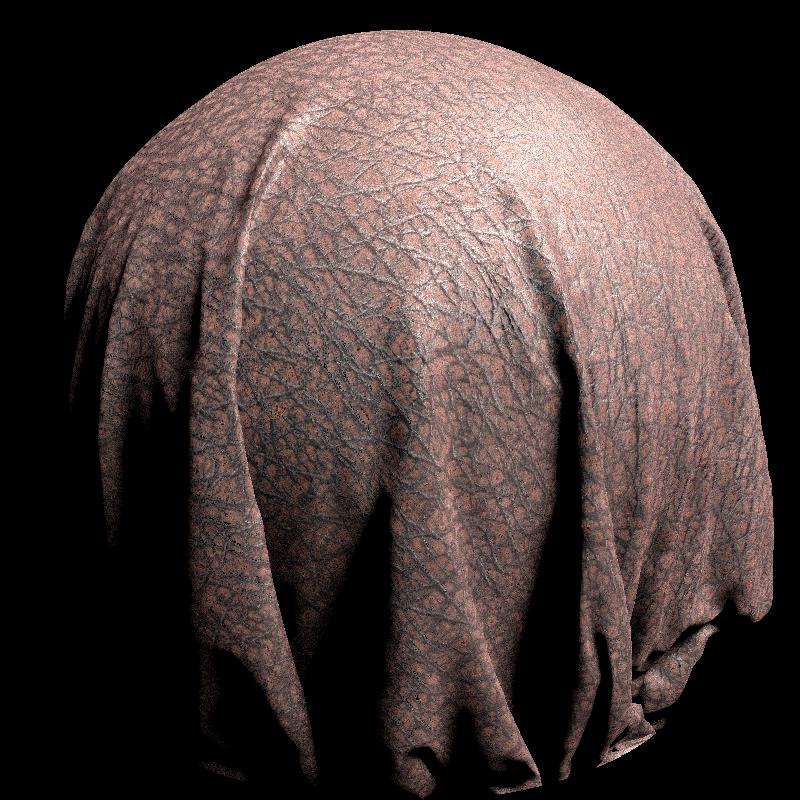}
                      \put(65, 2){\includegraphics[width=0.1\columnwidth]{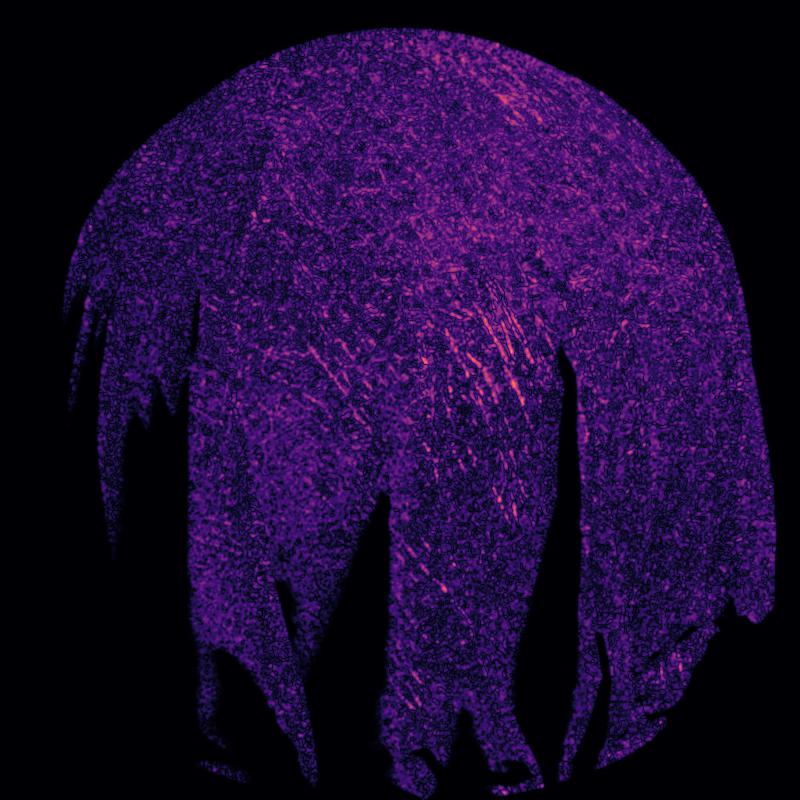}}
                   \put(2, 2){\color{white}\small{ 0.10227}}
            \end{overpic} &
   	\begin{overpic}[width=0.32\columnwidth]{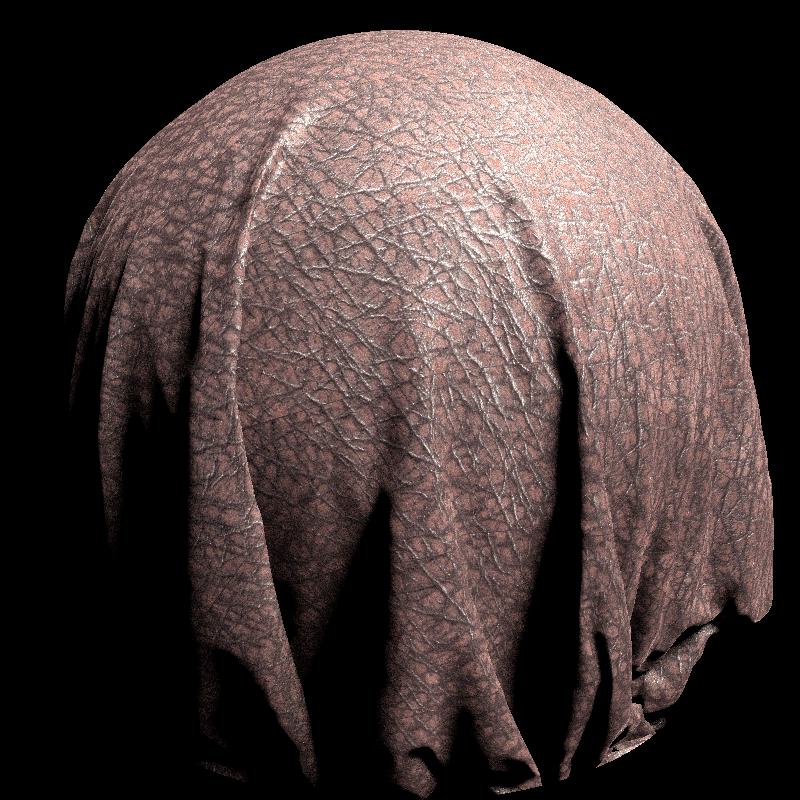}
            \put(65, 2){\includegraphics[width=0.1\columnwidth]{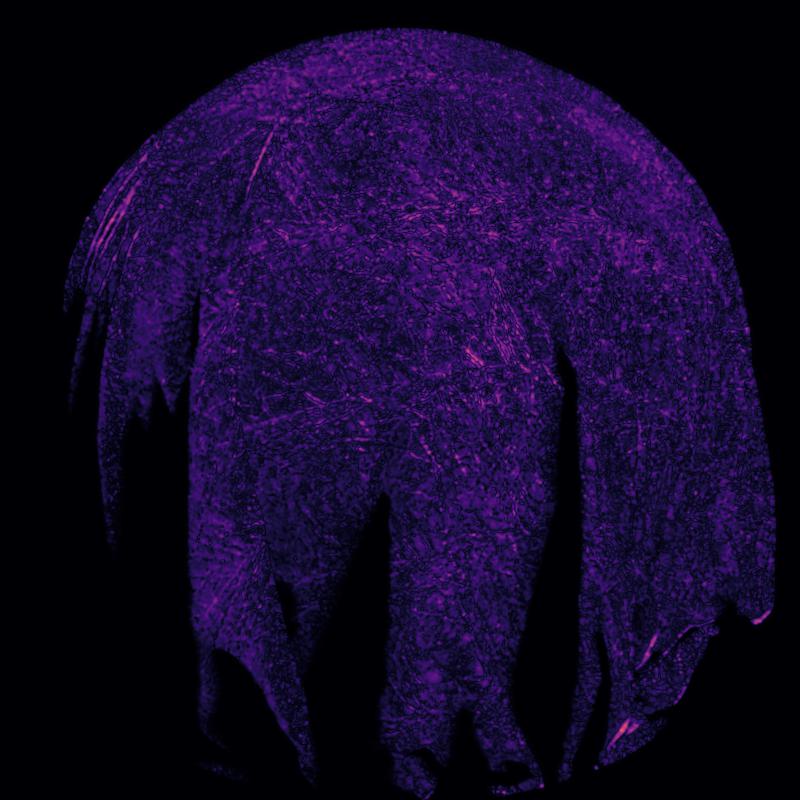}}
            \put(2, 2){\color{white}\small{ 0.07840}}
            \end{overpic} &
            \begin{overpic}[width=0.32\columnwidth]{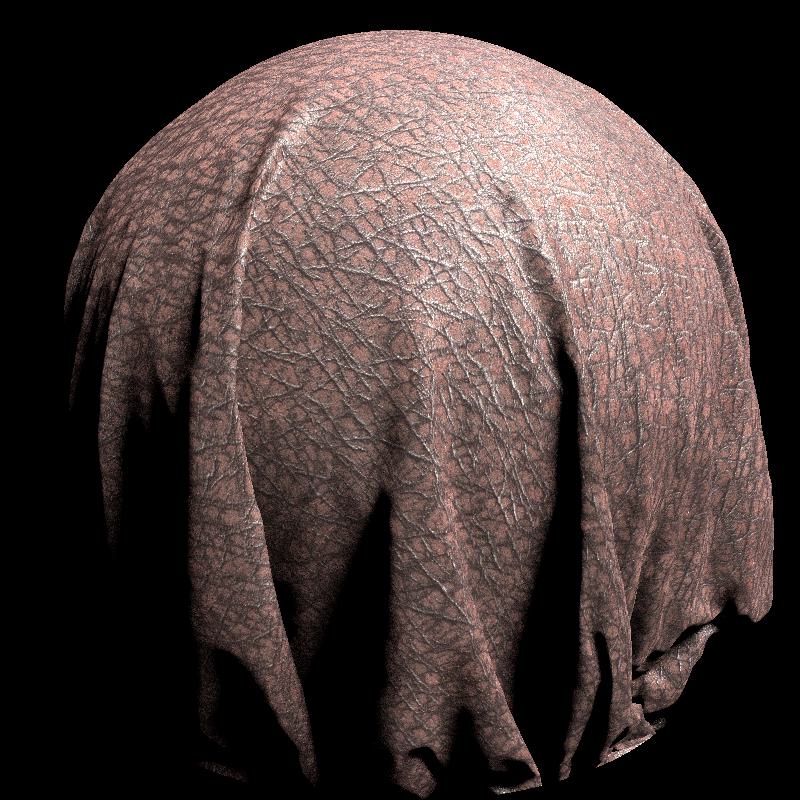}
            \put(74, 74){\includegraphics[width=0.08\columnwidth]{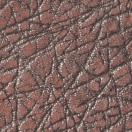}}
            \end{overpic}  
	\\
         & NeuMIP &  Triple Plane &  Reference  
	\end{tabular}
        \vspace{-3mm}
	\caption{\label{fig:comp_gt}
        Comparison with NeuMIP~\cite{kuznetsov:2021:neumip} and reference without applying synthesis, i.e., use repetitive tiling. The reference image is generated by interpolating the original BTF data. We show the \FLIP error$\downarrow$ and the error image on the bottom. Triple Plane is closer to the reference with more accurate highlights and patterns, but NeuMIP is also good. Therefore, both can be used in our scheme.
        }
        \vspace{-3mm}
\end{figure}

In Fig.~\ref{fig:comp_gt}, we compare our Triple Plane with a modified NeuMIP~\cite{kuznetsov:2021:neumip} using repetitive tiling, i.e., no synthesis is performed. The reference image is generated by interpolating the original BTF data.
Our Triple Plane shows more accurate highlights in \textsc{Wood06} and clearer stretch patterns in \textsc{Leather08} compared to NeuMIP. We believe the quality improvement is because of the novel dimension decomposition scheme which has a clear separation of the position and direction dimensions.

\begin{figure}[tb]
	\centering
	\addtolength{\tabcolsep}{-3.5pt}
	\small
	\begin{tabular}{ccc}
    	\begin{overpic}[width=0.32\columnwidth]{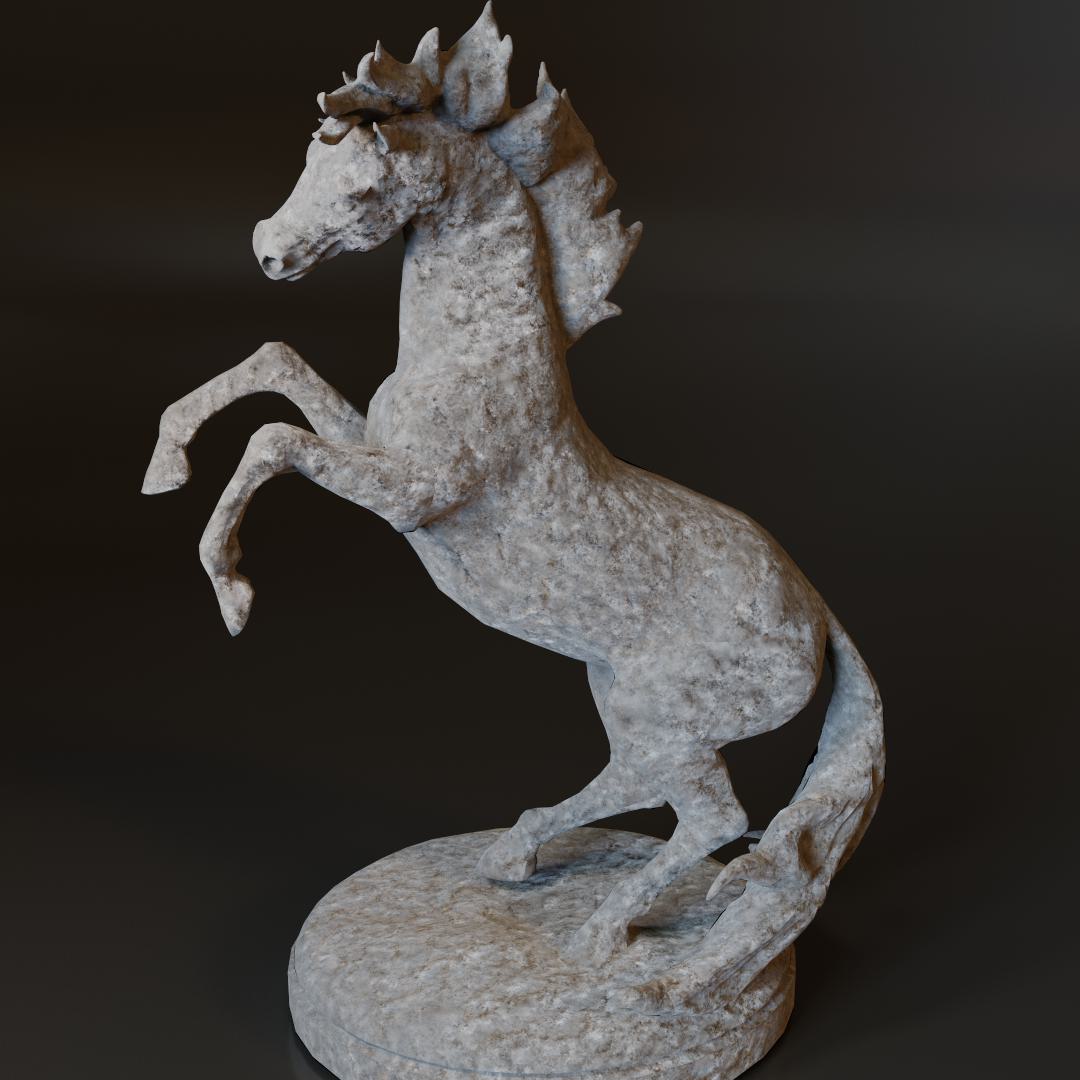}
            \put(2, 2){\color{white}\small{\textsc{Stone10}}}
              \put(82,2){\color{white}\small{$8 \times$}}
            \end{overpic} &
            \begin{overpic}[width=0.32\columnwidth]{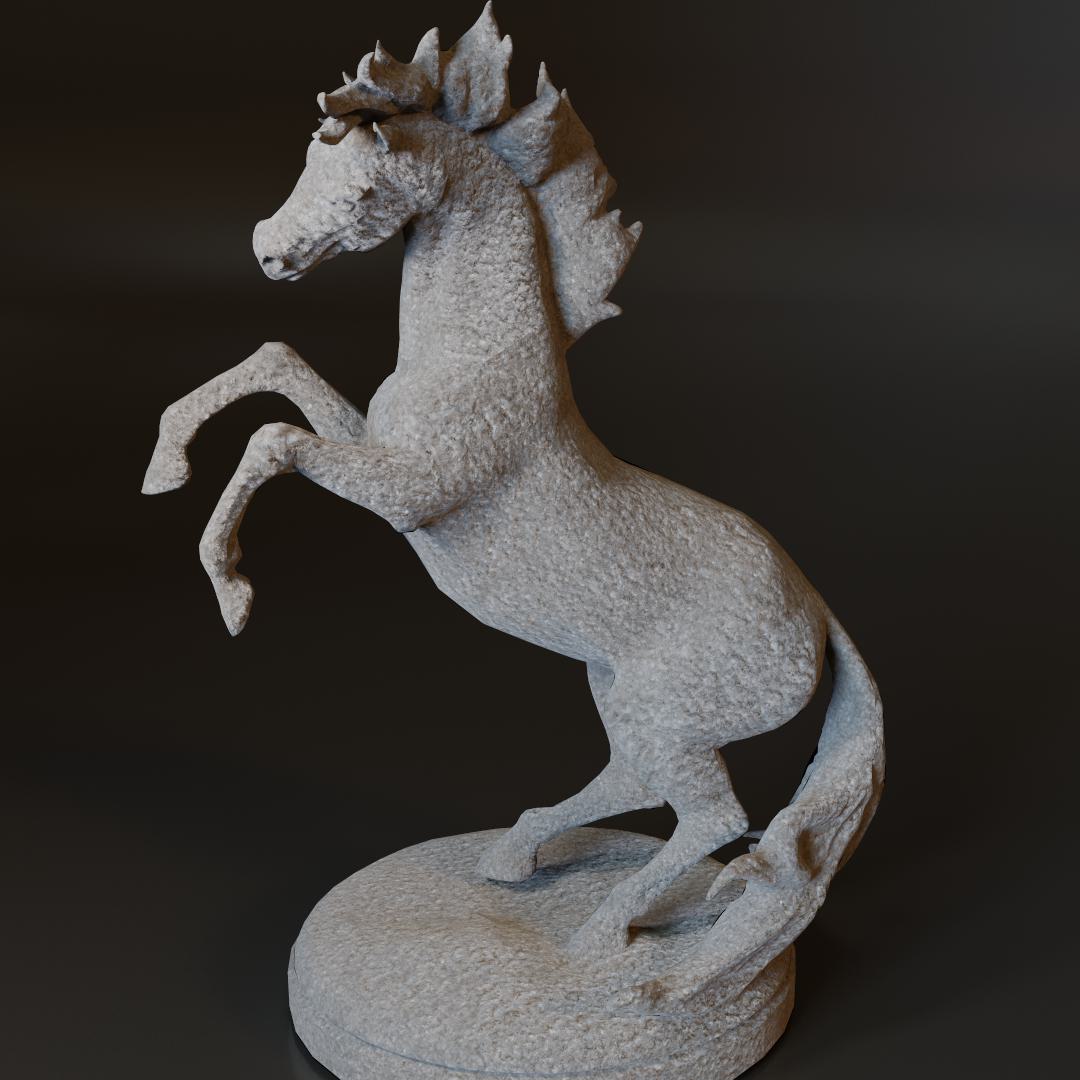}
            \put(80,2){\color{white}\small{$25 \times$}}
            \end{overpic} &
            \begin{overpic}[width=0.32\columnwidth]{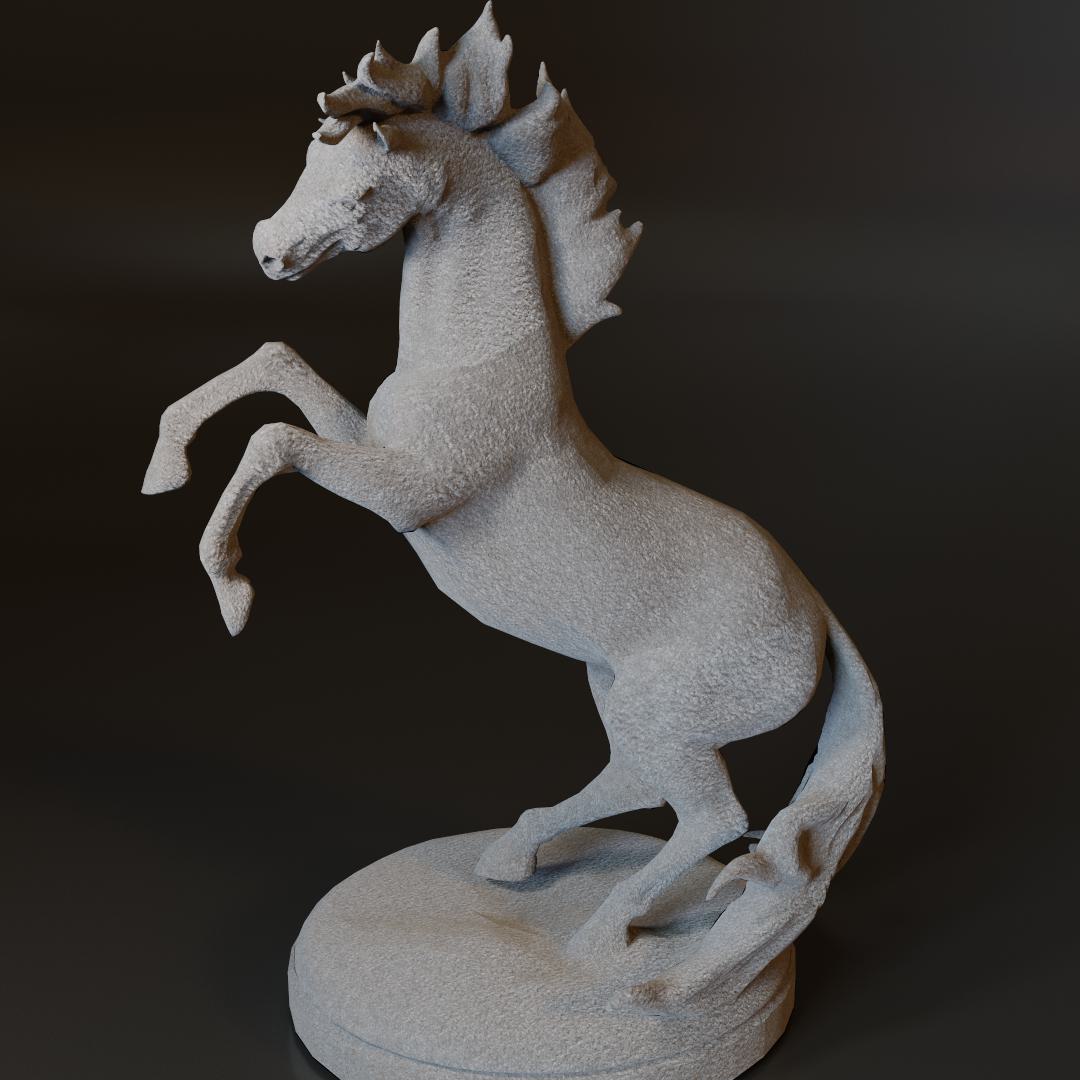}
            \put(74, 74){\includegraphics[width=0.08\columnwidth]{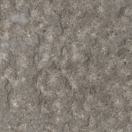}}
            \put(80,2){\color{white}\small{$45 \times$}}
            \end{overpic} 
	\end{tabular}
        \vspace{-3mm}
	\caption{
            \label{fig:valid_syn}
            We validate our by-example BTF synthesis in different scales by scaling the UV coordinate, the scaling factor is marked on the bottom-right.
            Even on a very large scale ($45 \times$), our method faithfully maintains the accurate appearance of the BTF, which demonstrates our capability of generating an infinitely large, non-repetitive BTF.
        }
        \vspace{-4mm}
\end{figure}

\begin{figure}[t]
	\centering
        \addtolength{\tabcolsep}{-4.5pt}
	\begin{tabular}{cccc}
  	     \raisebox{0.5in}{\rotatebox[origin=c]{90} {\textsc{Leather04}}} &
    	\begin{overpic}[width=0.32\columnwidth]{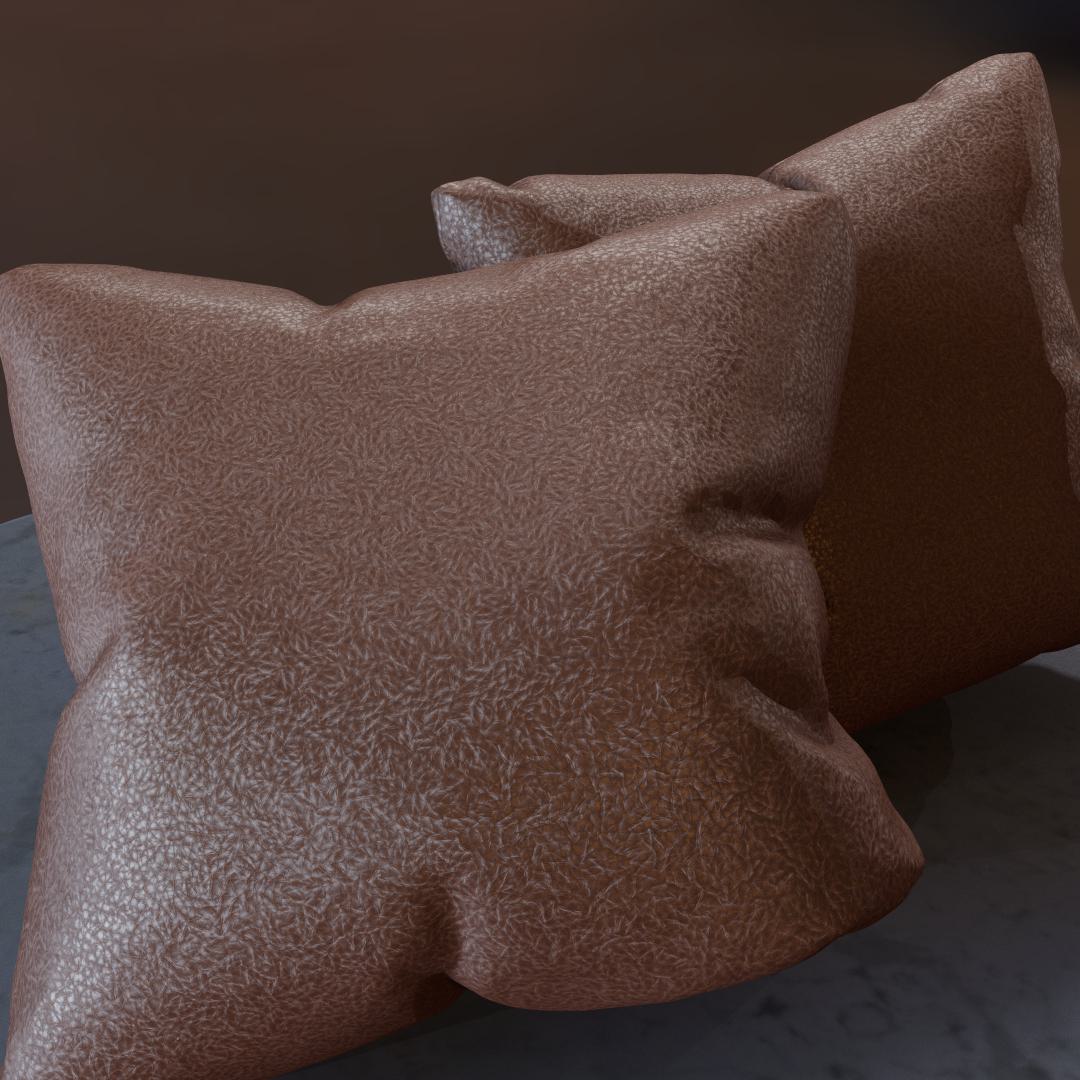}
            \put(45,2){\color{white}\small{$15 \times$ (10 MB)}}
            \end{overpic} &
            \begin{overpic}[width=0.32\columnwidth]{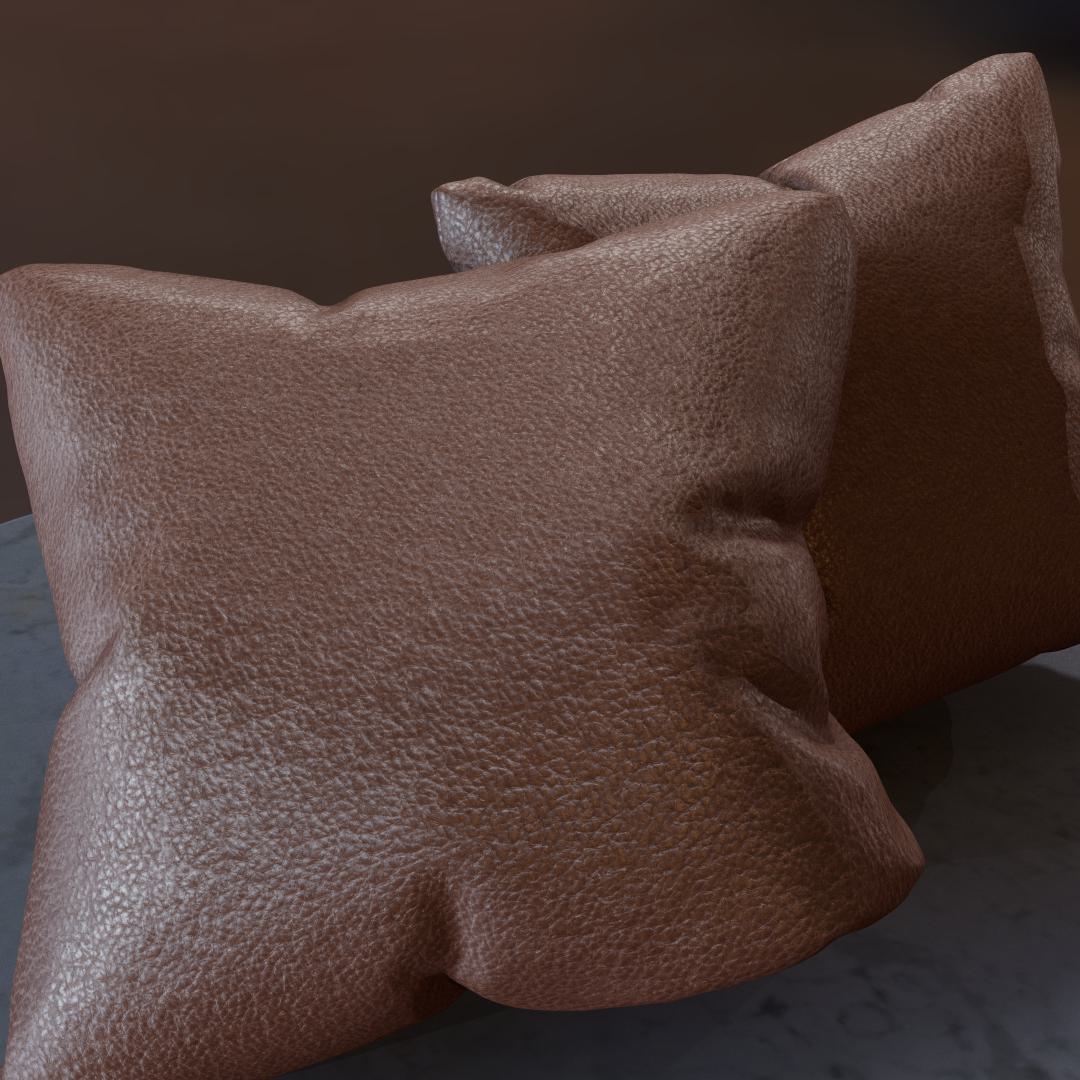}
            \put(45,2){\color{white}\small{$15 \times$ (10 MB)}}
            \end{overpic} &
            \begin{overpic}[width=0.32\columnwidth]{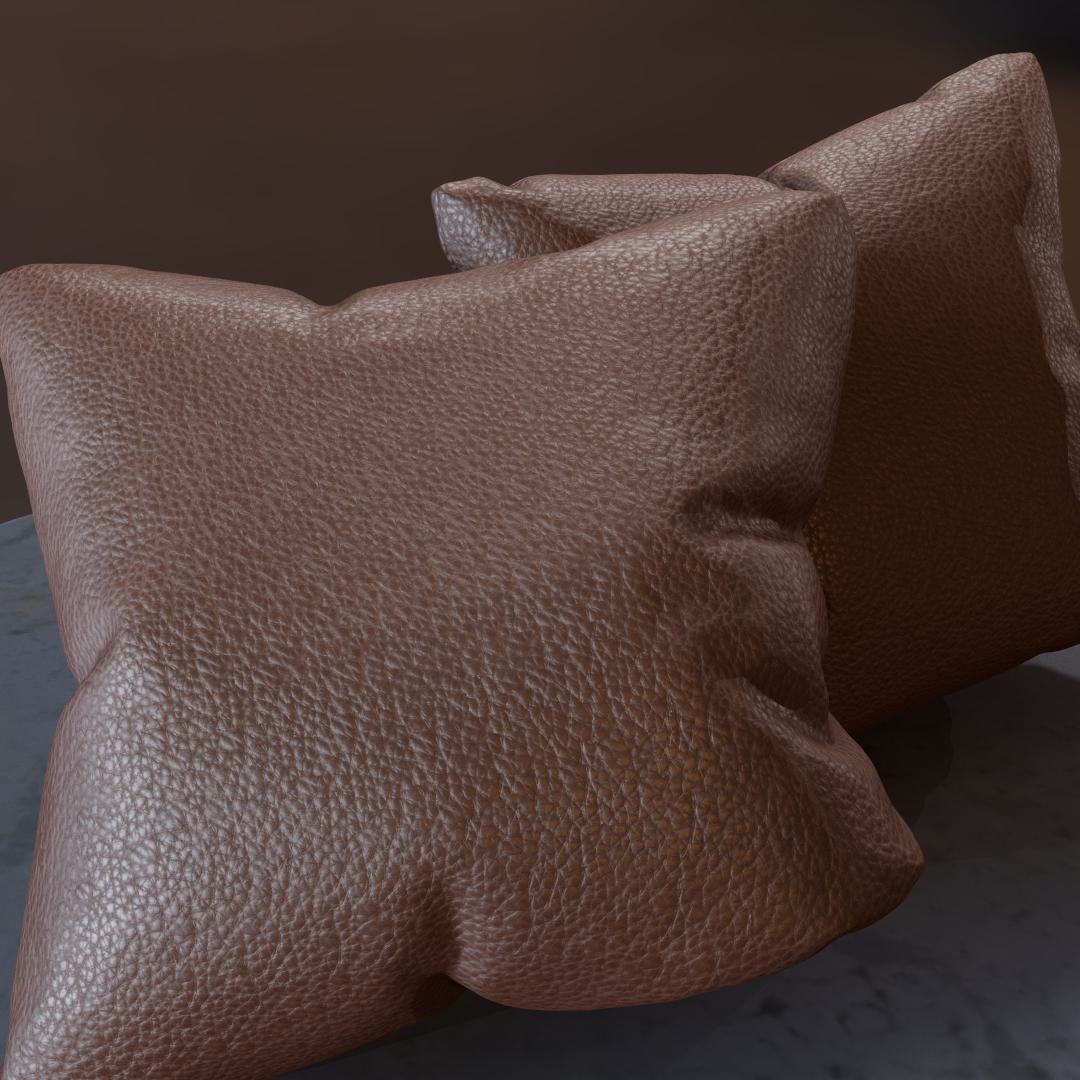}
            \put(74, 74){\includegraphics[width=0.08\columnwidth]{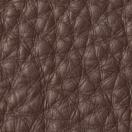}}
            \put(40,2){\color{white}\small{$15 \times$ }\color{red}\small{(2.15 GB)}}
            \end{overpic} 
	\\
    	     \raisebox{0.5in}{\rotatebox[origin=c]{90} {\textsc{Leather10}}} &
    	\begin{overpic}[width=0.32\columnwidth]{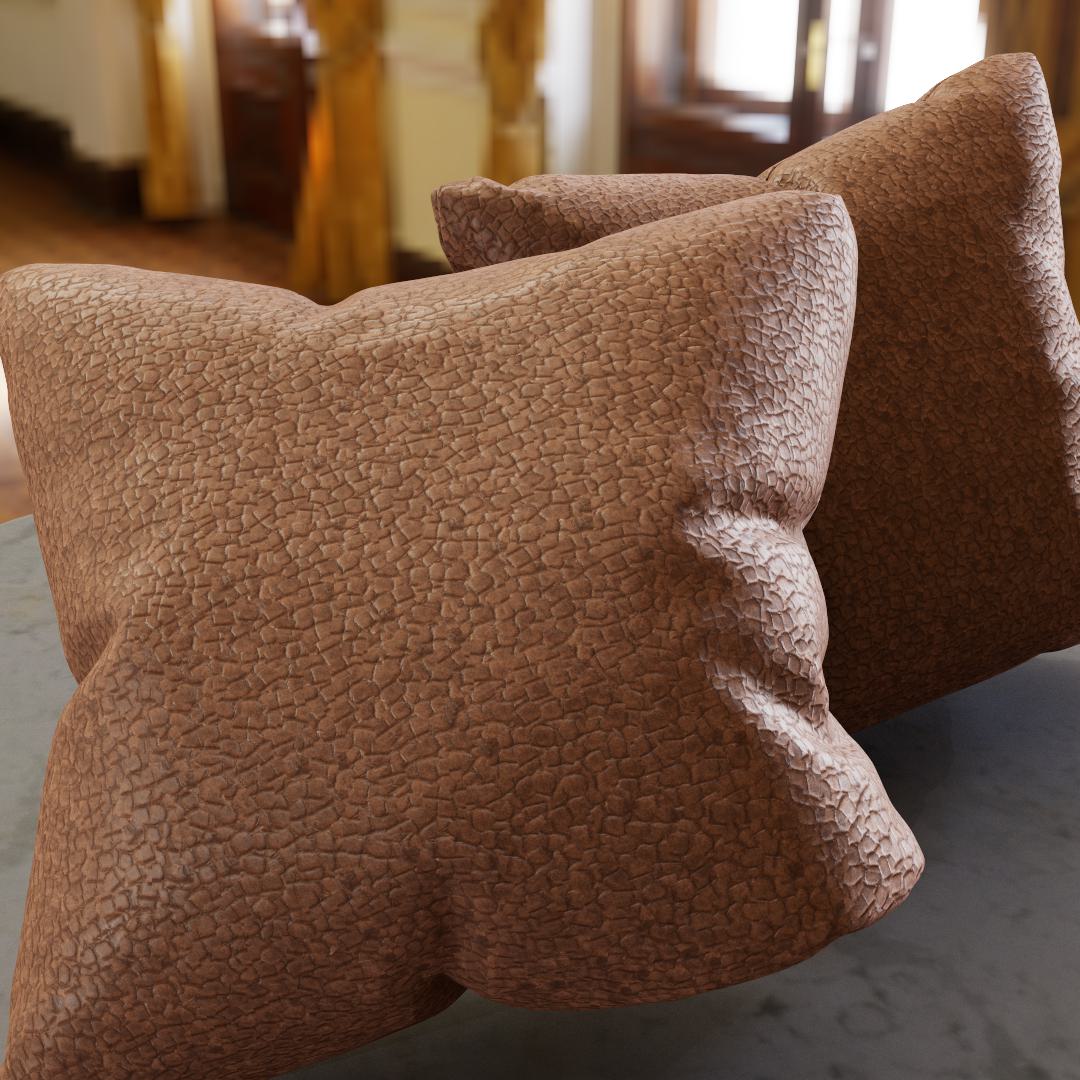}
            \put(45,2){\color{white}\small{$15 \times$ (10 MB)}}
            \end{overpic} &
            \begin{overpic}[width=0.32\columnwidth]{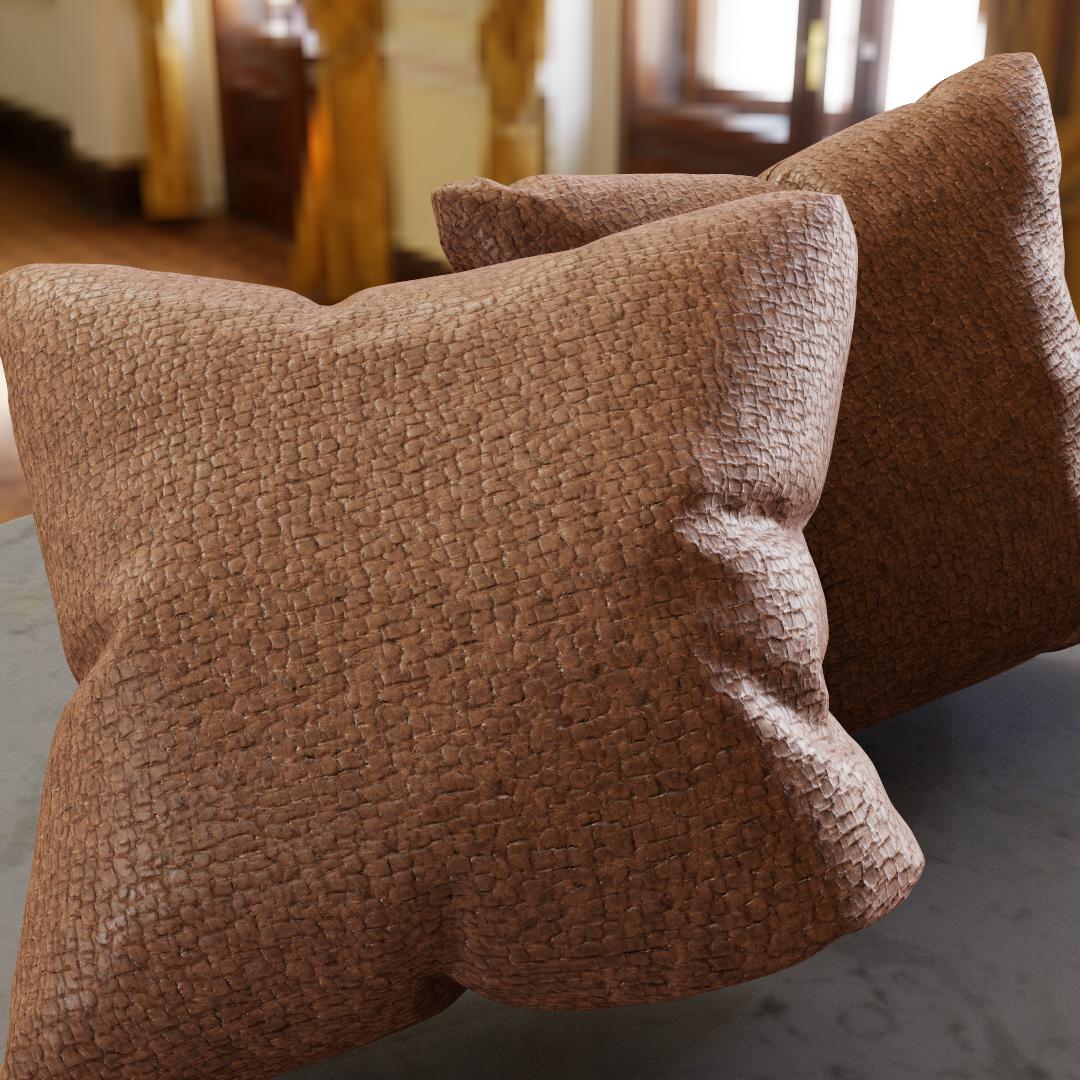}
            \put(45,2){\color{white}\small{$15 \times$ (10 MB)}}
            \end{overpic} &
            \begin{overpic}[width=0.32\columnwidth]{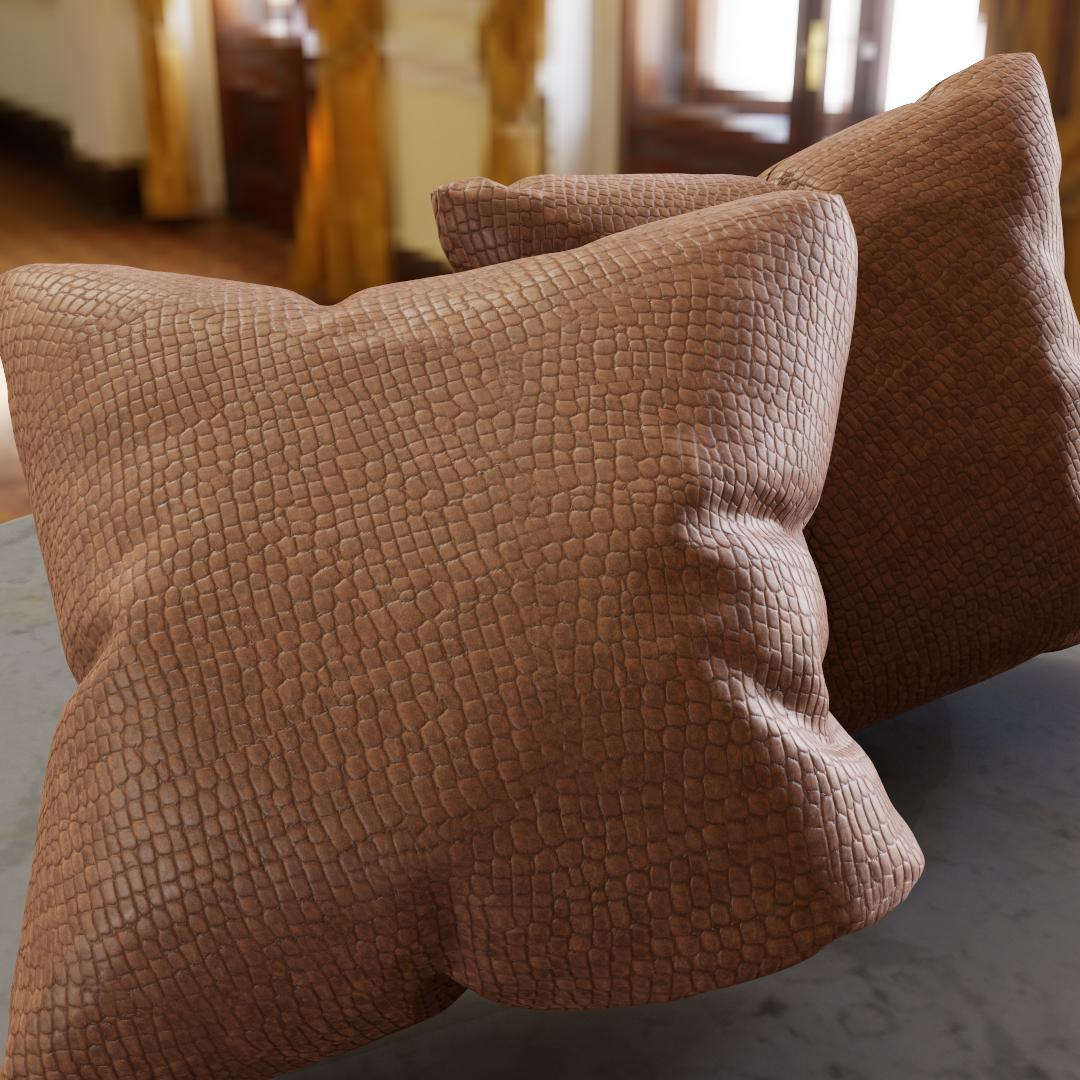}
            \put(74, 74){\includegraphics[width=0.08\columnwidth]{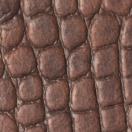}}
            \put(40,2){\color{white}\small{$15 \times $ }\color{red}\small{(2.15 GB)}}
            \end{overpic} 
        \\
         & Hex-Tiling & Hist. preserving & Texture Quilting 
        \\
         &  & blending & (offline searched)
	\end{tabular}
         \vspace{-3mm}
	\caption{\label{fig:valid_hex}
        Our BTF synthesis scheme is a general idea that is compatible with different texture synthesis methods. To demonstrate that, we implement another dynamic by-example texture synthesis approach Hex-Tiling~\cite{mikkelsen:2022:hextile} (middle), and a non-dynamic texture quilting~\cite{Efros:2001:quilting} (left). With  $15 \times$ UV scaling, the synthesized BTF is equivalent to have 6K resolution. Quilting produces the most visually pleasing result because it finds the best match patches by offline searching. However, once generated, the texture can not be changed, and even after our dimensional decomposition, the synthesized texture still takes GBs storage. 
	}
        \vspace{-3mm}
\end{figure}

\subsection{Validation and Comparison of Texture Synthesis}
Our approach treats the 2D positional feature plane similarly to standard 2D textures, enabling dynamic synthesis. We employ histogram preserving blending~\cite{heitz:2018:histo} as our primary solution. 
In Fig.~\ref{fig:valid_syn}, we first demonstrate our method's capability of dynamically synthesizing a non-repetitive BTF on an arbitrary scale. Even on a very large scale ($45 \times$ UV scaling), our method faithfully maintains the accurate appearance of the BTFs.

Our method can also benefit from other texture synthesis methods. To demonstrate that, we employ two additional texture synthesis methods.
The first one is Hex-Tiling~\cite{mikkelsen:2022:hextile}, another dynamic by-example texture synthesis approach, and the second one is a non-dynamic texture quilting technique~\cite{Efros:2001:quilting}, a classic method in patch-based quilting. 
As shown in Fig.~\ref{fig:valid_hex}, our synthesis scheme supports all these synthesis methods, and quilting produces the most visually pleasing result because it finds the best match patches by an offline search. However, it requires a pre-generation of the synthesized BTF. The synthesized BTF is with $15 \times$ UV scaling (equivalent to 6K resolution). Even if the quilting BTF has already been decomposed into 2D functions, storing the neural textures still takes about 2GB. The histogram-preserving blending and Hex-Tiling provide a balance between dynamic query and structured quality (further discussion in Sec.~\ref{sec:limitation}).

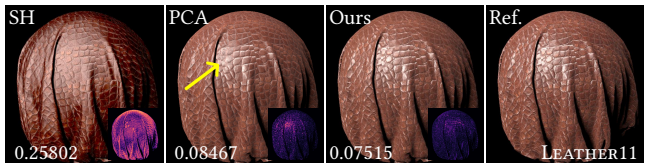
\begin{figure}[t]
	\centering
	\addtolength{\tabcolsep}{-4.5pt}
	\small
	\begin{tabular}{cccc}
    	\begin{overpic}[width=0.245\columnwidth]{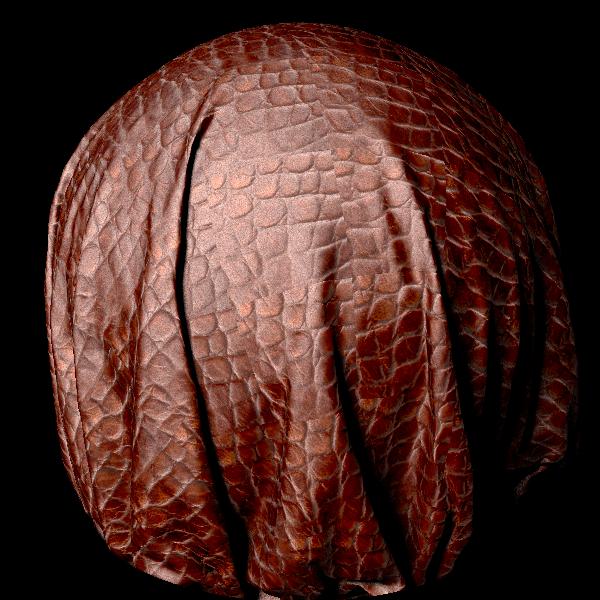}
             \put(2, 88){\color{white}\small{SH}}
            \put(65, 2){\includegraphics[width=0.08\columnwidth]{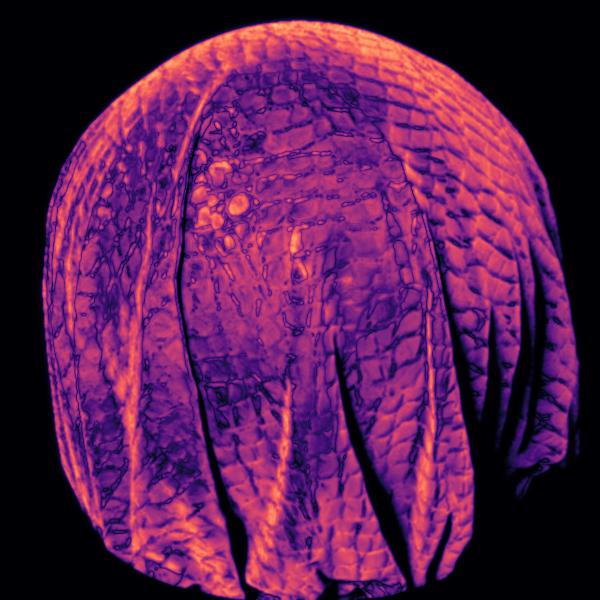}}
            \put(2, 2){\color{white}\small{ 0.25802}}
            \end{overpic} &
            \begin{overpic}[width=0.245\columnwidth]{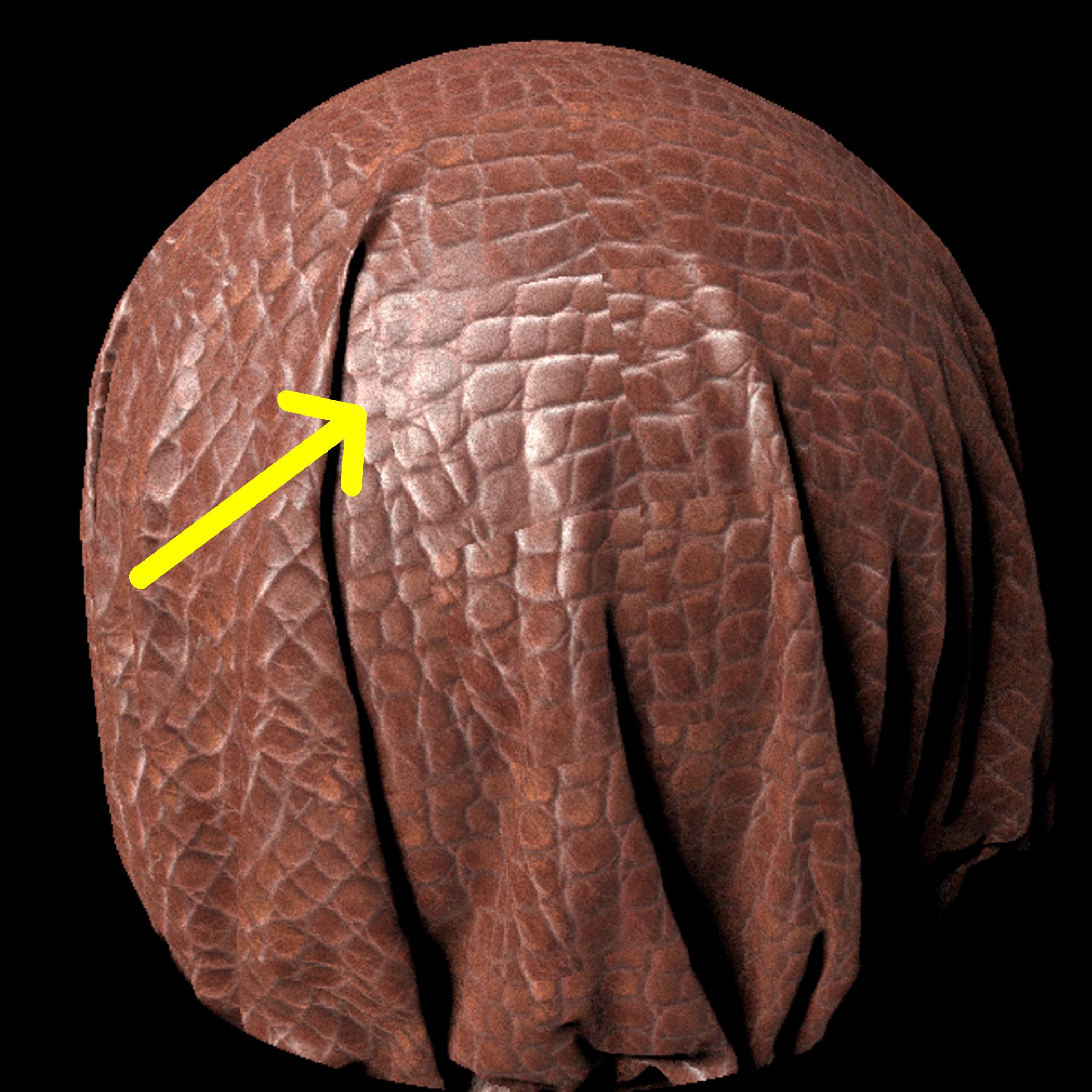}
            \put(65, 2){\includegraphics[width=0.08\columnwidth]{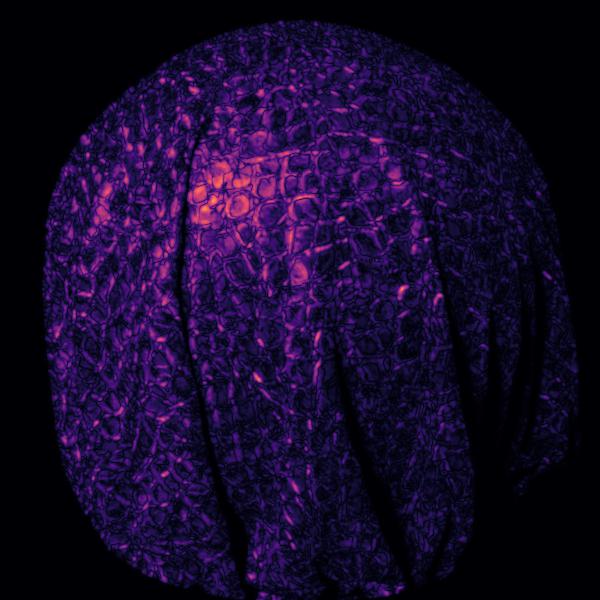}}
            \put(2, 88){\color{white}\small{PCA}}
            \put(2, 2){\color{white}\small{ 0.08467}}
            \end{overpic} &
            \begin{overpic}[width=0.245\columnwidth]{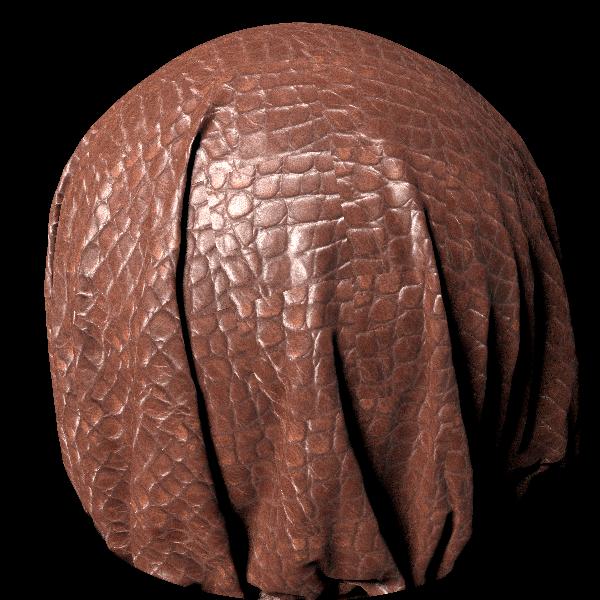}
            \put(65, 2){\includegraphics[width=0.08\columnwidth]{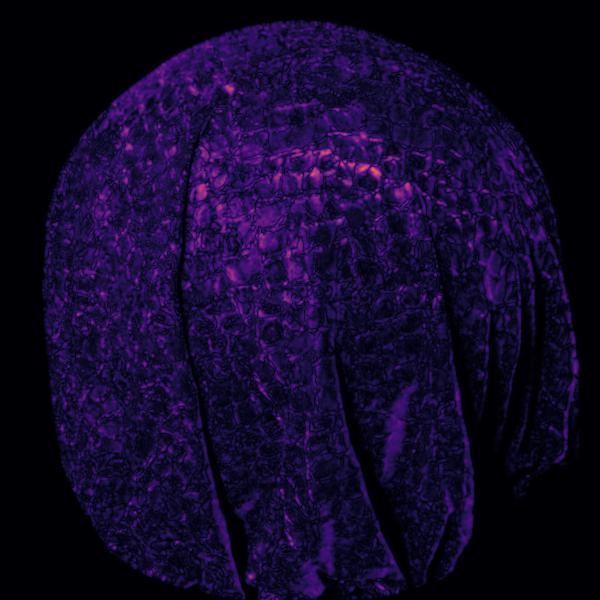}}
            \put(2, 2){\color{white}\small{0.07515}}
            \put(2, 88){\color{white}\small{Ours}}
            \end{overpic} &
            \begin{overpic}[width=0.245\columnwidth]{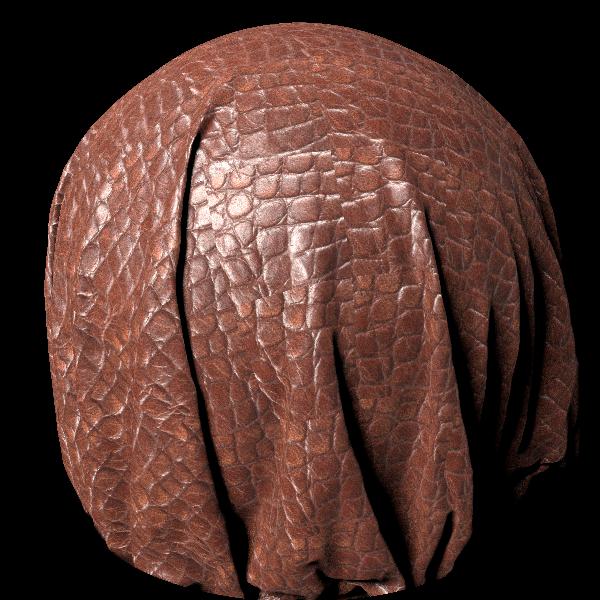}
            \put(2, 88){\color{white}\small{Ref.}}
            \put(35, 2){\color{white}\small{\textsc{Leather11}}}
            \end{overpic} 
	\end{tabular}
          \vspace{-3mm}
	\caption{\label{fig:comp_compression}
            Comparison with classical BTF compression methods: PCA~\cite{koudelka:2003:acquisition} and SH~\cite{kawasaki:2005:patch}. Since the spatial storage is the overhead of non-dynamic BTF synthesis, to have a relatively fair comparison, we employ a similar number of coefficients for each method. 
            For PCA, we keep 15 eigenvalues (5 eigenvalues for each RGB channel). For SH, we compute a discrete table of different view directions for each BTF textel. Each item in the table stores 27 SH coefficients (3 levels SH for each RGB channel) that fit the corresponding fixed view 2D BRDF. The \FLIP error$\downarrow$ and the error image are shown on the bottom. Our Triple Plane faithfully captures the appearance and the structure of the BTF, while SH, even only representing 2D angular distributions, loses most of the high-frequency signals, and PCA fails to preserve sharp highlights.
	}
         \vspace{-3mm}
\end{figure}

\subsection{Comparison of BTF Synthesis}
\label{sec:res:compBTF}

The existing non-neural-based BTF synthesis methods described in Sec.~\ref{subsec:btfsyn} are all quilting-based. Comparing with them is \emph{unfair} to ours since we prioritize the \emph{dynamic} property, whereas these established methods are \emph{non-dynamic}. 
Non-dynamic methods may excel in structure preservation, but generating very large BTFs is impractical, as demonstrated in Fig.\ref{fig:valid_hex}.


Most existing BTF synthesis methods are derived from the quilting method, which we use for comparison in Fig.~\ref{fig:valid_hex}. These quilting methods essentially rearrange the texels without changing the original content of the BTF (i.e., no blending is performed). As our method employs a general scheme, it can utilize the same quilting approach for benchmarking, with differences only in representation capabilities—for instance, our Triple Plane versus PCA/SH. Thus, in combination with Fig.~\ref{fig:valid_hex}, the comparisons of our Triple Plane with PCA and SH in Fig.~\ref{fig:comp_compression} effectively reflect comparisons with earlier BTF synthesis techniques.

Moreover, as analyzed in Sec.~\ref{sec:intro}, many of the neural-based appearance synthesis works aim at estimating 2D parametric maps, primarily reconstructing or estimating BTFs from a few images, which diverges from our objectives.
Additionally, some methods (e.g., ~\cite{zhou:2023:PhotoMat}), generate partial, 4D isotropic BTFs. These are inherently different from our approach and are not suitable for direct comparison with our full 6D BTF synthesis scheme.

\begin{figure*}[tb]
	\centering
	\addtolength{\tabcolsep}{-3.5pt}
	\small
	\begin{tabular}{ccccc}
     	\raisebox{0.8in}{\rotatebox[origin=c]{90} {\textsc{Fabric12}}} &
    	\begin{overpic}[width=0.23\textwidth]{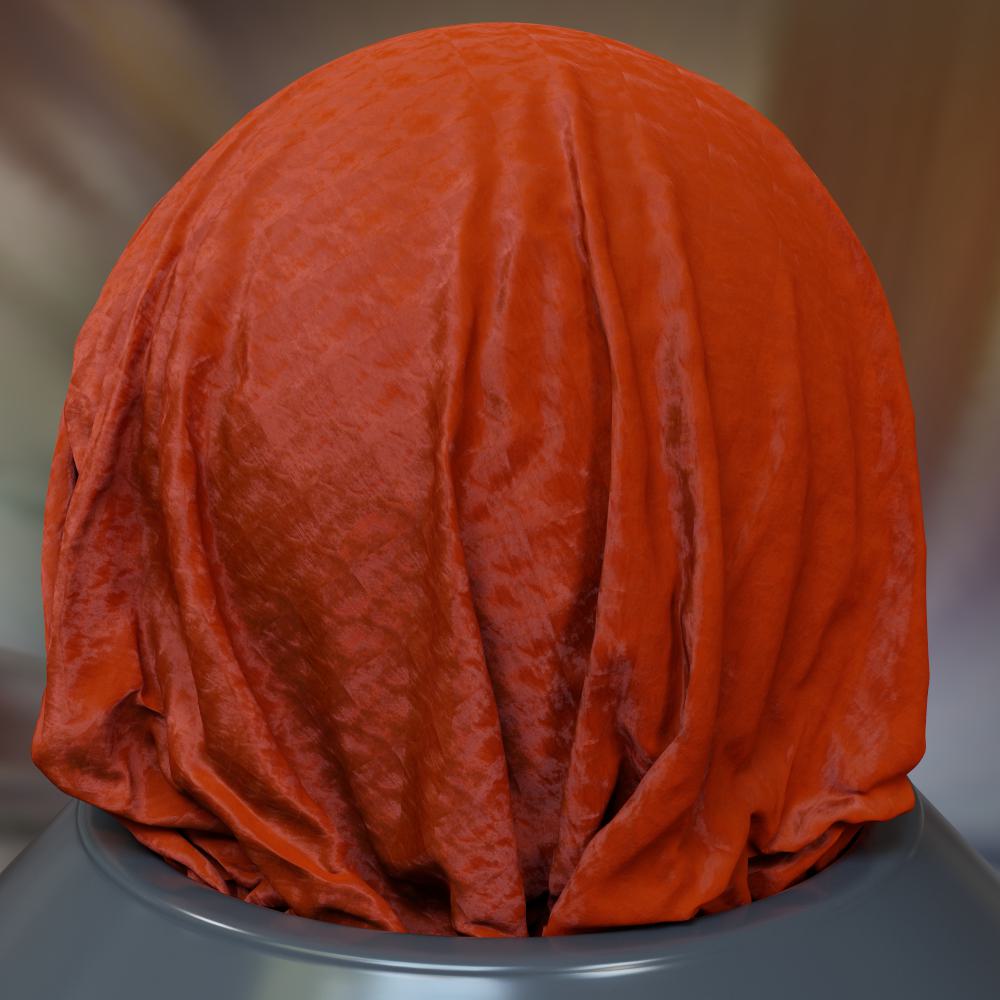}
            \end{overpic} &
            \begin{overpic}[width=0.23\textwidth]{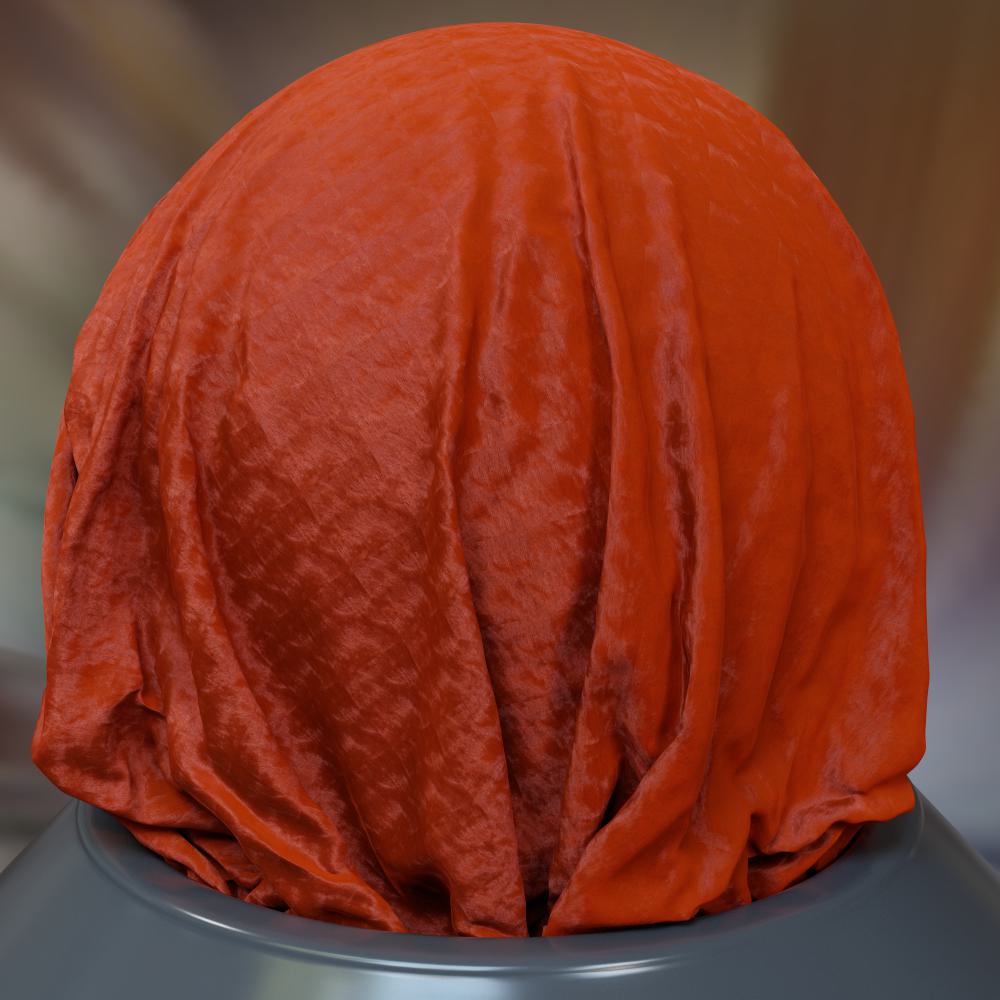}
            \end{overpic} &
            \begin{overpic}[width=0.23\textwidth]{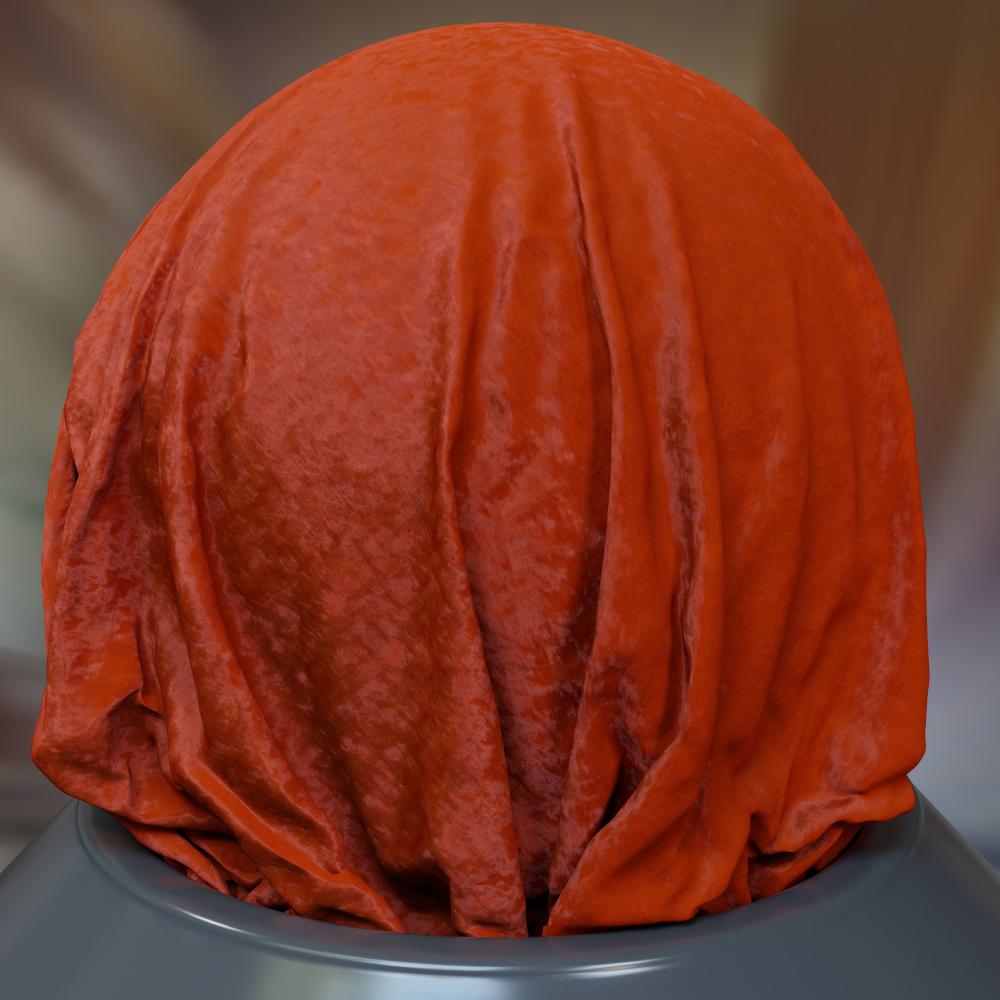}
            \end{overpic} &
            \begin{overpic}[width=0.23\textwidth]{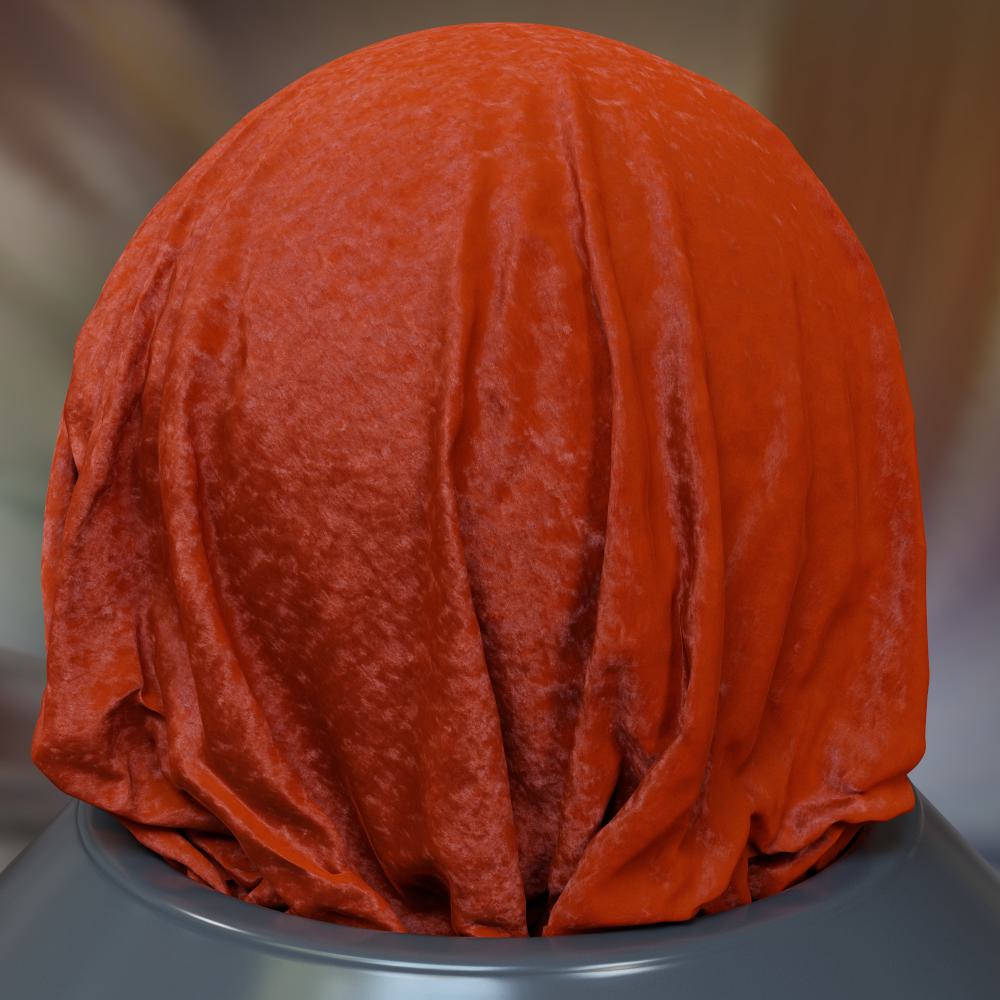}
            \end{overpic}
	\\
            &
            \begin{overpic}[width=0.24\columnwidth]{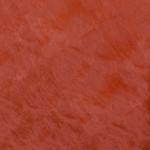}\end{overpic} 
            \begin{overpic}[width=0.24\columnwidth]{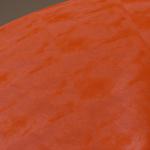}\end{overpic} &
            \begin{overpic}[width=0.24\columnwidth]{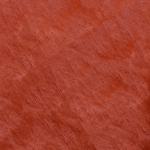}\end{overpic} 
            \begin{overpic}[width=0.24\columnwidth]{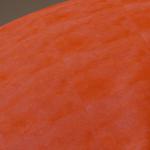}\end{overpic} &
            \begin{overpic}[width=0.24\columnwidth]{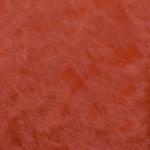}\end{overpic} 
            \begin{overpic}[width=0.24\columnwidth]{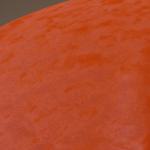}\end{overpic} &
            \begin{overpic}[width=0.24\columnwidth]{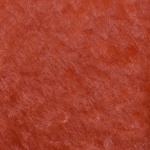}\end{overpic} 
            \begin{overpic}[width=0.24\columnwidth]{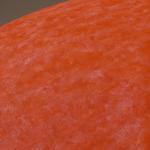}\end{overpic}
        \\
    	\raisebox{0.8in}{\rotatebox[origin=c]{90} {\textsc{Leather03}}} &
    	\begin{overpic}[width=0.23\textwidth]{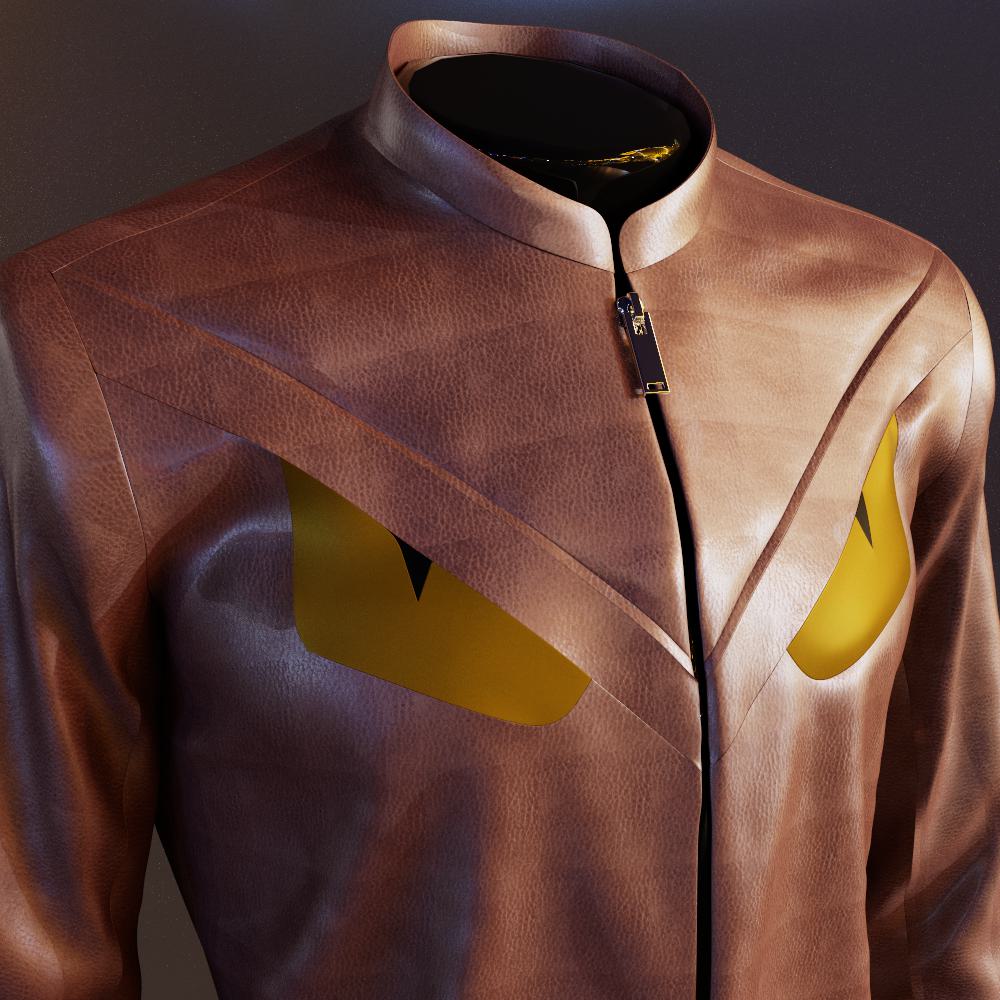}
            \end{overpic} &
            \begin{overpic}[width=0.23\textwidth]{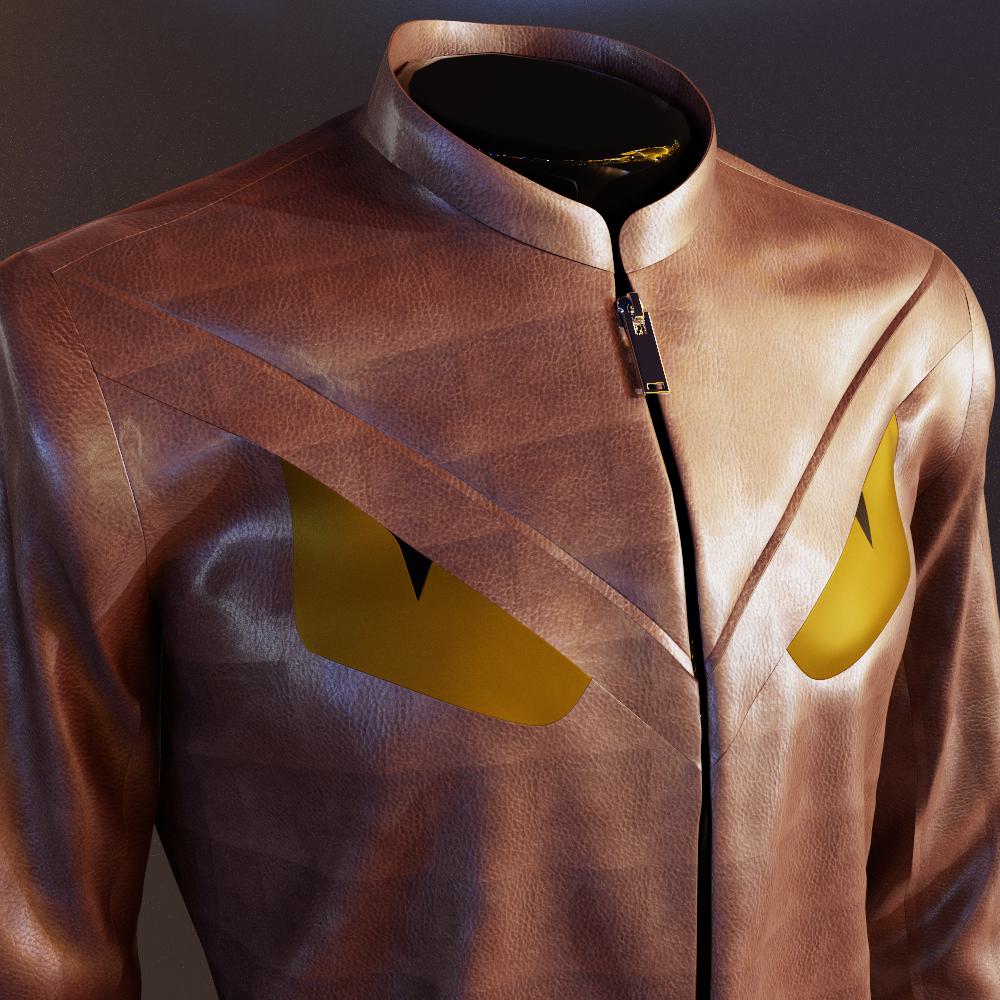}
            \end{overpic} &
            \begin{overpic}[width=0.23\textwidth]{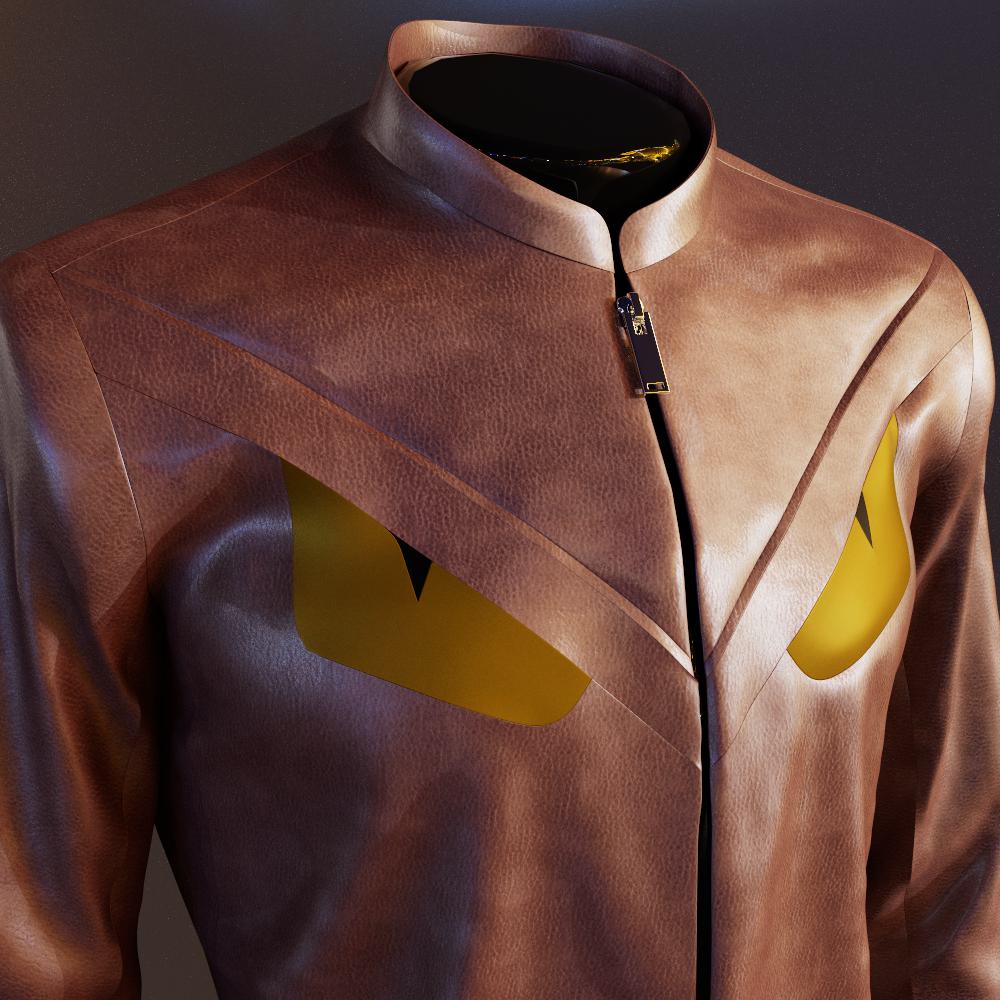}
            \end{overpic} &
            \begin{overpic}[width=0.23\textwidth]{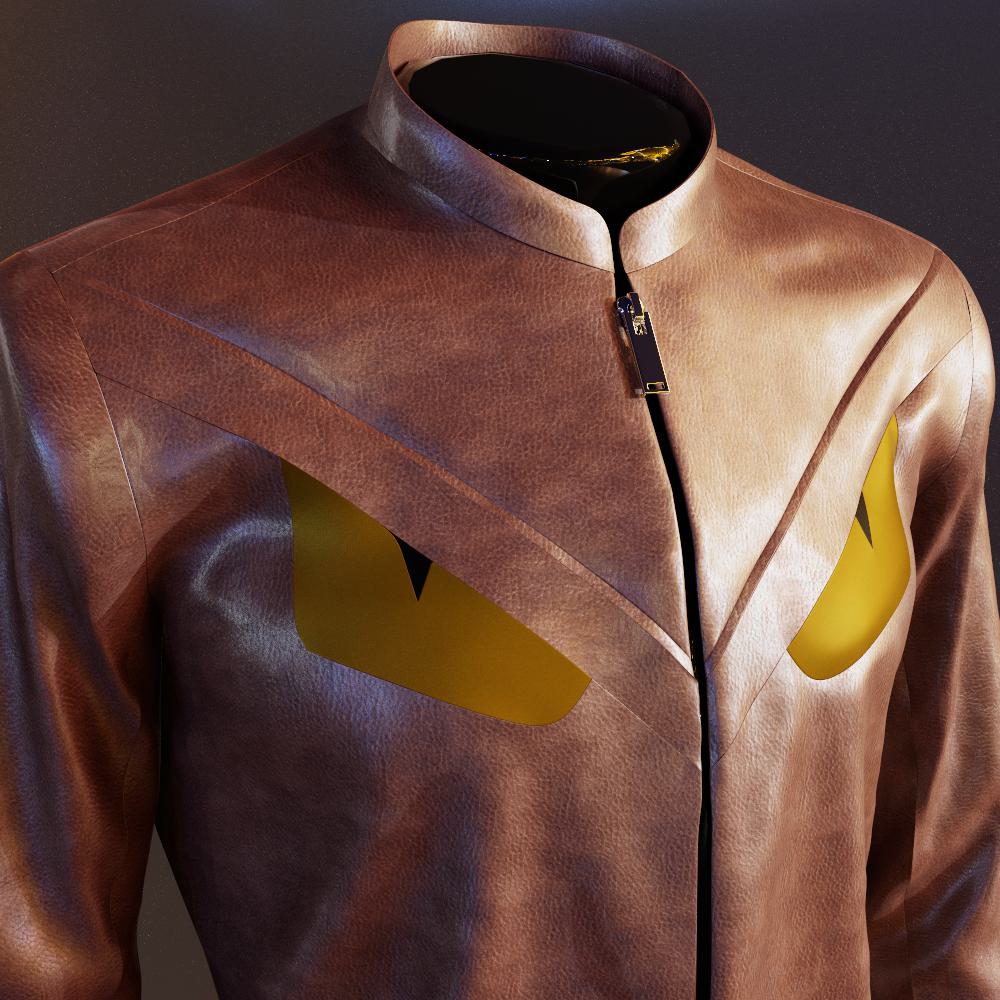}
            \end{overpic}
	\\
            &
            \begin{overpic}[width=0.24\columnwidth]{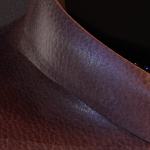}\end{overpic} 
            \begin{overpic}[width=0.24\columnwidth]{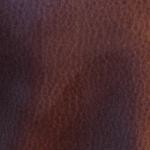}\end{overpic} &
            \begin{overpic}[width=0.24\columnwidth]{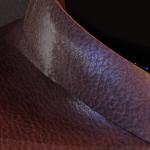}\end{overpic} 
            \begin{overpic}[width=0.24\columnwidth]{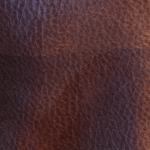}\end{overpic} &
            \begin{overpic}[width=0.24\columnwidth]{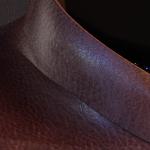}\end{overpic} 
            \begin{overpic}[width=0.24\columnwidth]{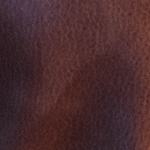}\end{overpic} &
            \begin{overpic}[width=0.24\columnwidth]{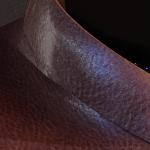}\end{overpic} 
            \begin{overpic}[width=0.24\columnwidth]{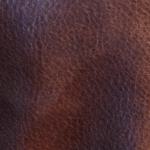}\end{overpic}
        \\
        \raisebox{0.8in}{\rotatebox[origin=c]{90} {\textsc{Stone04}}} &
    	\begin{overpic}[width=0.23\textwidth]{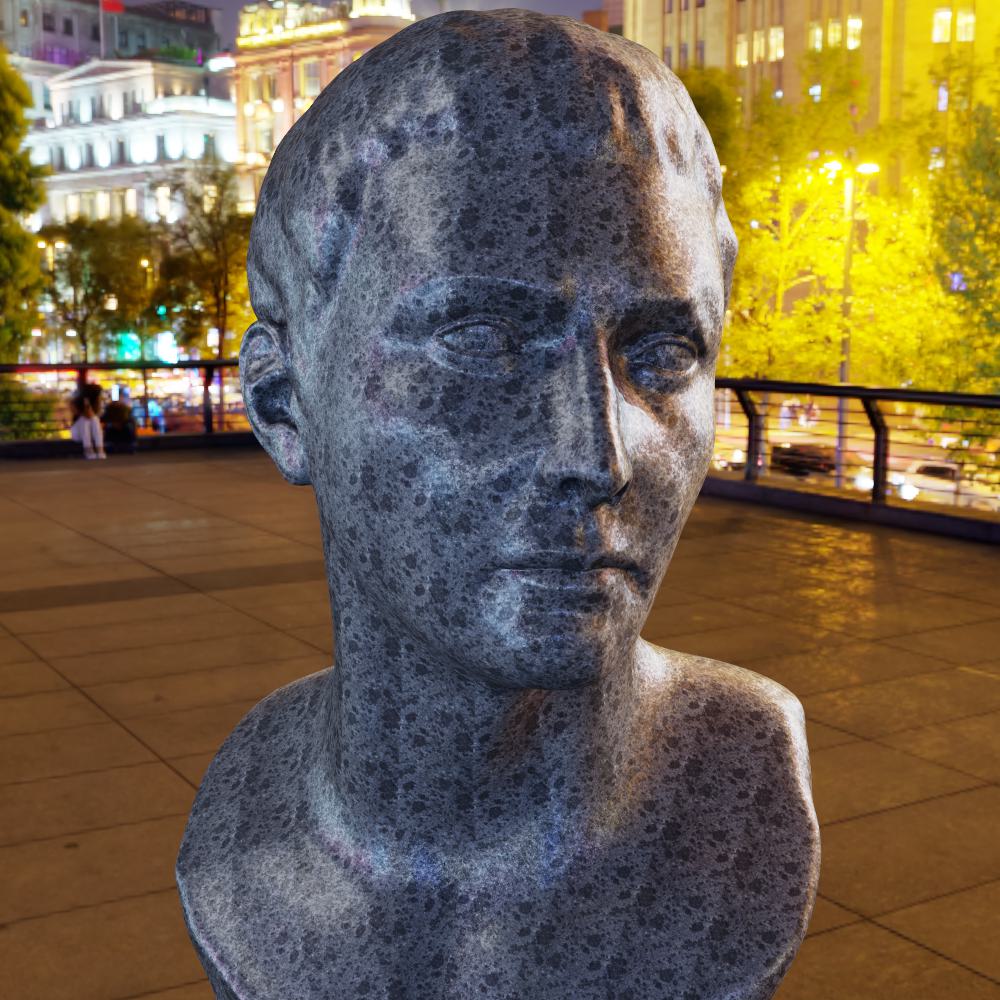}
            \end{overpic} &
            \begin{overpic}[width=0.23\textwidth]{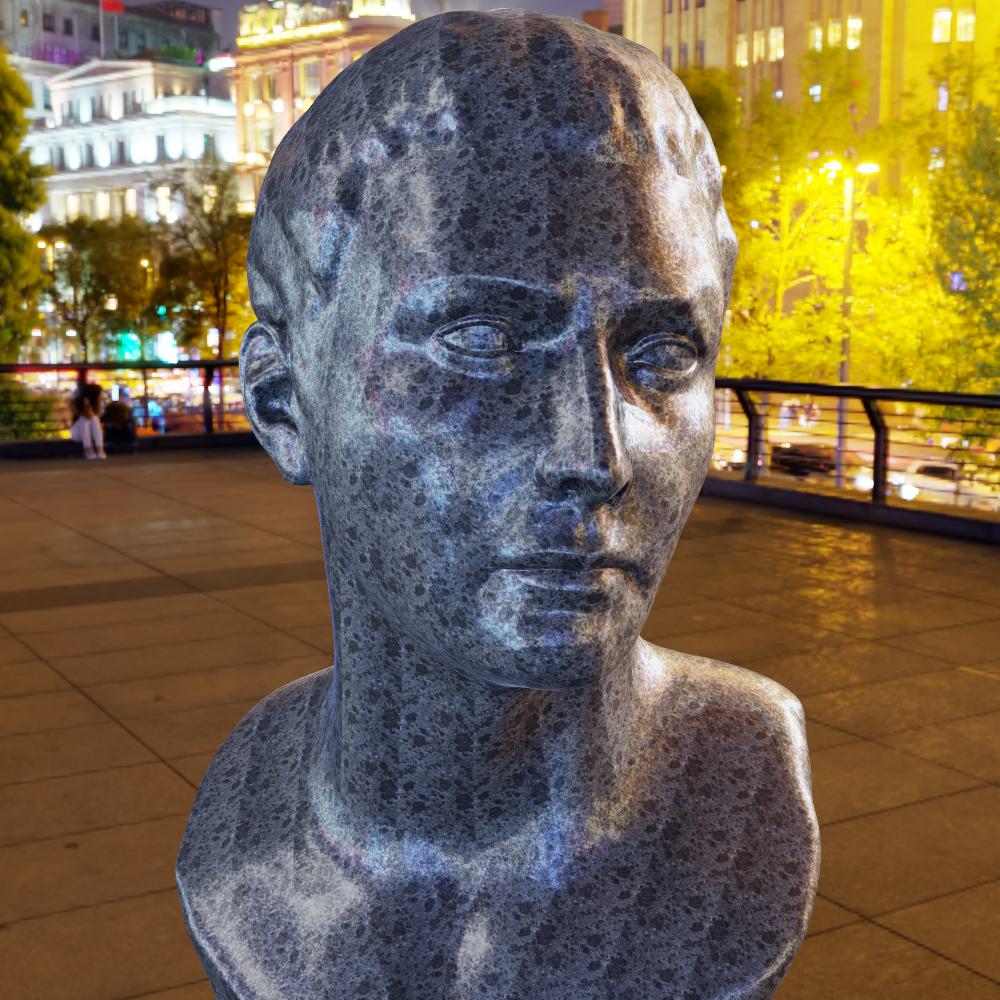}
            \end{overpic} &
            \begin{overpic}[width=0.23\textwidth]{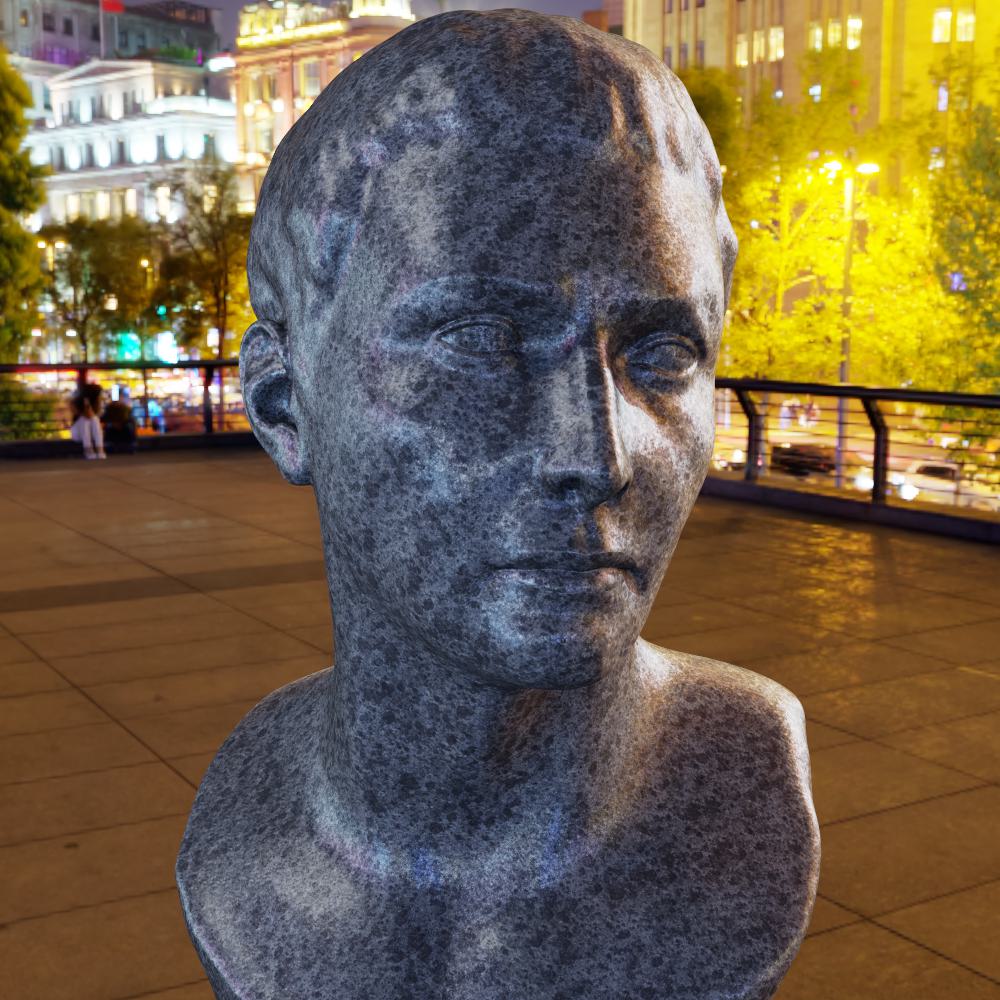}
            \end{overpic} &
            \begin{overpic}[width=0.23\textwidth]{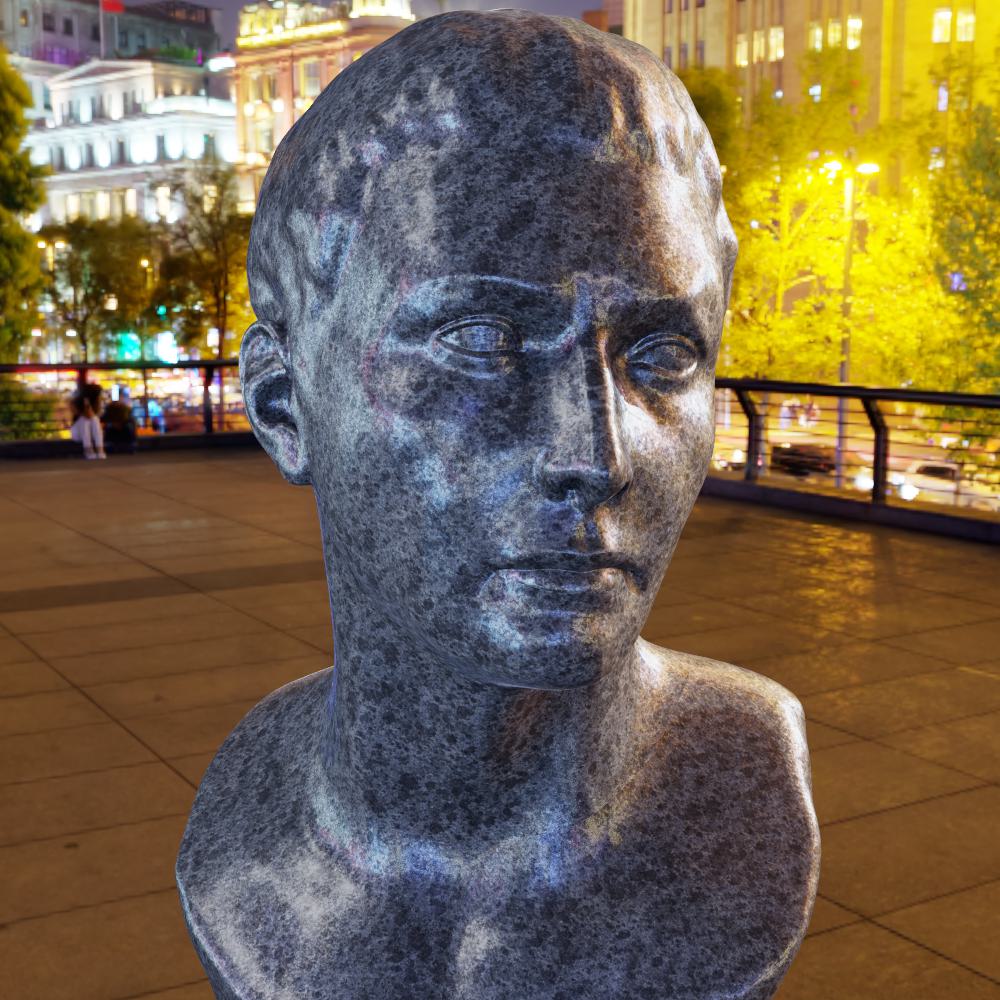}
            \end{overpic}
        \\
            &
            \begin{overpic}[width=0.24\columnwidth]{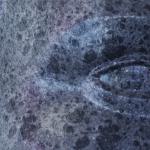}\end{overpic} 
            \begin{overpic}[width=0.24\columnwidth]{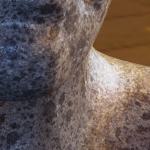}\end{overpic} &
            \begin{overpic}[width=0.24\columnwidth]{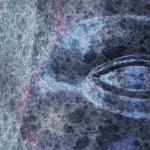}\end{overpic} 
            \begin{overpic}[width=0.24\columnwidth]{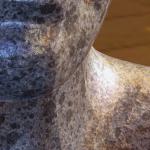}\end{overpic} &
            \begin{overpic}[width=0.24\columnwidth]{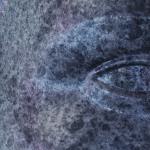}\end{overpic} 
            \begin{overpic}[width=0.24\columnwidth]{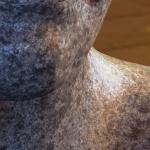}\end{overpic} &
            \begin{overpic}[width=0.24\columnwidth]{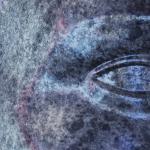}\end{overpic} 
            \begin{overpic}[width=0.24\columnwidth]{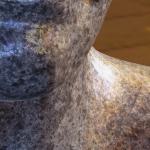}\end{overpic}
	\\
        & NeuMIP w/ repetitive tiling & Triple Plane w/ repetitive tiling & NeuMIP w/ hist. preserving &  Triple Plane w/ hist. preserving
	\end{tabular}
	\caption{
            \label{fig:comp_syn}
            We implement NeuMIP with histogram preserving blending (third column) to demonstrate our BTF synthesis scheme is general to any BTF dimension reduction method with an accessible positional feature plane.
            Our Triple Plane (second and fourth columns) shows great effectiveness in both representation and synthesis. NeuMIP, however, notably lacks high-frequency details, resulting in unsharp highlights (\textsc{Stone04}) and blurred reflections (\textsc{Leather03}). 
            %
        }\vspace{-3mm}
\end{figure*}

Based on the aforementioned considerations, we believe that the most appropriate solution is to compare with NeuMIP~\cite{kuznetsov:2021:neumip}, with and without using synthesis (repetitive tiling). As shown in Fig.~\ref{fig:comp_syn}.
Our Triple Plane (second and fourth columns) shows great effectiveness in both representation and synthesis. NeuMIP (first and third columns), however, notably lacks high-frequency details, resulting in unsharp highlights (\textsc{Stone04}) and blurred reflections (\textsc{Leather03}).
This is possible because our dimensional decomposition method completely separates the positional dimensions from the directional dimensions, whereas NeuMIP requires angular information as a condition for input, which might result in the directional information being implicitly mixed and stored with the positional features.


It is very important to emphasize that our BTF synthesis method is the first one enabling dynamic by-example synthesis on BTFs. Therefore, there is no ground truth, or rather, the ground truth should be by-example synthesizing directly on 6D BTFs and without compression. However, this is impossible since the high-dimensional extension of the by-example texture synthesis method does not exist, and it needs exponentially increasing numbers of example patches. Even if we only perform blending in the 2D positional domain, the remaining 4D angular dimensions may contain sharp lobes that blend to ghosting artifacts unless leveraging costly methods such as optimal transport~\cite{Bonneel:2011:ot}.

\subsection{Performance}
The evaluation time of our method via naive in-shader matrix multiplication takes 2.0 ms for 2,073,600 times queries ($1920 \times 1080$ resolution) on an RTX 4090 GPU. It includes both neural texture fetching cost and MLP inference cost with the overall cost scaling linearly with the number of evaluations. Currently, our MLP inference utilizes straightforward matrix multiplication in fp32 precision without specialized optimizations. Despite this, our method achieves the reported interactive performance. 


\section{Discussion \& Limitations}
\label{sec:limitation}
There are certain limitations to our scheme. As analyzed in Sec.~\ref{sec:overview}, our BTF synthesis scheme allows plug-and-play components for both representation and synthesis while we inherit the same limitations. We expect our method can directly benefit from a better choice of each component.

\paragraph{Representation Capability}
As shown in Fig.~\ref{fig:valid_rep} and Fig.~\ref{fig:comp_gt}, our Triple Plane produces slightly blurry results.
It demonstrates that the BTF with a complex appearance will be a challenge to our method. A better-designed decomposition approach may help solve this problem.

\paragraph{Structure-persevering Synthesis}
As shown in the second row of Fig.~\ref{fig:valid_hex}, with dynamic by-example texture synthesis (histogram-preserving blending~\cite{heitz:2018:histo} and Hex-Tiling~\cite{mikkelsen:2022:hextile}), one can not handle a highly structured BTF since the used texture synthesis methods can not.
We hope to find a dynamic synthesis method that has spatial variations and does not strictly repeat, e.g., dynamic Wang Tiling~\cite{wang:2020:example} or another by-example synthesis method that better preserves the structure.

\section{Conclusion \& Future Work}
In this paper, we have presented a by-example BTF synthesis scheme, allowing the dynamic synthesis of an infinitely large non-repetitive BTF from a small example BTF. We first introduce a novel dimension decomposition method (Triple Plane) that decomposes the high-dimensional 6D BTF into a series of 2D functions (including a 2D positional function and two 2D angular functions). Then, we design a simple but effective scheme that performs by-example texture synthesis on the decomposed 2D positional function as if it is a 2D texture. Finally, a lightweight MLP is employed to recover the synthesized BTF reflectance. Our results faithfully preserved the accurate appearance of the synthesized BTFs, and our method is robust in various types of BTFs.

A specifically designed importance sampling strategy is also in high demand, and one potential extension could be integrating normalizing flow~\cite{Xu:2023:neuSample} or histograms~\cite{Xu:2023:neuSample,zhu:2021:complexLum}.
Furthermore, in the future, it might be possible to explore a novel synthesis strategy that supports blending on structured patterns, which could further enhance our method.
In addition, we anticipate that runtime performance could be greatly enhanced through targeted optimizations such as batching inference, quantization, or utilizing hardware acceleration (e.g., tensor cores).



\bibliographystyle{ACM-Reference-Format}
\bibliography{combine, btf}

\end{document}


\newcommand{\todo}[1]{{\color{red}[TODO: #1]}}
\newcommand{\revise}[1]{{\color{blue}#1}}
\newcommand{\zilin}[1]{{\color{blue}#1}}
\newcommand{\dd}{\,\mathrm{d}}

\newcommand{\btf}{\mathbf{BTF}}
\newcommand{\bH}{\mathbf{H}}
\newcommand{\bU}{\mathbf{U}}
\newcommand{\bD}{\mathbf{D}}

\newcommand{\bx}{\mathbf{x}}
\newcommand{\by}{\mathbf{y}}
\newcommand{\bz}{\mathbf{z}}
\newcommand{\bh}{\mathbf{h}}
\newcommand{\bd}{\mathbf{d}}

\newcommand{\R}{\mathbb{R}}
\newcommand{\wi}{\mathbf{\bm{\omega}_i}}
\newcommand{\wo}{\mathbf{\bm{\omega}_o}}
\newcommand{\uv}{\mathbf{u}}

\newcommand{\FLIP}{\protect\reflectbox{F}LIP\xspace}

\title{Supplementary Material}

\maketitle

\begin{figure}[t]
	\centering
	\includegraphics[width=0.44\textwidth]{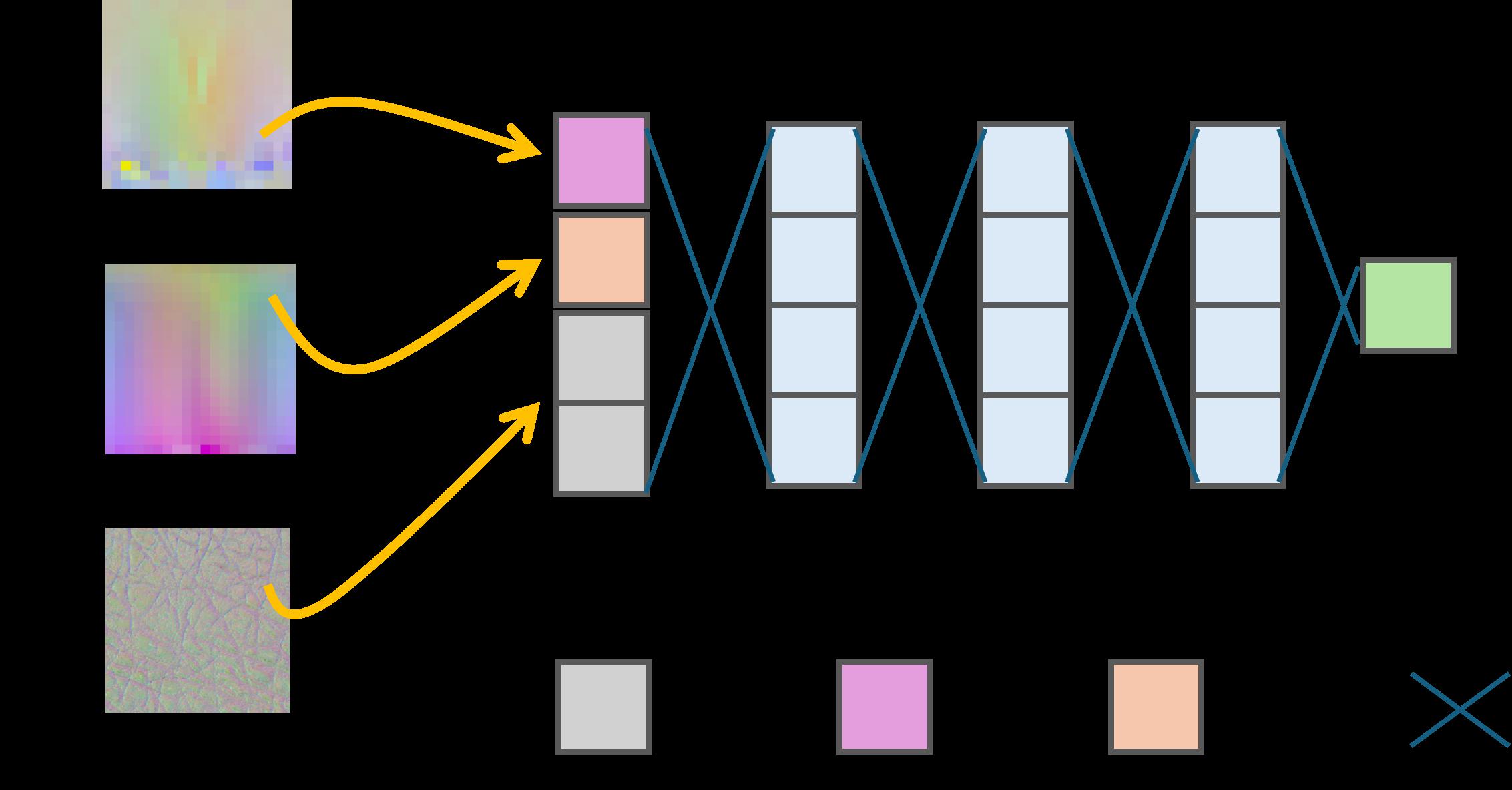}
	\caption{\label{fig:network}
        Our Triple Plane's network structure. It is a lightweight MLP with 4 fully connected (FC) layers. It takes an input of positional feature and directional features then outputs the RGB reflectance. We apply Leaky ReLU activation function to each layer.
        }
\end{figure}

\section{Implementation}
\paragraph{Training} We implement our network training on PyTorch~\cite{Paszke:2019:PyTorch}. The network structure is shown in Fig.~\ref{fig:network}. We apply AdamW with the default setting for the optimization process. The learning rate for the feature planes is $1e^{-3}$ and $3e^{-4}$ for the MLP. In Table~\ref{tab:dims}, we summarize the size of each feature plane, which follows Biplane~\cite{fan:2023:BTF}'s setting. Although the $20 \times 20$ resolution for directional planes may seem small, it is sufficient for UBO2014~\cite{weinmann:2014:ubo14} dataset with $151 \times 151$ angular resolution. The initialization strategy for the feature plane does not have a great impact on the result and we use the method from He et al.~\shortcite{he:2015:delving}. 
We circularly pad the phi axis of the angular feature planes ($f^{(\bH)}$ and $f^{(\bD)}$) to avoid discontinuity in the angular domain.
We use Leaky ReLU for all layers in the MLP.
In each iteration, we use a batch of $2,560,000$ ($16 \times 400 \times 400$) samples, i.e., 16 different BTF images, and each BTF image is under the same viewing and lighting condition. We train the model for 50 epochs, and after each epoch, the learning rate will decrease by a factor of $0.9$. The whole training takes about 2 hours on a single RTX 4090 GPU. Although the loss will continue decreasing after the 50th epoch, the improvement gain is minimal. Thus, stopping at the 50th epoch is sufficient.

\paragraph{Rendering Integration}
Our rendering is performed on the top of the NVIDIA Falcor~\cite{Kallweit:2022:falcor} renderer using its PathTracer pass with our modified BTF material. The inference of MLPs is implemented in Slang~\cite{he:2018:slang} shading language. We reorganized the matrix multiplication of MLP inference into a series of $\mathrm{float4x4} \times \mathrm{float4}$ inside the shader. The MLP's biases are also stored as $\mathrm{float4}$. Each neural texture is stored in a 2D $\mathrm{float4}$ texture array. The detailed performance is summarized in Table~\ref{tab:performance}.
We need to mention that our MLP inference is done inside the shader without the use of batching or hardware acceleration for neural network inference.
Integrating batch inference or utilizing GPU hardware inference acceleration structures, such as Tensor Cores, still has the potential to improve runtime performance.

\paragraph{Importance Sampling}
Importance sampling is crucial to efficient rendering. However, we are not focusing on developing a new importance sampling strategy, as we consider it orthogonal to our main objective. Instead, we employ a straightforward cosine-weighted hemisphere sampling to our method. Previous work as predicting a histogram distribution~\cite{zhu:2021:complexLum,Xu:2023:neuSample} or using normalizing flows~\cite{Xu:2023:neuSample} can be adapted to our method. 

\paragraph{Parallax Effects}
Incorporating offset~\cite{kuznetsov:2021:neumip} into neural materials provides a method for achieving a pseudo-3D effect on 2D surfaces. 
To obtain a parallax effect, previous work either implicitly estimates a 4D offset from BTFs~\cite{kuznetsov:2021:neumip,fan:2023:BTF} or explicitly creates a 7D synthetic SBTF (Silhouette BTF) dataset for training~\cite{kuznetsov:2022:neumip2}. 
Considering the measured BTF datasets do not include the offset data, obtaining the additional parallax effect involves an unstable, unsupervised learning process~\cite{kuznetsov:2021:neumip}. Consequently, we chose to exclude parallax effects from our method, a decision that does not impact the validity of our conclusions.

\begin{figure}[t]
	\centering
	\includegraphics[width=0.49\textwidth]{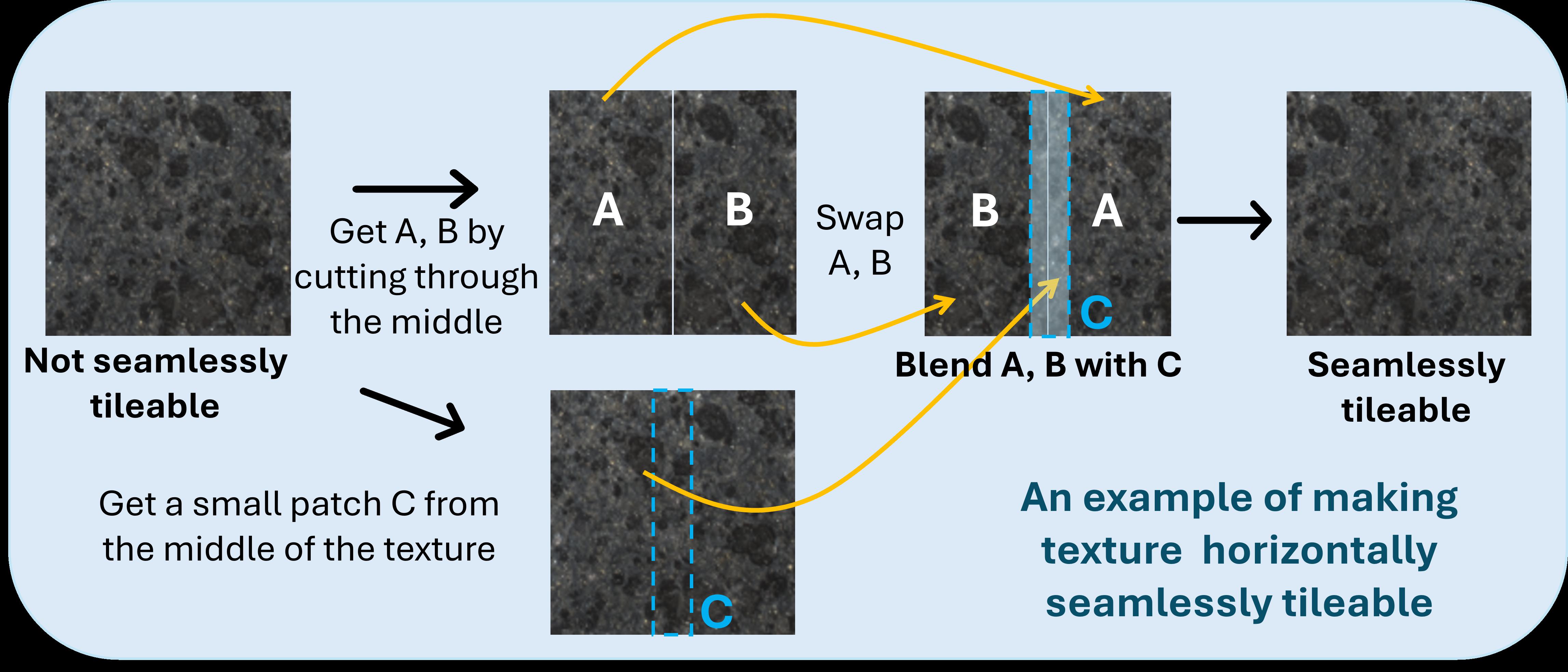}
	\caption{\label{fig:tileable}
         We first make BTF seamlessly tileable using a simple but effective method. This is an example of making texture horizontally seamlessly tileable, but in order to make BTF fully seamlessly tileable, it needs to be performed twice --- first horizontally and then vertically.
        }
\end{figure}

\paragraph{Avoiding Cuts in Synthesis}
By-example texture synthesis doesn't require the input texture to be seamlessly tileable. However, for some specific \emph{implementation}, which may require seamlessly tileable 2D texture. Otherwise, it will create some discontinuous cuts because the synthesis process may choose some patches that cross the example's edges. In our case, we need the input BTF to be seamlessly tileable to avoid these cuts.
We employ a trivial method that simply blends a small area around the edges to make BTF's positional plane seamlessly tileable without claiming the credit. as illustrated in Fig.~\ref{fig:tileable}. Moreover, making 6D BTF tileable in this way is only applicable within our proposed BTF synthesis scheme since blending the original 6D BTF is not trivial due to the high dimensionality.

\section{Limitation}
\begin{figure}[t]
	\centering
	\addtolength{\tabcolsep}{-3.5pt}
	\small
	\begin{tabular}{cc}
    	\begin{overpic}[width=0.49\columnwidth]{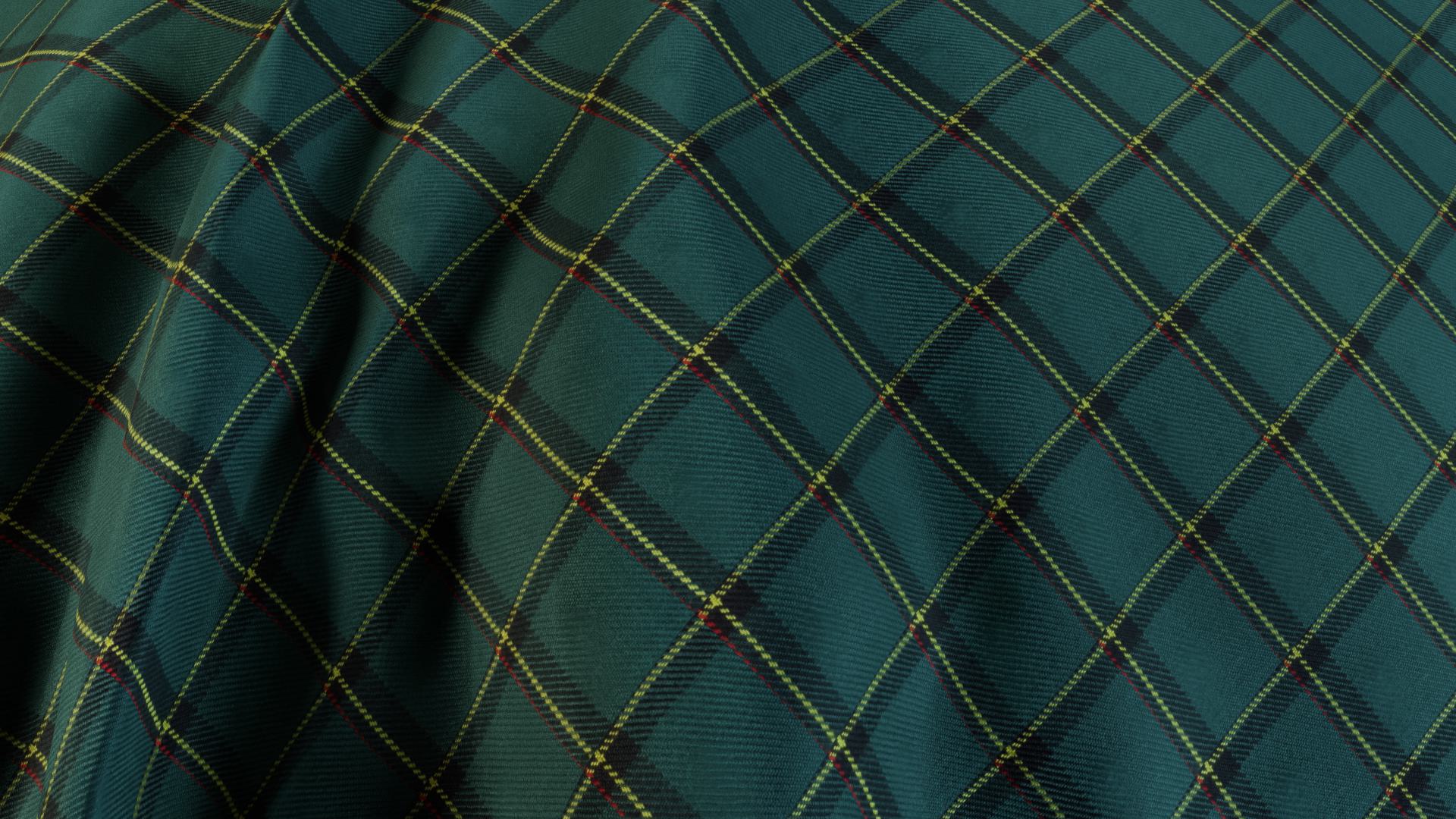}
            \put(2, 50){\color{white}\small{\textsc{Fabric09}}}
            \put(2, 2){\color{white}\small{Ours w/ repetitive tiling}}
            \end{overpic} &
            \begin{overpic}[width=0.49\columnwidth]{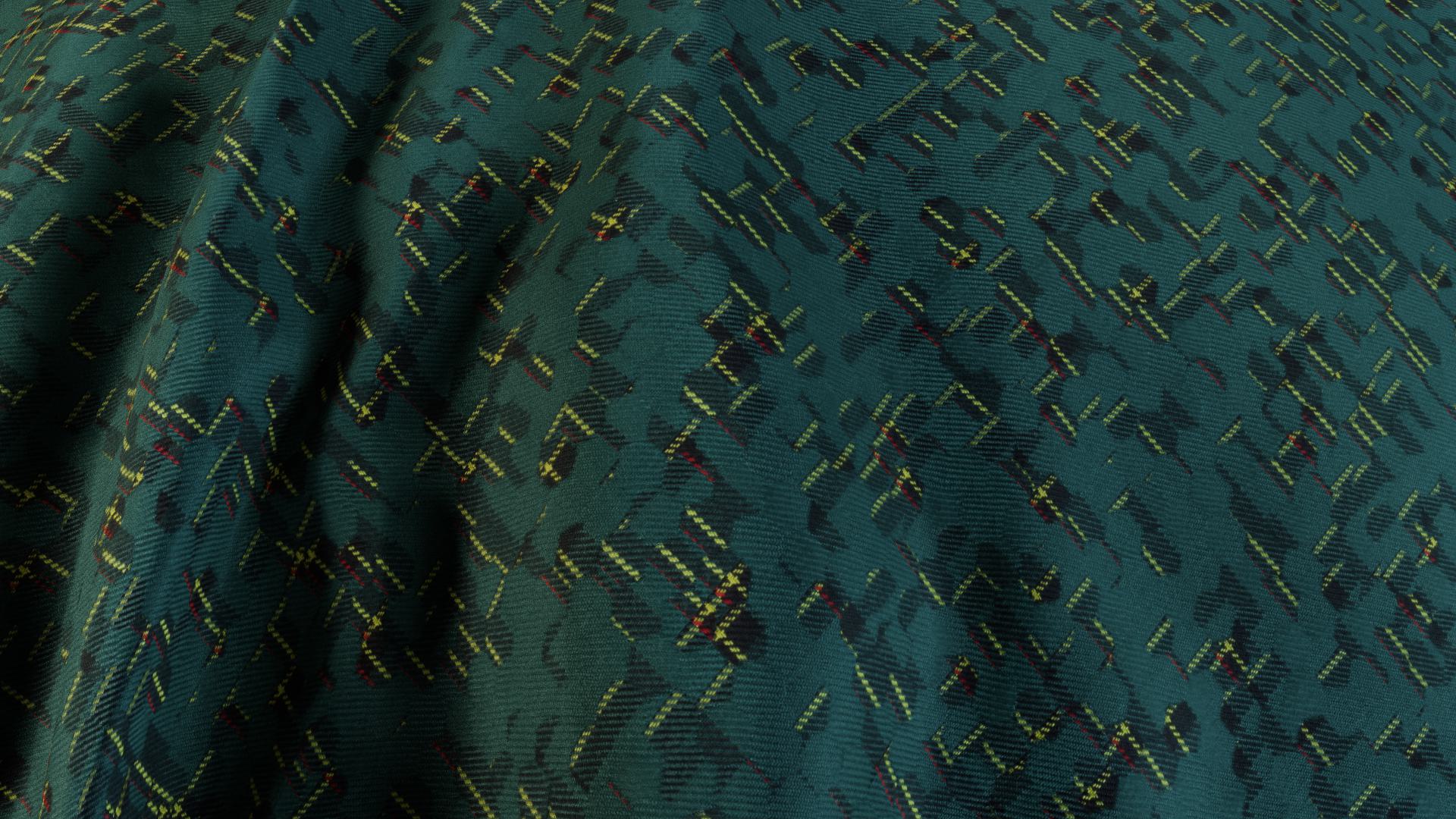}
             \put(2, 2){\color{white}\small{Ours w/ Hex-Tiling}}
                  \put(84, 40){\includegraphics[width=0.07\columnwidth]{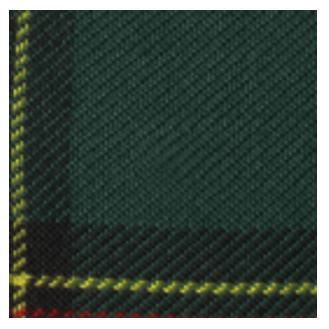}}
            \end{overpic} 
	\end{tabular}
	\caption{\label{fig:limitation}
        The example BTF may exhibit highly structured patterns. Applying dynamic texture synthesis methods sometimes ruins the structured patterns. This is an inherent limitation of dynamic texture synthesis. A better texture synthesis strategy may refine it.
	}
\end{figure}

\paragraph{Structure-persevering Synthesis}
One of the limitations we have is that we directly apply the existing by-example texture synthesis methods for the BTF synthesis. 
Therefore, we inherit the same limitations from the chosen texture synthesis approach. 
As shown in Fig.~\ref{fig:limitation}, our method can not handle a highly structured BTF since the used texture synthesis methods can not. Potential improvement may be obtained by applying a better texture synthesis strategy that can preserve the structures.
\begin{table}[t]
\caption{
    This table summarizes the dimensions of feature planes. The positional feature plane is set to match the dimension of BTF data~\cite{weinmann:2014:ubo14}. The directional plane's resolution follows Biplane~\cite{fan:2023:BTF}'s setting. Our decomposition method only takes $10$ MB storage for each 6D BTF, which originally needed hundreds of GBs without compression.
}
\begin{tabular}{cccc|c}
\toprule
& Height  & Width   & Channels  & Storage \\ \hline
$f^{(\bU)}$ & 400 & 400 & 16 & 10 MB \\ 
$f^{(\bH)}$ & 20  & 20  & 8  & 12.5 KB \\ 
$f^{(\bD)}$ & 20  & 20  & 8  & 12.5 KB \\ \hline
\multicolumn{3}{c}{Network parameters}   &   & 12.8 KB \\ \bottomrule
\end{tabular}

\label{tab:dims}
\end{table}

\begin{table}[t]
\caption{
    This table summarizes the performance of our method via a naive implementation of matrix multiplication inside the shader. The Eval. Time includes both neural texture fetching and MLP inference (once per pixel).
}
\begin{tabular}{cccc}
\hline
Resolution &   
        \begin{tabular}[c]{@{}c@{}} $1920 \times 1080$ \\  (1 SPP)\end{tabular} 
    &   \begin{tabular}[c]{@{}c@{}} $2560 \times 1440$ \\  (1 SPP)\end{tabular}  
    &   \begin{tabular}[c]{@{}c@{}} $3840 \times 2160$ \\  (1 SPP)\end{tabular}   \\ \hline
Eval. Time  & 2.0 ms & 4.8 ms & 8.4 ms \\
\hline
\end{tabular}

\label{tab:performance}
\end{table}

\bibliographystyle{ACM-Reference-Format}
\bibliography{combine, btf}